\documentclass[twocolumn,aps,prl,amsmath,floatfix,color]{revtex4}
\usepackage{graphicx}
\usepackage{colordvi}
\begin{document}
\title{Temperature and bath size in exact diagonalization \\ 
           dynamical mean field theory}
\author{Ansgar Liebsch$^1$ and Hiroshi Ishida$^2$}
\affiliation{$^1$Peter Gr\"unberg Institute and Institute of Advanced Simulation,
             Forschungszentrum J\"ulich, 52425 J\"ulich, Germany\\
             $^2$College of Humanities and Sciences, Nihon University,~Tokyo 156,
             Japan}  
\date{\today}
\begin{abstract}
Dynamical mean field theory (DMFT) combined with finite-temperature exact 
diagonalization is one of the methods to describe electronic properties 
of strongly correlated materials. Because of the rapid growth of the Hilbert space, 
the size of the finite bath used to represent the infinite lattice is severely 
limited. In view of the increasing interest in the effect of multi-orbital and 
multi-site Coulomb correlations in transition metal oxides, high-$T_c$ 
cuprates, iron-based pnictides, organic crystals, etc., it is appropriate to 
explore the range of temperatures and bath sizes in which exact diagonalization 
provides accurate results for various system properties. On the one hand, 
the bath must be large enough to achieve a sufficiently dense level 
spacing, so that useful spectral information can be derived, especially 
close to the Fermi-level. On the other hand, for an adequate projection   
of the lattice Green's function onto a finite bath, the choice of the 
temperature is crucial. The role of these two key ingredients in exact 
diagonalization DMFT is discussed for a wide variety of systems in order to 
establish the domain of applicability of this approach. Three criteria are used 
to illustrate the accuracy of the results: (i) the convergence of the self-energy 
with bath size, (ii)  quality of the discretization of the bath Green's function, 
and (iii) comparisons with complementary results obtained 
via continuous-time quantum Monte Carlo DMFT.
The materials comprise a variety of three-orbital and five-orbital systems, 
as well as single-band Hubbard models for two-dimensional triangular, square
and honeycomb lattices, where non-local Coulomb correlations are important.
The main conclusion from these examples is that a larger number of correlated 
orbitals or sites requires a smaller number of bath levels. Down to temperatures 
of 5 to 10 meV (for typical band widths $W\approx 2$~eV) two bath levels 
per correlated impurity orbital or site are usually adequate.\\
\mbox{\hskip1cm}   
\end{abstract}
\maketitle

\centerline{\bf Contents}
\vskip5mm

{    1. \ Introduction}

{    2. \ ED DMFT Formalism}

{    \mbox{\hskip5mm} 2.1. \ Multi-Band Correlations}

{    \mbox{\hskip5mm} 2.2. \ Multi-Site Correlations}

{    \mbox{\hskip5mm} 2.3. \ Size of Hilbert Space}  

{    \mbox{\hskip5mm} 2.4. \ Bath Discretization}

{    3. \ Results and Discussion}

{    \mbox{\hskip5mm} 3.1. \ Degenerate Three-Band Model}

{    \mbox{\hskip5mm} 3.2. \ Three-Band Materials}    

{    \mbox{\hskip5mm} 3.3. \ Degenerate Five-Band Model}

{    \mbox{\hskip5mm} 3.4. \ Triangular Lattice}

{    \mbox{\hskip5mm} 3.5. \ Square Lattice}

{    \mbox{\hskip5mm} 3.6. \ Honeycomb Lattice}

{    4. Summary and Outlook}

{    Acknowledgments} 

{    Appendix: \ Bath Discretization} 

{    \mbox{\hskip5mm} A.1. \ Degenerate Three-Band Model}

{    \mbox{\hskip5mm} A.2. \ Square Lattice}

{    References} 

\section{1. \ Introduction}

The study of strong Coulomb correlations is one of the central topics in 
condensed matter physics. A wide variety of phenomena, such as high-$T_c$
superconductivity in the cuprates, paramagnetic metal insulator transitions,
strong effective mass enhancements, and bad-metallic behavior  
are intimately related to electronic correlations not included in ordinary   
band theory. During the past two decades, dynamical mean field theory
\cite{metzner,hardtmann,georges,jarrell92,dmft,Phys.Today,RMP,held} (DMFT) 
has proved to be a highly useful scheme for the description of these kinds
of phenomena in many materials. The virtue of DMFT is that intricate 
single-electron properties, associated with complex lattice geometries and
unit cells, and many-electron interactions in partially filled, nearly 
localized atomic orbitals are treated on the same footing. Since the 
balance between kinetic energy and Coulomb repulsion depends in subtle
ways on important parameters, such as temperature, pressure, and 
chemical doping, a sufficiently detailed treatment of all key ingredients 
is essential. The conceptual advance of DMFT is that the problem 
of the correlated infinite lattice is mapped onto the problem of a 
correlated impurity immersed in a bath whose self-energy is determined via a 
self-consistent iterative procedure. The effective impurity may consist
of a single atom or a small cluster of atoms. Coulomb interactions within
this impurity are treated numerically exactly. Thus, dynamical fluctuations, 
which give rise to spectral weight transfer between low and high
energies, are fully taken into account. Spatial fluctuations, on the other
hand, are either neglected, as in local or single-site DMFT, or included
only at short distances, as in cluster extensions of DMFT 
\cite{hettler,lichtenstein,kotliar01,maier}.

The main computational task in DMFT lies in the solution of the quantum 
impurity problem for which techniques such as numerical renormalization 
group \cite{nrg,NRG} (NRG), quantum Monte Carlo \cite{qmc} (QMC), exact 
diagonalization \cite{ed} (ED), and other more approximate schemes can be 
used. For multi-band materials and clusters involving several atoms, the most 
accurate and versatile approaches are QMC and ED. The focus in the present 
work is on ED at finite temperatures
\cite{jaklic,aichhorn,prl2005,perroni,capone}. 
The key idea in ED is to discretize the continuum of the infinite lattice 
via a finite set of bath levels and to solve the many-body problem involving 
correlated impurity and finite bath exactly. 
Originally applied to the one-band Hubbard model \cite{ed},
it soon was realized that this approach is limited to a small number of bath levels   
because of the exponentially growing Hilbert space. The usefulness of ED DMFT for 
strongly correlated materials involving several orbitals or sites 
was therefore far from evident.

The purpose of this work is to demonstrate that finite-temperature ED DMFT is 
in fact a highly useful and accurate many-body scheme that is applicable to 
systems including (so far) up to five correlated orbitals and clusters up to 
six correlated sites. Typically, this implies two to three bath levels per 
orbital or site. In certain cases, only one bath level per orbital or site was 
found to give satisfactory results. This conclusion seems surprising in view
of the fact that, to obtain well converged results for the single-band Hubbard
model within local DMFT, at least three to five bath levels must be taken into 
account \cite{dmft}. 
An analogous five-orbital system would imply a total number of 20 to 30 levels, 
a virtually insurmountable and impractical task. The main point of our study 
is that an increasing number of correlated orbitals or sites requires a 
decreasing number of bath levels, so that, for the systems discussed
in this work, overall sizes of 10 to 15 levels are adequate. 
Since computational times per iteration are of the order of minutes for 
10 levels and a few hours for 15 levels, this implies that a large variety of 
strongly correlated materials of interest can be studied within ED DMFT. 
Moreover, this approach is applicable at large Coulomb energies and for full 
Hund exchange. Also, it is free of sign problems and statistical errors. 
We therefore conclude that finite-temperature ED DMFT is complementary to 
continuous-time QMC DMFT \cite{rubtsov,werner,gull}. While the latter becomes
computationally more demanding at low temperature because of statistical noise, 
the former gets more involved at high temperature due to the increasing number 
of excited states. Moreover, the low-temperature, low-energy properties are 
affected by finite-size limitations.      

As discussed in Section 2, at each iteration step ED DMFT implies 
two projections: from the infinite lattice to the finite cluster, and back from 
the cluster to the lattice. Mathematically, these projections assume that  
continuous and discrete real-energy spectra of the self-energy and Green's 
function exist which have nearly identical representations along the Matsubara 
axis. To demonstrate the accuracy of this approach three criteria are available: 
(i) the convergence of key quantities, such as the self-energy, with 
increasing bath size, (ii) the accuracy of the discretization of the 
lattice bath Green's function in terms of a cluster consisting of impurity 
and finite bath (the impurity may be formed by several sites), and (iii) 
the consistency with analogous results derived within QMC DMFT. In the case
of three-band systems, calculations for two, three, and four bath levels 
per orbital have been performed. The comparisons for different bath sizes,
the quality of the Green's function discretization, and the comparison 
with QMC results illustrate that not only integrated quantities, such as 
orbital occupancies, but also dynamical features are usually well described 
with only two bath levels per orbital or site. 

The successful application of ED DMFT involves two main ingredients: (i) the 
size of the Hilbert space and (ii) the discretization of the lattice via a 
finite cluster. The size of the Hilbert space determines the level spacing of 
excited states. This spacing must be sufficiently small, so that the low-energy
behavior of the self-energy can be resolved. Of particular interest are
correlation-induced effective mass enhancements, deviations from Fermi-liquid 
properties related to finite low-energy scattering rates, and the opening of 
Mott gaps. The accuracy of the discretization, on the other hand, depends 
in a crucial manner on the temperature used along the imaginary Matsubara axis. 
Decreasing this temperature implies increasing sensitivity to low-energy 
features of the lattice Green's function which require baths of accordingly   
larger size. As the cluster Green's function is always gapped, it cannot describe 
metallic behavior at very low energies. The examples discussed in the subsequent 
sections indicate that, to achieve accurate results with two bath levels per 
orbital or site, the Matsubara temperature should not be much lower than 
$T\approx 5,\ldots,10$~meV (assuming typical band widths $W\approx2$~eV). 
                      
In the past, several groups have carried out multi-band or multi-site ED DMFT 
calculations at $T=0$ \cite{bolech,capone2004,civelli,kyung,merino,koch,sakai,senechal}.
In these studies, the impurity Green's function is evaluated using the Lanczos 
method for the ground state while the DMFT iteration procedure and the discretization 
of the bath Green's function are performed at a fictitious finite temperature 
to obtain sufficiently smooth distributions along the Matsubara axis. In the 
present work we focus mainly on ED at finite temperature in order to demonstrate 
the consistency of this approach with analogous results derived within 
continuous-time QMC DMFT. As in QMC, the temperature for the Matsubara grid is then 
the same as for the interacting impurity Green's function. At low temperatures this
function can in principle be obtained by generalizing the Lanczos scheme to finite 
$T$. This approach, however, is rapidly plagued by the loss of orthogonality of 
excited states \cite{capone}. Here we make use of the Arnoldi algorithm 
\cite{arnoldi} for large sparse matrices, which is ideally suited to evaluate 
those excited states that are physically important at low temperatures 
\cite{perroni}. Whereas finite-$T$ ED with full matrix diagonalization is 
feasible only up to about $8$ levels \cite{prl2005} (maximum Hamiltonian dimension 
$N=4900$), the Arnoldi scheme at present allows the consideration of $12$ to 
$15$ levels ($N\approx 10^6,\ldots,40\times 10^6$) without significantly increased
storage requirements \cite{FeAsLaO}. Thus, materials involving up to five 
correlated orbitals or up to six correlated sites can now be investigated.    

As the ED DMFT iterative procedure is carried out at imaginary frequencies, 
an additional step is required to extrapolate the converged Green's function 
and self-energy to real $\omega$. In principle, these quantities can,
of course, be evaluated directly at $\omega+i\gamma$, where $\gamma$ denotes 
an artificial broadening. This aspect represents an important advantage compared 
to QMC DMFT, where real-energy spectra are usually obtained via the Maximum
Entropy method \cite{maxent}. On the other hand, the ED cluster properties are 
by definition discrete, so that the imaginary parts of the Green's function and
self-energy consist of (broadened) $\delta$-functions. These cluster spectra are 
adequate, for instance, if a metallic phase with many closely spaced peaks is
compared to an insulating phase, which exhibits a clear gap. Mott transitions
can therefore readily be identified via cluster spectra. To investigate finer  
features, such as the distinction between Fermi-liquid and non-Fermi-liquid
behavior, it is useful (as in QMC) to analytically continue the lattice 
quantities from the Matsubara axis to real $\omega$. 
Although this problem is known to be ill-posed, the important low-energy 
properties are usually rather insensitive to details of the extrapolation 
method, while large uncertainties exist in the incoherent region farther 
away from the Fermi energy. To illustrate some of these issues, cluster 
and lattice spectra will be compared in Section 3 for several systems.               
                       
Multi-orbital and multi-site Coulomb correlations are treated in this work 
within a common perspective since they share many computational aspects. 
The similarity becomes particularly evident by changing from the site basis
in the case of cluster DMFT 
to a ``molecular-orbital'' or ``plaquette'' basis in which the onsite repulsion
is converted into intra-orbital and inter-orbital Coulomb and exchange   
interactions. An interesting outcome of this parallel treatment is that 
characteristic physical phenomena, such as the transition from a Mott 
insulating phase at half-filling to non-Fermi-liquid and Fermi-liquid 
metallic phases at finite electron or hole doping, are found in multi-orbital 
materials as well as single-band models of high-$T_c$ cuprates \cite{FeAsLaO}.
Thus, the existence of inter-orbital or inter-site interaction channels 
generates fascinating new physics not present in a single-band, single-site 
picture.      

The outline of this paper is as follows. In Section 2 we review the 
finite-temperature ED DMFT approach for the evaluation of multi-band and
multi-site Coulomb correlations. We also discuss the role of the two main
ingredients in this approach, namely, the overall size of the Hilbert 
space of the finite cluster used to represent the infinite lattice, and the 
discretization of the bath Green's function. Section 3 presents the 
results for a variety of applications: (1) the degenerate three-band model,
(2) several realistic three-band materials, such as Ca$_{2-x}$Sr$_x$RuO$_4$,
LaTiO$_3$, V$_2$O$_3$, and Na$_x$CoO$_2$, (3) the degenerate five-band 
model and its connection to iron-based pnictides and chalcogenides, the 
single-band Hubbard model for (4) the triangular lattice, (5) the square 
lattice, and (6) the honeycomb lattice. We have also applied finite-$T$ ED 
within inhomogeneous DMFT \cite{potthoff} to study both single-site and 
non-local correlations 
in various heterostructures \cite{inhdmft}, but these will not be discussed 
in this paper. Section 4 provides a summary and outlook. In the Appendix
the discretization of the bath Green's function is discussed in more detail
for the degenerate three-band model and the square lattice.        
Energy units are eV unless noted otherwise.

\section{2. \ ED  DMFT formalism}
\subsection{2.1. \ Multi-band Correlations}
 
In this section we specify the details of ED DMFT appropriate for multi-band   
materials. In particular, we consider transition metal oxides, such as 
V$_2$O$_3$, LaTiO$_4$, and  Ca$_2$RuO$_4$, where, as a result of crystal field
splitting, Coulomb correlations take place primarily within the partially 
occupied $t_{2g}$ bands. In addition to these three-band systems, we also 
consider the more general case in which all five $d$ bands are relevant near
the Fermi level. This applies to ordinary transition metals, oxides such as
MnO or LaMnO$_3$, and iron pnictides and chalcogenides, such as FeAsLaO and
FeSe. We focus on the influence of onsite Coulomb correlations and neglect 
spatial fluctuations. Within this multi-band single-site DMFT approach the 
complex self-energy accounts for dynamical correlation effects, such as spectral 
weight transfer between low-energy and high-energy regions and quasi-particle
broadening, while its momentum dependence is neglected.                

The single-particle properties are assumed to be given in terms of an effective  
tight-binding Hamiltonian $t({\bf k})$ which is derived from a fit to results 
obtained within density functional theory (DFT).
The interacting Hamiltonian is given by:
\begin{eqnarray}
   H &=& \sum_{mn{\bf k}\sigma} t_{mn}({\bf k}) c^+_{{\bf k}m\sigma}
                  c_{{\bf k}n\sigma}  
        + \sum_{im} U n_{im\uparrow} n_{im\downarrow}  \nonumber\\
     &-& \sum_{im\sigma} \mu n_{im\sigma} +    
  \sum_{i m< m'\sigma\sigma'}\!\!\!\! (U'-J\delta_{\sigma\sigma'}) 
                 n_{im\sigma} n_{im'\sigma'}                \nonumber\\
  &-& \sum_{im\ne m'}\!\!\! J [c_{im\uparrow}^+ c_{im\downarrow}
            c_{im'\downarrow}^+ c_{im'\uparrow}   
           + c_{im\uparrow}^+ c_{im\downarrow}^+    
            c_{im'\uparrow} c_{im'\downarrow}],  \nonumber\\
                                               \label{H}
\end{eqnarray}
where $c_{im\sigma}^{(+)}$ are annihilation (creation) operators 
for electrons on site $i$ in orbital $m$ with spin $\sigma$  and
$n_{im\sigma}=c_{im\sigma}^+ c_{im\sigma}$. $c_{{\bf k}m\sigma}^{(+)}$ 
are the corresponding Fourier components. $\mu$ is the chemical potential 
and the intra-orbital and inter-orbital Coulomb energies are denoted by 
$U$ and $U'$, respectively. The exchange integral is $J$, where $U'=U-2J$ 
because of rotational invariance \cite{sugano}.

For simplicity we limit the discussion to systems in which non-diagonal
components of the density of states vanish for symmetry reasons. The 
diagonal elements of the lattice Green's function are then given by
\begin{equation}
   G_{m} (i\omega_n)  = \sum_{\bf k} 
   [i\omega_n + \mu - t({\bf k})  - \Sigma(i\omega_n)]^{-1}_{mm} ,
                                                   \label{G}
\end{equation}
where $\omega_n=(2n+1)\pi/\beta$ are 
Matsubara frequencies ($\beta=1/T$), and $\Sigma$ is the self-energy matrix.  
Since local Coulomb interactions preserve the symmetry properties of $G$, 
$\Sigma$ is also diagonal in orbital space, with elements $\Sigma_m(i\omega_n)$. 
We consider only paramagnetic systems, so that the spin index of $G$ and 
$\Sigma$ can be omitted. Note that, because of the non-diagonal nature of 
$t({\bf k})$, the components of $G$ are influenced by all components of 
$\Sigma$. In the limit of large $\omega_n$, the lattice Green's function
components take the simple form
\begin{equation}
 G_{m}(i\omega_n) = [i\omega_n + \mu - \epsilon_m  - 
       \Sigma_m(i\omega_n)  ]^{-1} ,
\label{Gasym1}
\end{equation}
where $\epsilon_m  = [\sum_{\bf k} t({\bf k})]_{mm}$. 

For the purpose of the quantum impurity calculation within DMFT it is
necessary to remove the self-energy from the central site in order to 
avoid double-counting of Coulomb interactions. This step yields the 
so-called bath Green's function \cite{cavity}      
\begin{equation}
G_{0,m}(i\omega_n)=[G_m(i\omega_n)^{-1}+\Sigma_m(i\omega_n)]^{-1}.
                                                      \label{G0}
\end{equation} 
Note that $G_0$ is not the non-interacting Green's function obtained from 
Eq.~(\ref{G}) for $\Sigma=0$. Instead, $G_0$ accounts for the electronic motion 
from the central site throughout the lattice back to the central site, where 
$\Sigma$ is present at all sites except the central one.

Within ED DMFT the infinite lattice surrounding
the central site is approximated via a bath of finite size. Thus, the above 
Green's function is approximated in terms of an Anderson impurity model for a 
cluster consisting of $n_c$ impurity levels $\varepsilon_{m=1,\ldots, n_c}$ 
and $n_b$ bath levels $\varepsilon_{k=n_c+1,\ldots, n_s}$, which hybridize via 
hopping matrix elements $V_{mk}$. Here, $n_s=n_c+n_b$ defines the total number 
of cluster levels. The bath levels $\varepsilon_{k}$ and hopping terms $V_{mk}$ 
may be viewed as auxiliary quantities to achieve an optimal representation of 
the lattice bath Green's function in terms of a finite cluster.  
(The term ``cluster'' here refers to the finite set of impurity and bath levels 
and should not be confused with the ``correlated impurity cluster'' discussed 
in the next subsection.) Thus,
\begin{equation}
     G_{0,m}(i\omega_n) \approx  G^{cl}_{0,m}(i\omega_n), 
 \label{G0m}
\end{equation} 
where
\begin{equation}
   G^{cl}_{0,m}(i\omega_n)    = \Big(i\omega_n + \mu - 
         \varepsilon_m - \sum_{k=n_c+1}^{n_s} 
         \frac{ \vert V_{mk} \vert^2 }
         {i\omega_n - \varepsilon_{k}}\Big)^{-1}  .
                                                       \label{G0cl}
\end{equation} 
(The bath levels are defined relative to $\mu$.)  
Since the lattice Green's function is diagonal in the orbital index, 
we can assume that each impurity orbital couples only to its own bath. For 
instance, for $n_c=4$ and $n_b=8$ (i.e., two bath levels per impurity orbital), 
only the elements $V_{m,m+4}$, $V_{m,m+8}$, and their transpose elements 
are nonzero. All other elements vanish. These interactions are schematically 
indicated in the left panel of Figure 1. Details concerning the discretization 
indicated in Eq.~(\ref{G0cl}) will be discussed in Section 2.4.   

\begin{figure} 
\begin{center}
\includegraphics[width=3.5cm,angle=-90]{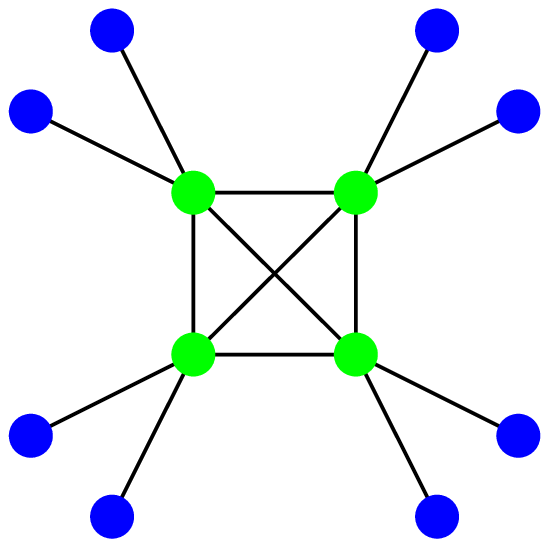}
\includegraphics[width=3.5cm,angle=-90]{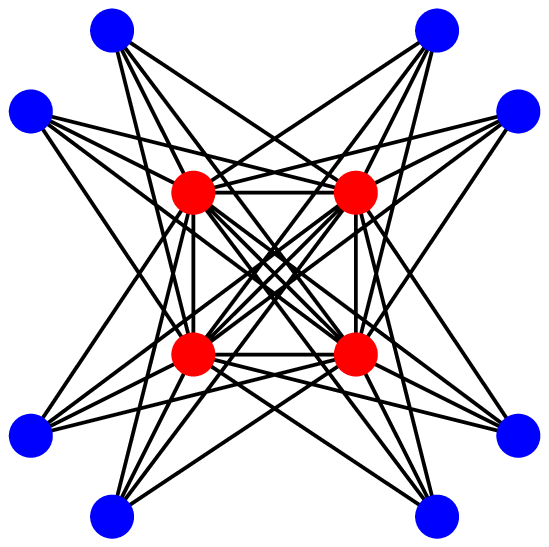}
\end{center}
\caption{(Color online)
Left panel: Impurity bath interactions within diagonal basis for four orbitals. 
Green dots: impurity levels ($m=1,\ldots,4$); blue dots: bath levels 
($k=5,\ldots,12$). Each impurity orbital interacts with its own bath via 
hopping terms $V_{mk}$. The impurity orbitals interact only via Coulomb and 
exchange terms since the system is diagonal in this basis.    
Right panel: Non-diagonal mixed basis involving impurity sites ($i=1,\ldots,4$) 
(red dots) and molecular-orbital bath levels (blue dots). Interactions among 
sites are determined by single-particle hopping, whereas the Coulomb interaction 
is purely onsite. Bath levels now hybridize with all impurity sites.    
The interactions indicated in the left and right panels are equivalent, 
since they differ only by a unitary transformation among impurity orbitals.     
}\label{bath}\end{figure}

Adding the onsite Coulomb interactions to the single-particle  
Hamiltonian specified by $\varepsilon_m$, $\varepsilon_k$, and $V_{mk}$,
the interacting cluster Green's function at finite $T$ is derived from the 
expression \cite{perroni,capone}
\begin{eqnarray}
 G^{cl}_{m}(i\omega_n)
   &=& \frac{1}{Z} \sum _{\nu\mu} e^{-\beta E_\nu}\Big( 
        \frac{\vert\langle \mu|c_{m\sigma}^+|\nu \rangle \vert^2}
                             {i\omega_n + E_\mu - E_\nu }  \nonumber \\  
  && \hskip10mm + \ \ \frac{\vert\langle \mu|c_{m\sigma}  |\nu \rangle \vert^2}
                             {i\omega_n + E_\nu - E_\mu }  \Big)\\  
   &=&  \frac{1}{Z} \sum_{\nu}\,e^{-\beta E_\nu}\,
      \!\![ G^{cl+}_{m,\nu}(i\omega_n) +  G^{cl-}_{m,\nu}(i\omega_n)],  
    \nonumber 
\label{Gclm}
\end{eqnarray}
where $E_\nu$ and $|\nu \rangle$  denote the eigenvalues and eigenvectors of 
the full Hamiltonian, and $Z=\sum_\nu {\rm exp}(-\beta E_\nu)$ is the 
partition function. $G^{cl\pm}_{m,\nu}(i\omega_n)$ are Green's functions 
derived from the excited states $\vert\nu\rangle$. $G^{cl}$ satisfies the 
same symmetry properties as $G$ and $G_{0}$, so that it is also diagonal in 
orbital space. The orbital components of the cluster self-energy can be 
obtained from an expression analogous to Eq.~(\ref{G0}):
\begin{equation}
\Sigma^{cl}_{m}(i\omega_n) = G^{cl}_{0,m}(i\omega_n)^{-1}
                            -G^{cl}_{m}(i\omega_n)^{-1} .
\label{Scl}
\end{equation} 

The key assumption in DMFT is that this cluster self-energy is 
a physically reasonable representation of the lattice self-energy. Thus, 
\begin{equation}
     \Sigma_{m}(i\omega_n) = \Sigma^{cl}_{m}(i\omega_n) ,  
\label{S}
\end{equation}
so that, as a result of Eq.~(\ref{G0m}),
\begin{equation}
     G_{m}(i\omega_n) \approx G^{cl}_{m}(i\omega_n) .  
\label{GmGmcl}
\end{equation}
In the next iteration step, these diagonal self-energy components are used 
as input in the lattice Green's function, Eq.~(\ref{G}). The iteration cycle
then has the schematic form
\begin{eqnarray}\Blue{
 \varepsilon_{m},\,\varepsilon_k,\,V_{mk} 
  &\rightarrow&  G^{cl}_{0,m},\,G^{cl}_{m},\,\Sigma^{cl}_{m}\nonumber\\ 
  &\rightarrow&  \Sigma_{m},\, G_{m},      \,G_{0,m} }       \\  \Blue{
  &\rightarrow&  \varepsilon_m,\,\varepsilon_{k},\,V_{mk} .}  \nonumber
\label{iteration}
\end{eqnarray}
In practice, the initial cluster parameters $\varepsilon_{m}$, $\varepsilon_{k}$, and 
$V_{mk}$ are obtained from a fit of the uncorrelated lattice Green's function
or from a previous solution for weaker interaction parameters, other doping 
concentrations, etc.  

The ED DMFT self-consistency procedure involves at each iteration 
two projections: from lattice to cluster, and back from cluster to lattice. 
By definition, lattice quantities such as $G$, $G_0$, and 
$\Sigma$ have continuous spectra at real energies, while the analogous
cluster quantities, $G^{cl}$, $G_0^{cl}$, and $\Sigma^{cl}$ are discrete.
In this sense, the cluster quantities are never equal to the corresponding 
lattice versions, except for an infinite number of bath levels. 
The advantage of working at finite $T$ is that, along the Matsubara axis at 
not too low temperatures, both sets of quantities become rather smooth, so 
that they can be meaningfully compared, even if only a finite number of bath
levels is employed.     

Mathematically, the lattice-to-cluster and cluster-to-lattice projections in 
Eqs.~(\ref{G0m}) and (\ref{S}) imply that continuous and discrete real-energy 
spectra exist which have nearly identical representations along the Matsubara 
axis. In particular, the success of ED DMFT as a method for describing strongly 
correlated properties depends crucially on the accuracy of the discretization 
of $G_0$, Eq.~(\ref{G0m}), which, in turn, depends on bath size and temperature. 
This will be discussed in more detail in Section 2.4. The quality of this 
discretization will also be illustrated in Section 3 for a variety 
of realistic systems. Additional details concerning the finite-temperature 
multi-orbital ED DMFT approach can be found in Ref.~\cite{perroni}.

Comparing the ED DMFT formalism outlined above to QMC DMFT, it is clear that
both schemes involve additional errors beyond the basic single-site DMFT
approximation: In the case of ED, the bath Green's function $G_0$ is discretized,
so that the interacting impurity Green's function suffers from finite-size effects.
In the case of QMC, the discretization is avoided, but the interacting impurity 
Green's function suffers from errors due to statistical uncertainties.     
As a result, the limit of very low temperatures and energies is inaccessible 
to both schemes. 
    
We also note that the iterative procedure indicated in Eq.~(\ref{iteration}) 
is not unique. For instance, in the definition of the cluster self-energy, 
Eq.~(\ref{Scl}), one could replace $G_{0,m}^{cl}$ by the more accurate lattice
function $G_{0,m}$. Nevertheless, Eq.~(\ref{Scl}) is found to yield a more
reliable convergence towards a self-consistent solution. In addition, instead
of finding the cluster parameters $\varepsilon_{m}$, $\varepsilon_{k}$, and 
$V_{mk}$ in the $(i+1)^{th}$ iteration step from the actual lattice bath Green's 
function $G_{0,m}$ as indicated in Eq.~(\ref{G0m}), it is more convenient to 
employ an admixture
\begin{equation}
   G^{cl,(i+1)}_{0,m} \approx  \alpha G_{0,m} + (1-\alpha) G^{cl,(i)}_{0,m},  
 \label{alfa}
\end{equation} 
where $G^{cl,(i)}_{0,m}$ is the cluster bath Green's function of the previous 
iteration step and  $\alpha\approx 0.5$. With these definitions, 5 to 10 
iterations are usually sufficient to achieve convergence. 

As mentioned above, we have assumed so far that the local Green's function and 
self-energy are diagonal in the orbital basis. Complex lattice geometries, however, 
can give rise to non-diagonal components which require a suitable generalization 
of the above formalism. This will be briefly addressed at the end of the next 
subsection.

\subsection{2.2. \ Multi-Site Correlations}

Let us consider now a single-band Hubbard model and include spatial correlations
among a finite number of sites ($n_c$). To be specific, we limit the discussion 
to two-dimensional systems, such as the square lattice with isotropic nearest 
and next-nearest neighbor coupling, and the triangular and honeycomb lattices 
with isotropic nearest-neighbor hopping integrals. The Coulomb interaction 
within the band is assumed to be a purely on-site interaction. Models of this
kind are believed to capture important features related to short-range 
correlations in high-$T_c$ cuprate superconductors, organic molecular 
crystals, and in graphene. The Hamiltonian of the system is given by
\begin{equation}
H=-\sum_{\langle ij\rangle\sigma} t_{ij}( c^+_{i\sigma} c_{j\sigma} + {\rm H.c.}) 
              + U \sum_i n_{i\uparrow} n_{i\downarrow} -
               \sum_{i\sigma} \mu n_{i\sigma} 
\label{Hcl}
\end{equation}
where the sum in the first term extends up to first or second neighbors.

\begin{figure} 
\begin{center}
\includegraphics[width=4cm,angle=-90]{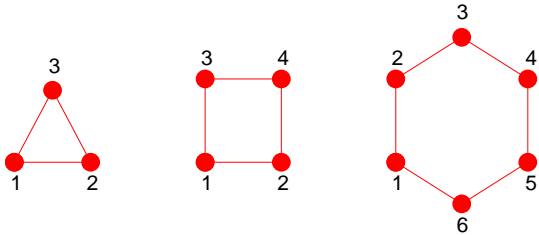}
\end{center}
\caption{(Color online)
Unit cells within CDMFT for triangular, square, and honeycomb lattices. 
}\label{unitcells}\end{figure}

Within the cluster extension \cite{kotliar01} of DMFT (CDMFT) the interacting 
lattice Green's function in the cluster site basis may be written as
\begin{equation}
     G_{ij}(i\omega_n) = \sum_{\bf k'} [ i\omega_n + \mu - t({\bf k'})- 
                   \Sigma(i\omega_n)]^{-1}_{ij} ,
\label{Gij}
\end{equation}
where the ${\bf k'}$ sum covers the reduced Brillouin Zone and $t({\bf k'})$ 
denotes the hopping matrix for the superlattice. $\Sigma_{ij}$ represents 
the cluster self-energy matrix. The site labels for the three lattices are 
indicated in Figure~\ref{unitcells}. For the purpose of the ED calculation, 
it is convenient to change from the non-diagonal site basis to a 
``molecular-orbital'' basis, in which the lattice Green's function is diagonal. 
For the triangular lattice, the diagonal components $G_m$ are given by:    
\begin{eqnarray}
 G_1       &=&  G_{11} + 2 G_{12}  \nonumber \\ 
 G_2 = G_3 &=&  G_{11} -  G_{13}. 
\label{G3}
\end{eqnarray}
For the square lattice, the diagonal components are: 
\begin{eqnarray}
 G_{1,2}  &=& G_{11} \pm 2 G_{12}+ G_{14} \nonumber \\ 
 G_3=G_4  &=& G_{11}-G_{14}.
\label{G4}
\end{eqnarray}
For the honeycomb lattice, they are given by: 
\begin{eqnarray}
 G_{1,2}         &=& (G_{11}+2G_{13})\pm (G_{14}+2G_{12}) \nonumber \\ 
 G_{3,4}=G_{5,6} &=& (G_{11}- G_{13})\pm (G_{14}- G_{12}).
\label{G6}
\end{eqnarray}
The cluster self-energy can be diagonalized in the same manner, with elements 
$\Sigma_m$.  In the molecular-orbital basis, the lattice Green's function 
takes the form:
\begin{equation}
     G_{m}(i\omega_n) = \sum_{\bf k'} [ i\omega_n + \mu - t'({\bf k'})- 
                   \Sigma(i\omega_n)]^{-1}_{mm} ,
\label{Gm}
\end{equation}
where $t'({\bf k'})=\bar T^{-1} t({\bf k'}) \bar T$ and $\bar T =[\bar T_{im}]$ 
denotes the 
unitary transformation from the site basis to the molecular-orbital basis.
For the triangular, square and honeycomb lattices, the matrices $\bar T$
and the molecular-orbital components $\rho_m(\omega)$ of the uncorrelated 
density of states can be found in Refs.~\cite{PRB78,PRB80,honey}. 
The asymptotic behavior of the Green's function components is given by
\begin{equation}
 G_{m}(i\omega_n) = [i\omega_n + \mu - \varepsilon_m  - 
       \Sigma_m(i\omega_n)  ]^{-1} ,
\label{Gasym2}
\end{equation}
where $\varepsilon_m  = [\sum_{\bf k'} t'({\bf k'})]_{mm}$. 

Evidently, with the transformation from lattice sites to molecular 
orbitals, the ED CDMFT procedure becomes analogous to the multi-orbital 
formalism discussed in the previous subsection. In particular, the removal 
of the self-energy from the ``impurity cluster'' yields the bath 
Green's function components $G_{0,m}(i\omega_n)$ given by Eq. (\ref{G0}),
which are discretized according to Eq. (\ref{G0cl}) by replacing the 
host continuum by a finite set of levels $\varepsilon_k$. Thus, cluster
molecular orbitals and bath form a ``super-cluster'' of size $n_s=n_c+n_b$ 
whose many-body properties can be solved exactly. (In the following we 
use the term ``cluster'' mainly for the unit cells depicted in Figure~2.
From the context it will be clear whether the larger supercluster is 
implied instead.) 

Assuming independent baths for the impurity molecular orbitals, the
hybridization interaction takes the form as schematically indicated in the 
left diagram of Figure~1 for the case $n_c=4$. The advantage of this diagonal 
basis is that, as in the multi-orbital case discussed in the previous 
subsection, the bath Green's function components $G^{cl}_{0,m}(i\omega_n)$  
can be discretized independently, rather than fitting the non-diagonal Green's 
function matrix $G^{cl}_{0,ij}(i\omega_n)$ in the site basis in one step 
\cite{koch,PRB78}.

The evaluation of the interacting cluster Green's function $G^{cl}$ 
can be carried out within the diagonal molecular-orbital basis, 
yielding  the components $G^{cl}_m(i\omega_n)$ defined in Eq.~(\ref{Gclm}). 
In this basis, the onsite Coulomb interaction is a non-diagonal matrix, 
with intra-orbital and inter-orbital elements that are analogous to the 
Coulomb and exchange interactions of the single-site, multi-orbital 
systems discussed in the previous subsection. For instance, in the case
of a $2\times2$ cluster representing a square lattice, one finds  
$\langle m_1m_2\vert\vert m_3m_4\rangle = U/4$ for $64$ of the $256$ 
matrix elements, where $U=\langle ii\vert\vert ii\rangle$ is the on-site
Coulomb interaction. All other elements vanish. An equivalent scheme
consists in transforming the cluster molecular orbitals back to the 
site basis, by using the transformation $\bar T$ introduced above.
Thus, the bath molecular-orbital levels remain unchanged and the Coulomb 
interaction is purely onsite. In this mixed site molecular-orbital picture 
the non-diagonal cluster Green's function is evaluated from the expression      
\begin{eqnarray}
 G^{cl}_{ij}(i\omega_n) &=& \frac{1}{Z} \sum_{\nu\mu}\,e^{-\beta E_\nu}\, 
          \Big(\frac{\langle\nu\vert c_{i\sigma}  \vert\mu\rangle 
                     \langle\mu\vert c_{j\sigma}^+\vert\nu\rangle}
                                  {i\omega_n + E_\mu - E_\nu}  \nonumber\\
       &&\hskip9mm + \ \ \frac{\langle\nu\vert c_{i\sigma}^+\vert\mu\rangle 
                               \langle\mu\vert c_{j\sigma}  \vert\nu\rangle}
                                  {i\omega_n + E_\nu - E_\mu } \Big) \\
   &=&  \frac{1}{Z} \sum_{\nu}\,e^{-\beta E_\nu}\,
      \!\![G^{cl+}_{ij,\nu}(i\omega_n) +  G^{cl-}_{ij,\nu}(i\omega_n)],  
        \nonumber
     \label{Gclij}
\end{eqnarray}
where $G^{cl\pm}_{ij,\nu}(i\omega_n)$ are Green's functions corresponding
the excited state $\vert\nu\rangle$.
Since $G^{cl}_{ij}$ satisfies the same symmetry properties as $G_{ij}$, it 
is diagonal within the molecular-orbital basis, whose elements $G^{cl}_{m}$
coincide with those derived via Eq.~(\ref{Gclm}). Within the non-diagonal 
site basis, the cluster bath hybridization takes the form indicated in the 
right diagram of Figure~1. The molecular-orbital bath levels do not mix within 
this representation, but hybridize with all cluster sites. Further details 
concerning the finite-temperature multi-site ED DMFT approach are provided in 
Ref.~\cite{PRB80}.

The lattice examples considered above are special in the sense that a diagonal
molecular-orbital basis exists. More general cases, such as the non-isotropic 
triangular or square lattices, require suitable generalizations of the above
formalism. As long as direct coupling between bath levels is excluded, the 
most general hybridization between impurity and bath levels is still represented 
by the right-hand diagram in Figure~1, but the hopping terms $V_{ik}$ obey weaker 
symmetry relations than before. As a result, the back-transformation to the 
molecular-orbital basis does not yield a diagonal Green's function. Because
of the off-diagonal elements, the diagonal components can no longer be fitted
independently of one another. Instead, the full bath Green's function matrix 
(or a suitable subblock) must be discretized in a single sweep   
\cite{civelli,merino,sakai,koch,PRB79,PRB82}.
Similarly, complex lattice geometries in multi-band single-site cases 
can give rise to non-diagonal inter-orbital components. Since the impurity 
bath hybridization is then also of the form as indicated in the right diagram 
of Figure~1, the discretization must be carried out in one step, rather than using
independent fits of the $G_{0,m}$ as in the diagonal case.

\subsection{2.3. \ Size of Hilbert Space}

An important criterion for the accuracy of ED DMFT is the level spacing
related to the finite size of the cluster consisting of impurity and bath.  
The dimension of the Hilbert space for $n_s=n_c+n_b$ cluster levels is 
$2^{2n_s}$. Since different spin sectors of the Hamiltonian do not couple,
the largest sector for $n_s=9$ has dimension $N=(9!/(4!5!))^2=15876$. 
For $n_s=12$, the dimension is $(12!/(6!6!))^2= 853776$, while 
for $n_s=15$, it is $(15!/(7!8!))^2= 41409225$. Fortunately,
the cluster Hamiltonian is extremely sparse, with typically no more than 
20 to 30 finite off-diagonal elements per row. Thus, using $32$ parallel 
processors for $n_s=15$, storage requirements per processor are about the 
same as for full matrix diagonalization for $n_s=8$ levels ($N=4900$). 
Moreover, because of the Boltzmann factor in the cluster Green's function, 
Eqs.~(\ref{Gclm}) and (\ref{Gclij}), only a limited number of excited states 
is required at finite $T$. The quantum impurity 
calculation can therefore be carried out very efficiently by using the Arnoldi 
algorithm \cite{arnoldi}, which generates a finite number of eigenvalues 
and eigenvectors of large sparse matrices. The contributions to the cluster 
Green's function are then given by a superposition of excited state Green's 
functions $ G^{cl\pm}_{m,\nu}(i\omega_n)$ or $ G^{cl\pm}_{ij,\nu}(i\omega_n)$
which can readily be calculated by applying the Lanczos method.    

The key point here is that for $n_s=9,\ldots,15$ the components of the 
interacting cluster Green's function, if evaluated at real $\omega$,  
contain vastly more poles than their non-interacting counterparts, 
$G^{cl}_{0,m}$. Thus, even if a particular impurity level $\varepsilon_m$
couples only to two bath levels $\varepsilon_k$, the indirect coupling 
to the remaining baths via internal Coulomb and exchange 
interactions within the impurity ensures a nearly continuous spectral
distribution. In the case of multi-site correlations, the diagonal
molecular-orbital basis implies many intra-orbital and inter-orbital 
Coulomb and exchange-like terms, whereas in the non-diagonal site basis 
the coupling between baths stems from the single-particle hopping between 
cluster sites.      

Dynamical mean field theory may be viewed as a scheme for evaluating the 
complex self-energy, which defines the redistribution of spectral weight 
arising from Coulomb interactions. According to Eq.~(\ref{Scl}), the cluster 
self-energy is governed by the spectral details of the interacting 
cluster Green's function. Thus, for $n_s=9,\ldots,15$ the nearly    
continuous spectral distributions of the components $G^{cl}_{m}$ 
ensure accordingly smooth distributions of $\Sigma^{cl}_{m}$. 
Moreover, because of the high quality of these functions along the 
Matsubara axis, analytic continuation to real $\omega$ close to the 
Fermi energy is usually found to be rather stable. In principle, the cluster 
Green's function and self-energy can be evaluated close to the real-energy 
axis which is a crucial advantage compared to QMC DMFT. 
On the other hand, as pointed out above, at each iteration in ED DMFT the 
cluster self-energy (with its discrete level spectrum) is assumed to provide 
a physically reasonable representation of the
continuous lattice self-energy, in the sense that along the Matsubara
axis both are nearly identical. Thus, as in QMC DMFT, analytic continuation 
of the self-energy is desirable and can be carried out using various methods.
(A detailed discussion of the analytic continuation of the self-energy is 
given in Ref.~\cite{wang}.) Since the single-particle properties are known
at real energies, no further approximations are then required. This should 
be contrasted to analytic continuation of the lattice Green's function, which
implies continuation of both self-energy and single-particle features. (The
limit of weak interactions makes the difference between these two methods
particularly striking.)      

For some of the examples discussed in the next section, we have employed the 
routine {\it ratint} \cite{ratint} for the extrapolation of the self-energy 
and lattice Green's function to real $\omega$. Typically, up to a few hundred 
Matsubara points are used, with a small frequency-dependent broadening. This 
procedure usually yields reliable results at low energies, while considerable 
uncertainties arise in the incoherent region at larger energies.

The multi-orbital and multi-site scenarios discussed above differ 
greatly from the simpler single-band, single-site case. Even if at 
some finite temperature the lattice bath Green's function can be 
discretized accurately via a non-interacting cluster Green's function 
using few bath levels, the small cluster size yields a very coarse level 
spectrum. It is therefore essential to include more bath levels so that 
the larger Hilbert space guaranties better energy resolution. 
Nevertheless, it has been shown that integrated quantities, 
such as the critical Coulomb energies $U_{c1,c2}$ determining the $T/U$ 
phase diagram for the Mott transition at half-filling, are already well 
converged using only 3 or 4 bath levels \cite{prl2005} (see also \cite{capone}).

\subsection{2.4. \ Bath Discretization} 
    
The second important criterion for the accuracy of ED DMFT concerns the 
projection of the bath Green's function components $G_{0,m}(i\omega_n)$ 
onto to a finite cluster, as indicated by the discretization in Eqs.~(\ref{G0m}) 
and (\ref{G0cl}). 
Evidently, the quality of this discretization depends sensitively on temperature 
and bath size. Along the Matsubara axis the spectrum of $G_{0,m}(\omega)$ is 
effectively sampled with Lorentzian functions of width $\omega_n=(2n+1)\pi T$:
\begin{equation}
    {\rm Im}\,G_{0,m}(i\omega_n) = \frac{1}{\pi}\int d\omega 
                 \frac{\omega_n}{\omega^2+\omega_n^2}\,
    {\rm Im}\, G_{0,m}(\omega) . \label{lorentz}
\end{equation} 
Thus, choosing a very low temperature gives rise 
to fine structure in $G_{0,m}(i\omega_n)$ that can be fitted only with a 
sufficiently large number of bath levels. For the description of multi-band 
or multi-site correlations with no more than two bath levels per impurity 
orbital it is therefore important to employ a Matsubara grid for not too low $T$. 
As will be discussed in more detail in the next section, applications to a 
variety of multi-band and multi-site correlation problems suggest that a 
temperature in the range $T\approx 5,\ldots,20$~meV (for typical band widths
of the order of $W=2$~eV) permits discretizations of reasonable accuracy. 
(Note that at very small $\omega_n$ a sufficiently large number 
of ${\bf k}$ points must be used to get accurate lattice Green's functions, 
Eqs.~(\ref{G}) and (\ref{Gij}). See also Eq.~(\ref{lorentz}).)   
    
To determine the bath levels $\varepsilon_k$  and hopping terms $V_{mk}$
we minimize the difference 
\begin{equation}
 {\rm Diff}_m = \sum_{n=0}^M  W_n
       \vert G_{0,m}(i\omega_n) - G^{cl}_{0,m}(i\omega_n)\vert^2  ,
\label{diff}
\end{equation}
where $M\approx 2^{10}$ is the total number of Matsubara points and the weight 
function  $W_n=1/\omega_n$ is introduced to give more weight to the low-frequency 
region \cite{capone2004}. Since $G^{(cl)}_{0,m}\rightarrow 1/i\omega_n$ at large
$\omega_n$, it is appropriate to optimize the low-energy region. To carry out 
the above fit, we use the routine {\it minimize} provided in Ref.~\cite{dmft}. 
To start the iterative procedure, we use bath parameters obtained for the 
uncorrelated system, or from a converged solution for nearby Coulomb and exchange 
energies. The resulting $\varepsilon_k$ and $V_{mk}$ are usually very stable against 
(not too large) variations of initial conditions.

We have also tested other weight functions, such as $W_n=1$ and 
$W_n=1/\omega_n^2$, as well as minimization of \cite{capone2004}       
\begin{equation}
 {\rm Diff}'_m = \sum_{n=0}^M  W_n
       \vert G_{0,m}(i\omega_n)^{-1} - G^{cl}_{0,m}(i\omega_n)^{-1}\vert^2  ,
\label{diff'}
\end{equation}
which represents the discretization of the lattice hybridization function
\begin{equation}
\Delta_m(i\omega_n)= i\omega_n + \mu-\varepsilon_m - G_{0,m}(i\omega_n)^{-1}
\label{Delta}
\end{equation}
via the corresponding cluster expression, i.e., the last term in Eq.~(\ref{G0cl}).
Thus,
\begin{equation}
\Delta_m(i\omega_n) \approx \Delta_m^{cl}(i\omega_n) = 
     \sum_{k=n_c+1}^{n_s} \frac{ \vert V_{mk} \vert^2 }
         {i\omega_n - \varepsilon_{k}},
\label{Delta0}
\end{equation}
so that 
\begin{equation}
 {\rm Diff}'_m = \sum_{n=0}^M  W_n
       \vert \Delta_{m}(i\omega_n) - \Delta^{cl}_{m}(i\omega_n)\vert^2  .
\label{diffw'}
\end{equation}
It can easily be shown that, in the limit of large $\omega_n$,
\begin{equation}
  \vert G_{0,m}(i\omega_n) - G^{cl}_{0,m}(i\omega_n)\vert \rightarrow 
   \frac{1}{\omega_n^2}
  \vert \Delta_{m}(i\omega_n) - \Delta^{cl}_{m}(i\omega_n)\vert  .
\label{diffasym}
\end{equation}
Thus, minimization of $\vert G - G^{cl}\vert$ ensures greater accuracy at small 
$\omega_n$ than minimization of $\vert 1/G - 1/G^{cl}\vert$. Experience has shown  
that the ED DMFT iteration procedure, using the distance function defined in 
Eq.~(\ref{diff}) with $W_n=1/\omega_n$, yields reliable and efficient convergence.    
Nonetheless, as discussed in the Appendix, other weight functions, as well as 
utilization of Eq.~(\ref{diff'}), also provide accurate discretization of the 
bath Green's function.   
(For a recent discussion of various choices of weight functions, mainly for 
the one-dimensional Hubbard model, see Ref.~\cite{senechal}.) 

In principle, the asymptotic behavior of $G_{0,m}(i\omega_n)$ and 
$G^{cl}_{0,m}(i\omega_n)$ is governed by the impurity levels $\varepsilon_m$,
which follow from the single-particle properties of the multi-band or 
multi-site Hamiltonians, as indicated in Eqs.~(\ref{Gasym1}) and (\ref{Gasym2}).
Since this asymptotic behavior is only weakly affected by the precise 
value of $\varepsilon_m$, a more accurate discretization of $G_{0,m}(i\omega_n)$ 
in the low-frequency region may be achieved by allowing $\varepsilon_m$ to vary.
In effect, $\epsilon_m$ is then included among the auxiliary parameters used 
to obtain the optimal fit of the lattice bath Green's function $G_{0,m}(i\omega_n)$ 
via the cluster bath Green's function $G^{cl}_{0,m}(i\omega_n)$,     
even though this implies a (small) inconsistency between cluster and lattice
properties. As a consequence of the finite bath size, slight inconsistencies 
of this kind are an intrinsic feature of ED quantum impurity calculations. 
They appear, for example, in the orbital occupancies, as calculated from the 
cluster and lattice Green's functions. In most cases  we have found 
the freely determined impurity levels to be very close to the nominal bulk 
values specified in  Eqs.~(\ref{Gasym1}) and (\ref{Gasym2}). Thus, the variation
of $\varepsilon_m$ is not essential for the discretization procedure.

For a typical situation with two bath 
levels per impurity orbital, each complex function $G_{0,m}(i\omega_n)$
is then fitted with five parameters: two bath levels $\varepsilon_k$, two 
hopping terms $V_{mk}$, and $\varepsilon_m$. For instance, in the case of 
$2\times2$ clusters representing the square lattice with $n_c=4$ and $n_b=8$, 
we have a total of 15 fit parameters for the three independent Green's function 
components (12 fit parameters for fixed $\varepsilon_m$). Naturally, this parameter 
set provides a more accurate bath discretization than in the site basis, 
where for symmetry reasons only four fit parameters are usually taken into account. 
In the special case of particle-hole symmetry at half-filling, the diagonal basis 
employs 8 to 10 parameters, compared to 2 parameters in the site basis, unless 
additional hybridization terms are incorporated as indicated in the right panel 
of Figure~1.

\section{3. \ Results and Discussion}
\subsection{3.1. \ Degenerate Three-Band Model}

As a first application of multi-band ED DMFT we consider a three-band Hubbard model
consisting of identical subbands with a Bethe lattice density of states 
($t=1/2$, band width $W=2$). The subbands are coupled only via Coulomb and 
exchange interactions. Single-particle hybridization among bands is neglected.
Despite its simplicity, this model is highly non-trivial. As shown by
Werner {\it et al.}~\cite{prlwerner}, it exhibits metal insulator 
transitions at integer occupancies and a spin freezing transition 
within an intermediate range of occupancies. At the latter 
transition, the self-energy changes from Fermi-liquid to non-Fermi-liquid 
behavior and the spin-spin correlation function changes from Pauli to 
Curie-Weiss behavior. We discuss this model here in some detail in order
to demonstrate the convergence of the ED results with bath size, the quality
of the discretization of the bath Green's function $G_0$, and the consistency 
with analogous continuous-time QMC results reported in Ref.~\cite{prlwerner}.     
We also discuss briefly that, for sufficiently large Hund's rule coupling, 
the self-energy reveals a collective mode due to spin fluctuations, which leads 
to a pseudogap in the density of states. A similar collective mode is found 
in the five-band model discussed in Section 3.3.   

\begin{figure} [t!] 
\begin{center}
\includegraphics[width=4.5cm,height=6.5cm,angle=-90]{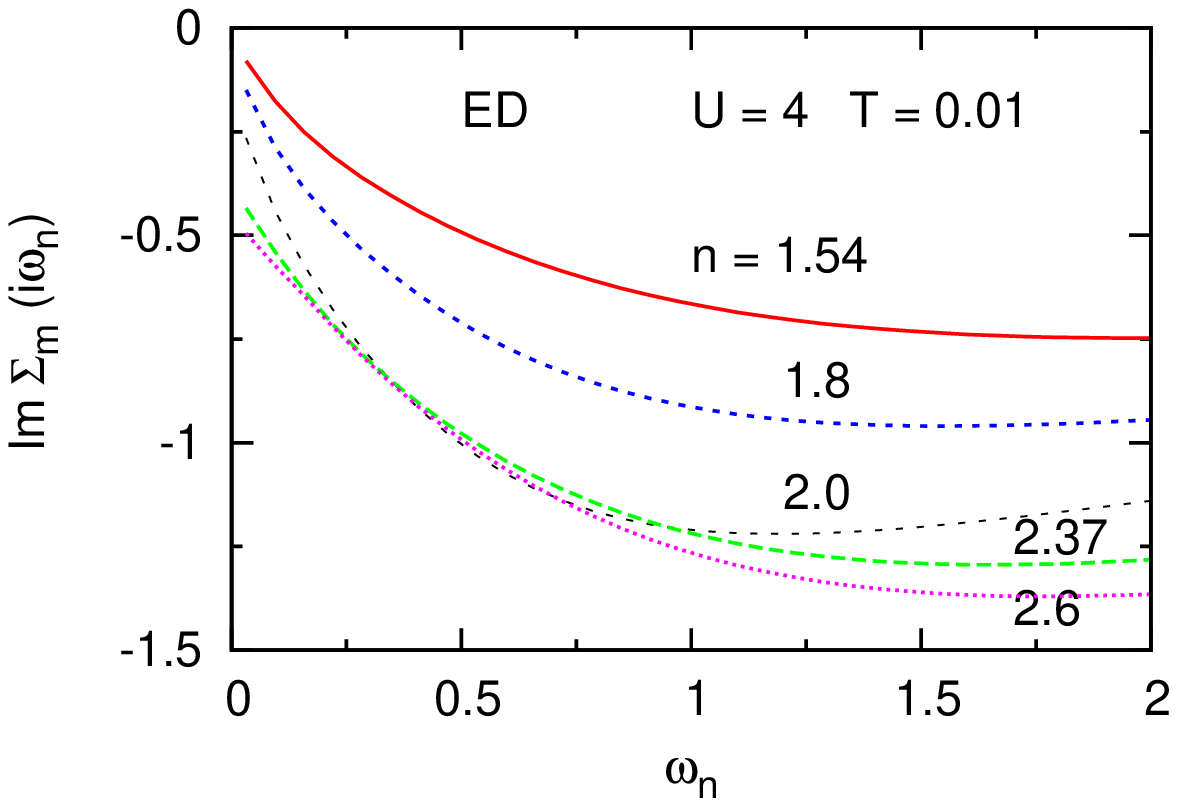} 
\includegraphics[width=4.5cm,height=6.5cm,angle=-90]{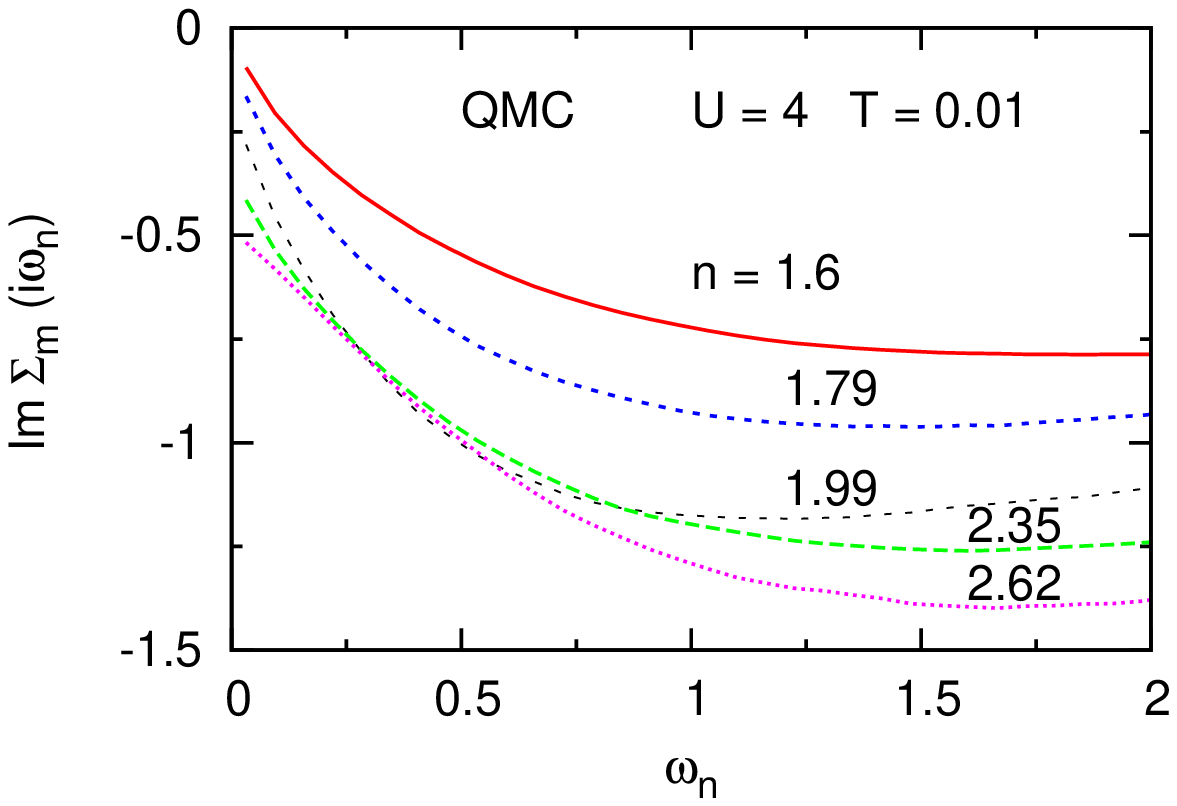} 
\end{center}
\caption{(Color online)
Self-energy of degenerate three-band model for $U=4$, $J=U/6$, $T=0.01$
at several occupancies. Upper panel: ED DMFT results for 
$n_s=12$ (3 bath levels per impurity orbital). 
Lower Panel: continuous-time QMC DMFT results for similar 
occupancies (adapted from Figure~3 of \cite{prlwerner}). 
}\label{3Sigma}\end{figure}

Figure \ref{3Sigma} (upper panel) shows the subband self-energy as a function of 
Matsubara frequency for various occupancies.  The cluster size is $n_s=12$, 
with three bath levels per orbital. Full Hund exchange is used, with $J=U/6$, 
and the Coulomb energy $U=2W=4$ is chosen so that at half-filling the system 
is a Mott insulator ($U_c\approx 1.5$ \cite{wernerprb79}). The corresponding 
QMC results are given in the lower panel for similar occupancies. The comparison 
shows that there is nearly quantitative agreement between the ED and QMC results, 
in particular, in the important low-energy region. At $n\approx1.6$, the system 
is a Fermi liquid, with Im\,$\Sigma_m(i\omega_n) \rightarrow \omega_n$ in the 
limit of small $\omega_n$. Near $n\approx2$, Im\,$\Sigma_m$ 
begins to exhibit a finite onset as a result of local-moment formation. 
At larger occupancies, this onset continues to increase until the Mott phase is 
reached and,  at half-filling, Im\,$\Sigma_m(i\omega_n) \rightarrow 1/\omega_n$ 
at small $\omega_n$. The results in Figure~\ref{3Sigma} are for T=0.01. Similarly
good agreement is obtained for $T=0.005$ (not shown here).   

\begin{figure}  [t!] 
\begin{center}
\includegraphics[width=4.5cm,height=6.5cm,angle=-90]{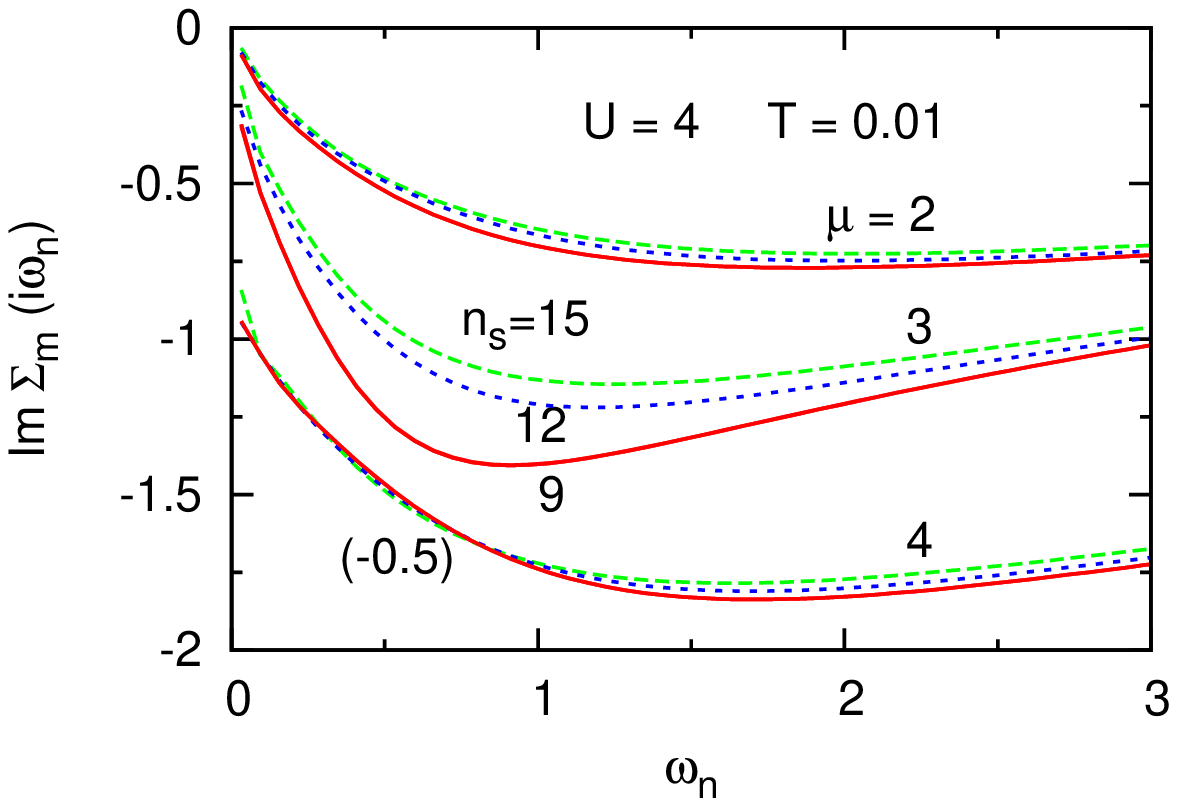} 
\includegraphics[width=4.5cm,height=6.5cm,angle=-90]{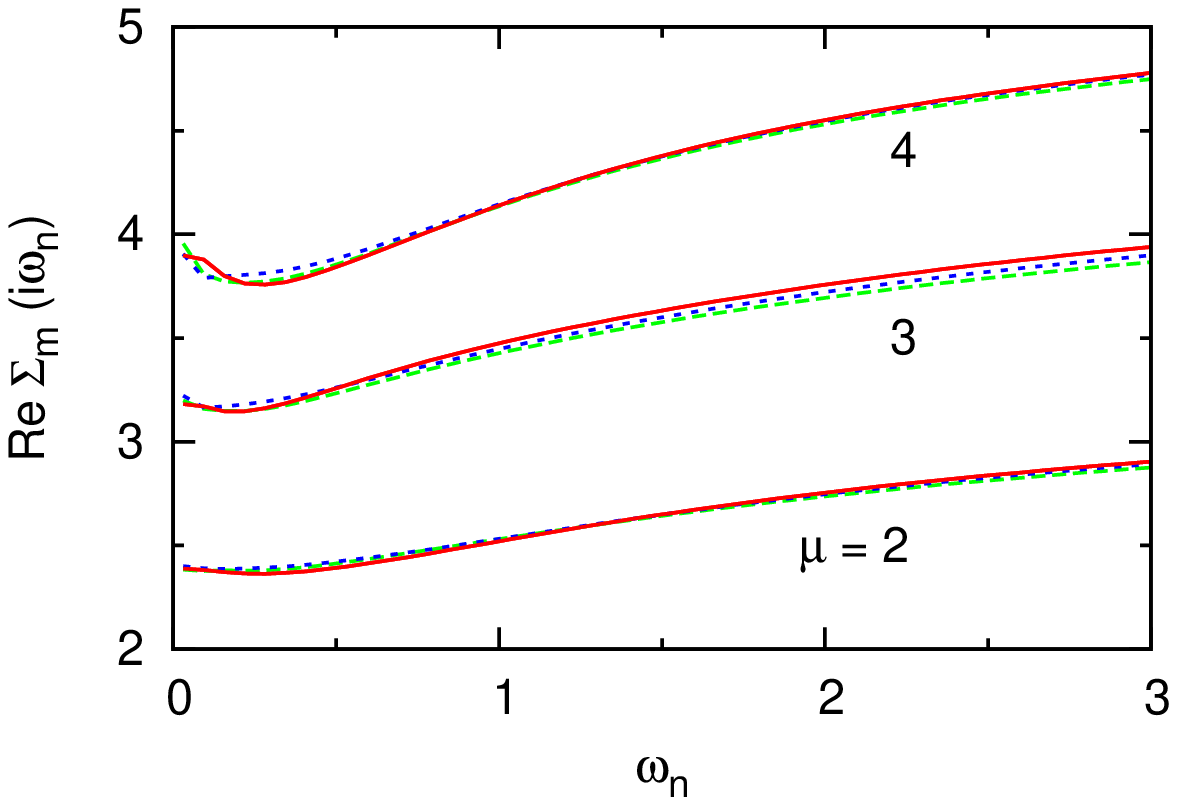} 
\end{center}
\caption{(Color online)
Convergence of self-energy for three-band model with bath size at different 
chemical potentials. Solid red curves: $n_s=9$, dashed blue curves: $n_s=12$, 
long-dashed green curves: $n_s=15$, corresponding to two, three, and four bath 
levels per impurity orbital, respectively. For clarity, the results in the upper 
panel for Im\,$\Sigma_m$ at $\mu=4$ are shifted down by $0.5$. $U=4$, $J=U/6$, 
$T=0.01$. 
}\label{3Sigma.ns}\end{figure}

Analogous results for only two bath levels per orbital ($n_s=9$) show 
qualitatively the same behavior as for $n_s=12$. In fact, at $\mu=2$ 
($n\approx1.6$) and $\mu=4$ ($n\approx2.5$), the self-energies are nearly 
the same as for the larger cluster.
Only in the transition region near $\mu=3$ ($n\approx2.0$), larger differences
are found, including a stronger dependence on temperature. To illustrate 
the convergence of the self-energy with bath size more systematically, 
we show in Figure~\ref{3Sigma.ns} results of ED DMFT calculations for 
$n_s=9,\ 12,$ and $15$. 
The calculations for 2 and 3 bath levels per orbital are carried out at $T=0.01$, 
whereas for computational reasons the ones for $n_s=15$ are performed at $T=0$ 
in the evaluation of the cluster Green's function, Eq.~(\ref{Gclm}). 
The Matsubara grid employed in the DMFT self-consistency procedure 
corresponds to $T=0.01$ in all cases. For $\mu=2$ and $4$, both 
the real and imaginary parts of $\Sigma_m$ are in good agreement
for the different cluster sizes. In particular, in the Fermi liquid 
range for $\mu=2$, Im\,$\Sigma_m\sim \omega_n$ in the limit of small 
$\omega_n$, with an effective mass enhancement of $m^*/m\approx3$. 
In contrast, for $\mu=4$, Im\,$\Sigma_m$ approaches a constant, reflecting 
the finite scattering rate due to the presence of frozen moments. 
For $\mu=3$, the increase from $n_s=9$ to $15$ is seen to shift 
the minimum of Im\,$\Sigma_m(i\omega_n)$ to larger $\omega_n$ and to make 
it more shallow, with a larger change from $n_s=9$ to $12$ than from 
$n_s=12$ to $15$. The reason for the stronger variation with bath size 
at $\mu=3$ is that this chemical potential nearly coincides with the phase 
boundary of the spin freezing transition. Thus, small variations due to the 
finite cluster size can give rise to larger changes of the self-energy.  
 
\begin{figure}  [t!] 
\begin{center}
\includegraphics[width=4.5cm,height=6.5cm,angle=-90]{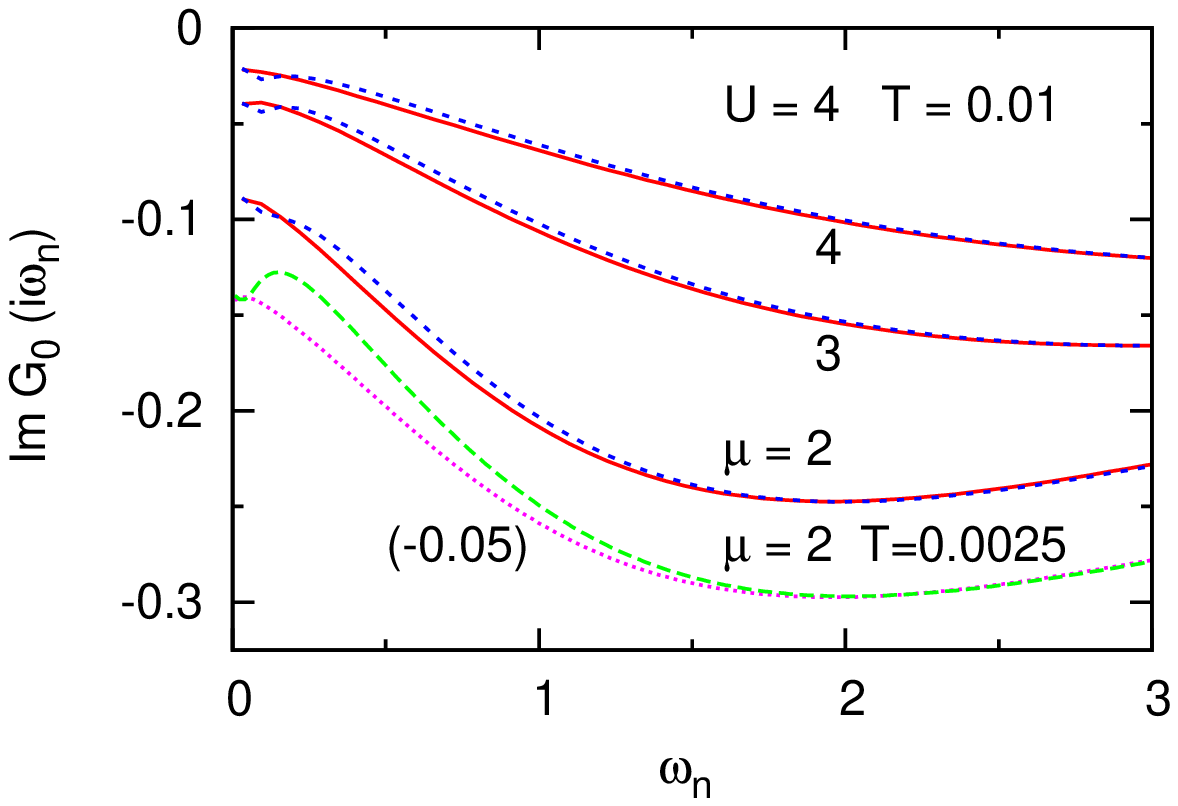} 
\includegraphics[width=4.5cm,height=6.5cm,angle=-90]{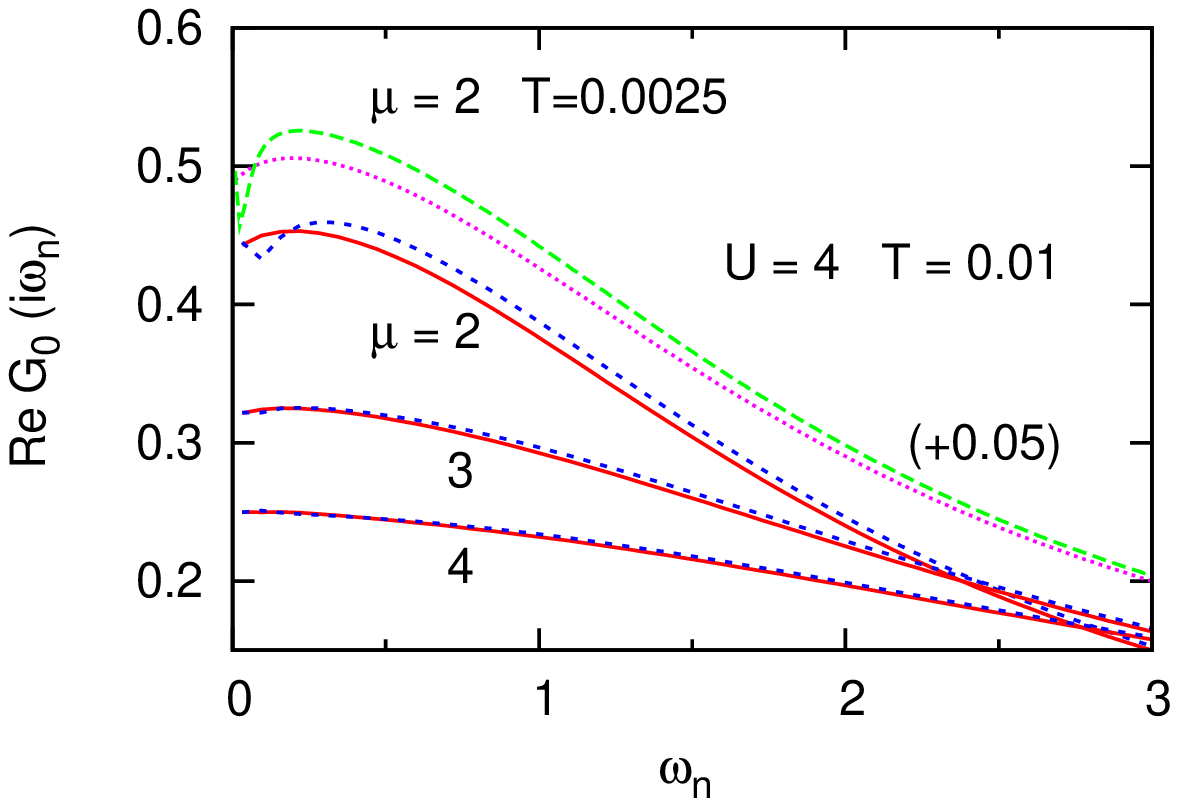} 
\end{center}
\caption{(Color online)
Discretization of bath Green's function: lattice $G_{0,m}(i\omega_n)$ 
(solid red curves) and cluster $G_{0,m}^{cl}(i\omega_n)$ 
(dashed blue curves) for 2 bath levels per impurity orbital ($n_s=9$) 
at several chemical potentials; $U=4$, $J=U/6$, $T=0.01$.
The dashed (magenta and green) curves show the fit for $\mu=2$ at $T=0.0025$.
(For clarity the latter results are shifted vertically by $\pm0.05$).
}\label{3G0fit}\end{figure}

We point out that the self-energy shown in Figure~\ref{3Sigma.ns} as a 
function of bath size does not exhibit any oscillatory behavior related to
the even or odd number of bath levels per impurity orbital. This kind of 
dependency on bath size occurs in more special cases, such as particle-hole 
symmetric bands at half-filling. (See, for example, the two-band model 
discussed in Ref.~\cite{prl2005}.) The presence or absence of bath 
levels at the Fermi energy then tends to favor metallic or insulating behavior, 
leading to a slow convergence with the number of bath levels. In the present 
degenerate three-band model, the bath levels away from half-filling do not 
obey any particular symmetry rules, so that oscillatory variation with bath 
size does not arise.

\begin{figure}  [t!] 
\begin{center}
\includegraphics[width=4.5cm,height=6.5cm,angle=-90]{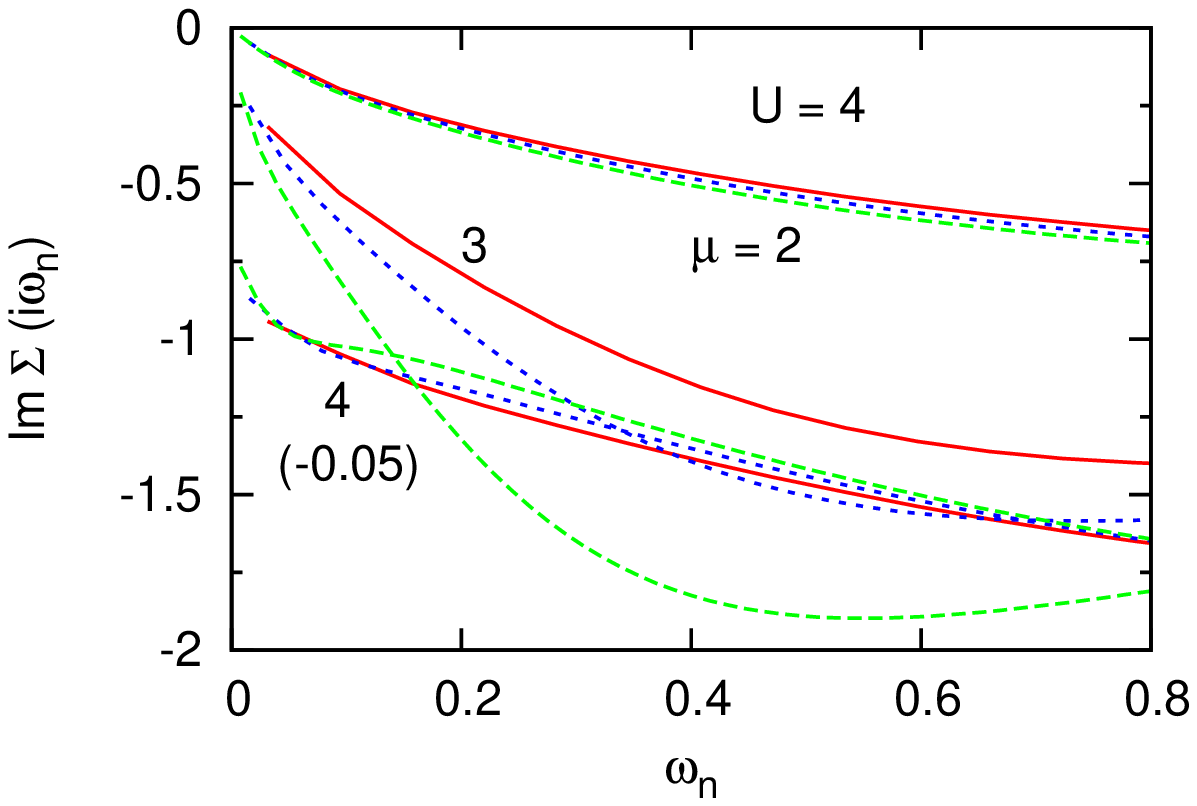} 
\includegraphics[width=4.5cm,height=6.5cm,angle=-90]{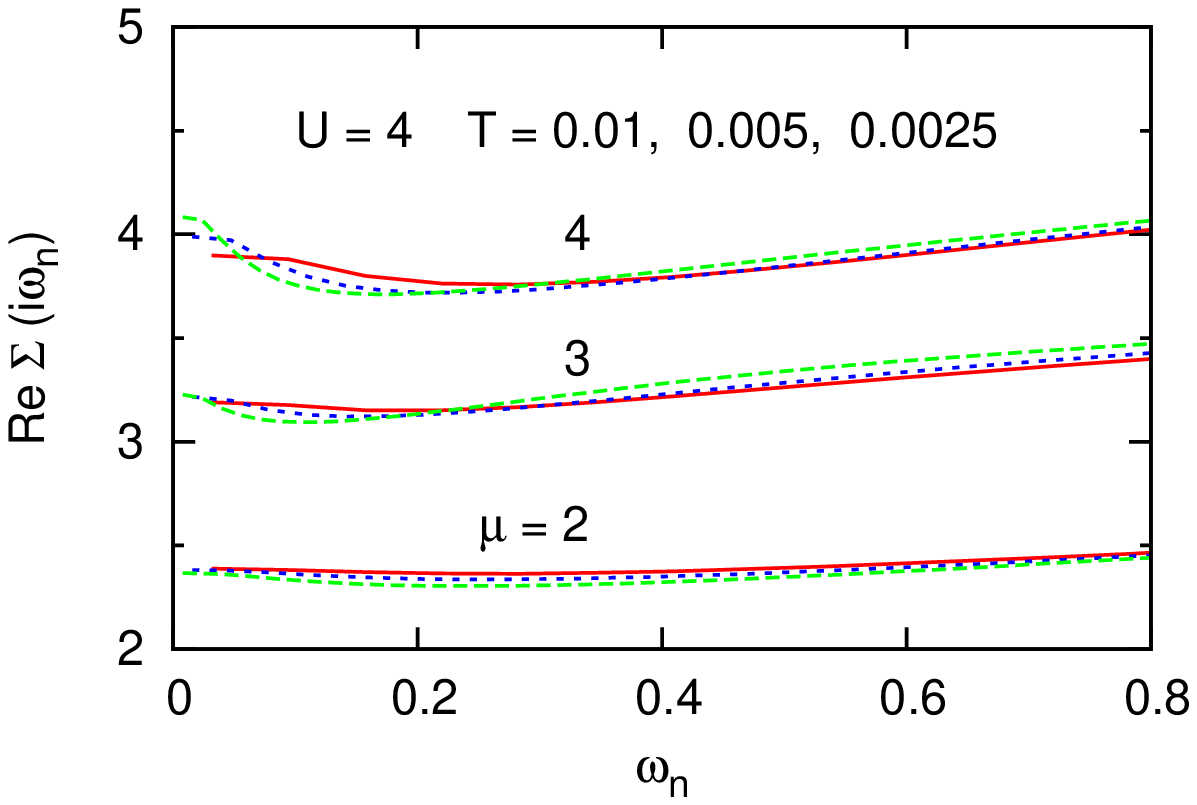} 
\end{center}
\caption{(Color online)
Low-energy region of self-energy of three-band model for several chemical 
potentials at temperatures $T=0.01$ (red solid curves), $T=0.005$ (short-dashed 
blue curves), and $T=0.0025$ (long-dashed green curves); $U=4$,  $J=U/6$, $n_s=9$ 
(two bath levels per impurity orbital). For clarity, the curves in the upper panel 
for $\mu=4$ are shifted down by $0.05$.   
}\label{3Sigma.T}\end{figure}

As discussed in the previous section, a crucial criterion for the convergence 
with bath size in ED DMFT is the quality of the discretization of the bath 
Green's function $G_{0,m}(i\omega_n)$, Eq.~(\ref{G0}), via the  
cluster Green's function $G_{0,m}^{cl}(i\omega_n)$, Eq.~(\ref{G0cl}). 
These functions are compared in Figure~\ref{3G0fit} for several chemical potentials,
assuming two bath levels per orbital. Evidently, at $T=0.01$, the fit of 
the bath Green's function with 5 parameters is already remarkably good.
Including more bath levels improves the fit at low $\omega_n$ even further.
Also shown is the discretization of $G_{0,m}$ at $T=0.0025$ for $\mu=2$. 
In this case the fit is much less satisfactory, indicating that more bath levels
are required at this low temperature.

The results in Figure 5 are based on minimization of $\vert G_0 - G_0^{cl}\vert$
as in Eq.~(\ref{diff}), using the weight function $W_n=1/\omega_n^N$, where $N=1$.
In Appendix A.1 it is shown that discretizations of similarly good quality
are obtained for $N=0$ and $N=2$, as well as for minimization of 
$\vert 1/G_0 - 1/G_0^{cl}\vert$, as in Eq.~(\ref{diff'}), with $N=0,1,2$.    

It should be noted that the small kinks in Im\,$G_{0,m}^{cl}(i\omega_n)$ at low 
$\omega_n$ are related to the fact that, because of the finite size of the cluster, 
the cluster spectrum is always gapped. Thus, Im\,$G_{0,m}^{cl}(i\omega)=0$ in the 
limit $\omega\rightarrow0$, giving rise to large deviations from the lattice bath 
Green's function (see also Section 3.5 and Figure~\ref{DOPE4.Fig2}).  
These deviations occur primarily at very low frequencies near or below $\omega_0$. 
Naturally, they are much less pronounced if the system is insulating. 
To achieve an accurate discretization of $G_{0,m}$ with a small 
number of bath levels it is therefore necessary to use a Matsubara grid for 
not too low temperatures. On the basis of the example shown in Figure~\ref{3G0fit} 
and various other systems involving multi-orbital and multi-site correlations, 
$T\approx 0.005,\ldots,0.01$ is a reasonable lower limit, if two or three 
bath levels per orbital are available.
 
Although the discretization of the bath Green's function becomes less accurate 
at low temperatures, the self-consistently determined self-energy can still be 
qualitatively correct. To illustrate this point, we show in Figure~\ref{3Sigma.T}
the self-energy at $T=0.01$, $0.005$, and $0.0025$. Evidently, all curves exhibit 
the same trend as a function of chemical potential, with Fermi-liquid behavior 
at low occupancy ($\mu=2$, $n\approx1.6$) and a pronounced low-energy scattering 
rate closer to half-filling ($\mu=4$, $n\approx2.5$). As in Figure~\ref{3Sigma.ns}, 
the results near the spin freezing phase boundary ($\mu=3$, $n\approx2$) are most 
sensitive and therefore show larger variation with $T$.       

\begin{figure}  [t!] 
\begin{center}
\includegraphics[width=4.5cm,height=6.5cm,angle=-90]{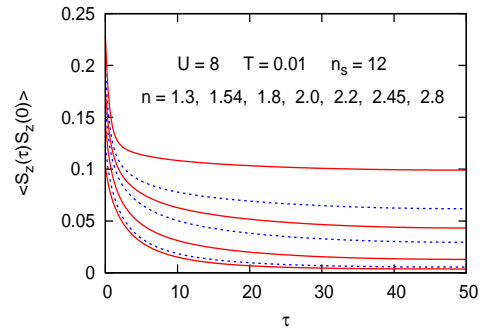} 
\end{center}
\caption{(Color online)
Spin-spin correlation function for degenerate three-band model at various  
occupancies (increasing from below) for $n_s=12$; $U=4$, $J=U/6$, $T=0.01$.
In the Fermi-liquid regime at small $n$, the susceptibility is Pauli-like, 
whereas at larger $n$ it reflects Curie-Weiss behavior.    
}\label{3chi}\end{figure}

As shown in Ref.~\cite{prlwerner}, the physical reason for the change of 
the self-energy from Fermi-liquid to non-Fermi-liquid behavior seen in 
Figure~\ref{3Sigma} is associated with a qualitative change of the spin-spin 
correlation function $C_{m}(\tau)=\langle S_{mz}(\tau)S_{mz}(0)\rangle$, 
where $\tau$ denotes imaginary time. In the Fermi-liquid regime $\mu<3$, 
this function decays with $\tau$, so that $C_{m}(\tau=\beta/2)$ is very 
small. The susceptibility
$\chi_m\sim \int_0^\beta d\tau\,\langle S_{mz}(\tau)S_{mz}(0)\rangle$ 
then is Pauli-like, i.e., independent of temperature. 
For $\mu>3$, however,  $C_{m}(\tau)$ approaches a finite value for 
$\tau\rightarrow\beta/2$, so that $\chi_m$ becomes proportional to $1/T$, 
as expected for Curie-Weiss behavior due to the formation of local moments.
Figure~\ref{3chi} shows the correlation function as calculated within ED DMFT.
The variation of $C_{m}(\tau)$ with doping found here agrees rather well 
with the continuous-time QMC results discussed in Ref.~\cite{prlwerner}. 

We note here that a similar sequence of phases, from a Mott insulator at 
half-filling towards non-Fermi-liquid and Fermi-liquid behavior with increasing
electron or hole doping, is also seen in the five-band Hubbard model discussed in 
Section 3.3 and in the non-local properties of the single-band Hubbard 
model discussed in Section 3.5.         

\begin{figure}  [t!] 
\begin{center}
\includegraphics[width=4.5cm,height=6.5cm,angle=-90]{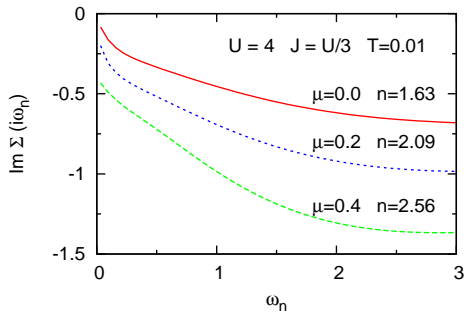} 
\end{center}
\caption{(Color online)
Imaginary part of self-energy of three-band model at $U=4$, $J=U/3$ for several 
chemical potentials. $T=0.01$, $n_s=12$. Note the onset and low-energy kink.
}\label{3Sigma.U=3J}\end{figure}

Apart from the spin freezing transition, the three-band model exhibits another 
interesting phenomenon not present in the single-band case. As shown in 
Figure~\ref{3Sigma.U=3J},    
for Hund exchange $J=U/3$ at $U=4$, the imaginary part of the self-energy 
reveals not only a finite low-energy scattering rate, but also a kink at 
about $\omega_n\approx 0.2$. Extrapolation of the self-energy to real 
$\omega$ yields a resonance in Im\,$\Sigma(\omega)$ near $\omega=-0.2$ 
(below $E_F=0$), which is weak in the Fermi-liquid region ($n<2$), but gets
very pronounced in the non-Fermi-liquid range ($n>2$). Moreover, the resonance
disappears at small $J$. The reason for this dependency on $J$ is the fact
that, at half-filling, the three-band Hubbard model is metallic at small $J$,
and becomes Mott insulating for sufficiently large $J$ \cite{wernerprb79,luca}.
Thus the resonance exists only in proximity to the Mott phase in the limit of 
half-filling. It can also be shown that the peak in Im\,$\Sigma(\omega)$
gives rise to a pseudogap in the density of states slightly below $E_F$ and 
that it is associated with a maximum in the local spin correlation function.        
The resonance in the self-energy induced via Hund coupling may therefore be
viewed as a collective mode caused by spin fluctuations. According to Figure
\ref{3Sigma.U=3J}, the non-Fermi-liquid behavior associated with the low-energy 
onset of Im\,$\Sigma(\omega)$ can then be interpreted as a consequence of the
appearance of this collective mode. A similar phenomenon occurs in the five-band 
Hubbard model discussed in Section 3.3. For a discussion of collective modes 
due to spin fluctuations in the single-band Hubbard model, see Ref.~\cite{raas}.   



\subsection{3.2. \ Three-Band Materials}

The multi-band ED  DMFT approach outlined in Section 2.1 has recently 
been used to study the role of local Coulomb interactions in several 
materials with partially filled $t_{2g}$ bands, where the $e_g$ subbands 
are pushed above the Fermi level as a result of crystal field effects.
In this subsection, we summarize the main results of these calculations,
illustrate the convergence of the self-energy with the size of the bath,
and point out the consistency with complementary QMC DMFT results. 

\bigskip
\centerline{\bf {3.2.1. \  Ca$_{2-x}$Sr$_x$RuO$_4$}}
\bigskip

A system that has been studied extensively is the layer perovskite 
Ca$_{2-x}$Sr$_x$RuO$_4$ which exhibits a remarkably rich phase diagram, 
ranging from unconventional superconductivity in the pure Sr limit to an 
antiferromagnetic Mott insulator at $x=0$, with a variety of exotic magnetic 
phases at intermediate concentrations \cite{maeno}.
As the substitution of Sr by the slightly smaller Ca ions is isoelectronic
at occupancy $4d^4$, the various electronic and magnetic phases are entirely 
the result of subtle hybridization changes among the $t_{2g}$ states caused by 
structural distortions.  
Because of the essentially two-dimensional structure 
of this material, the $d_{xy}$ band has a much larger band width than the 
more nearly one-dimensional $d_{xz,yz}$ bands. Thus, Coulomb interactions can 
be expected to modify these subbands differently. Moreover, LDA calculations
\cite{oguchi+singh} predict the $d_{xy}$ van Hove singularity for $x=2$ to lie 
only about 50 meV above the Fermi level, so that correlation induced charge 
transfer between the $t_{2g}$ bands could possibly shift this singularity 
below $E_F$, giving rise to a qualitative change of the topology of the $d_{xy}$ 
Fermi surface from electron-like to hole-like. 
Multi-band QMC and ED DMFT calculations for Sr$_2$RuO$_4$ do indeed predict 
a decrease of orbital polarization, i.e., a shift of charge from $d_{xz,yz}$ 
to $d_{xy}$ \cite{prl2000,anisimov,pchelkina,prl2007,mravlje}. 
The concomitant lowering of the $d_{xy}$ van Hove singularity for realistic 
interaction parameters, however, is not large enough to modify the topology 
of the Fermi surface.

\begin{figure}  [t!] 
\begin{center}
\includegraphics[width=4.5cm,height=6.5cm,angle=-90]{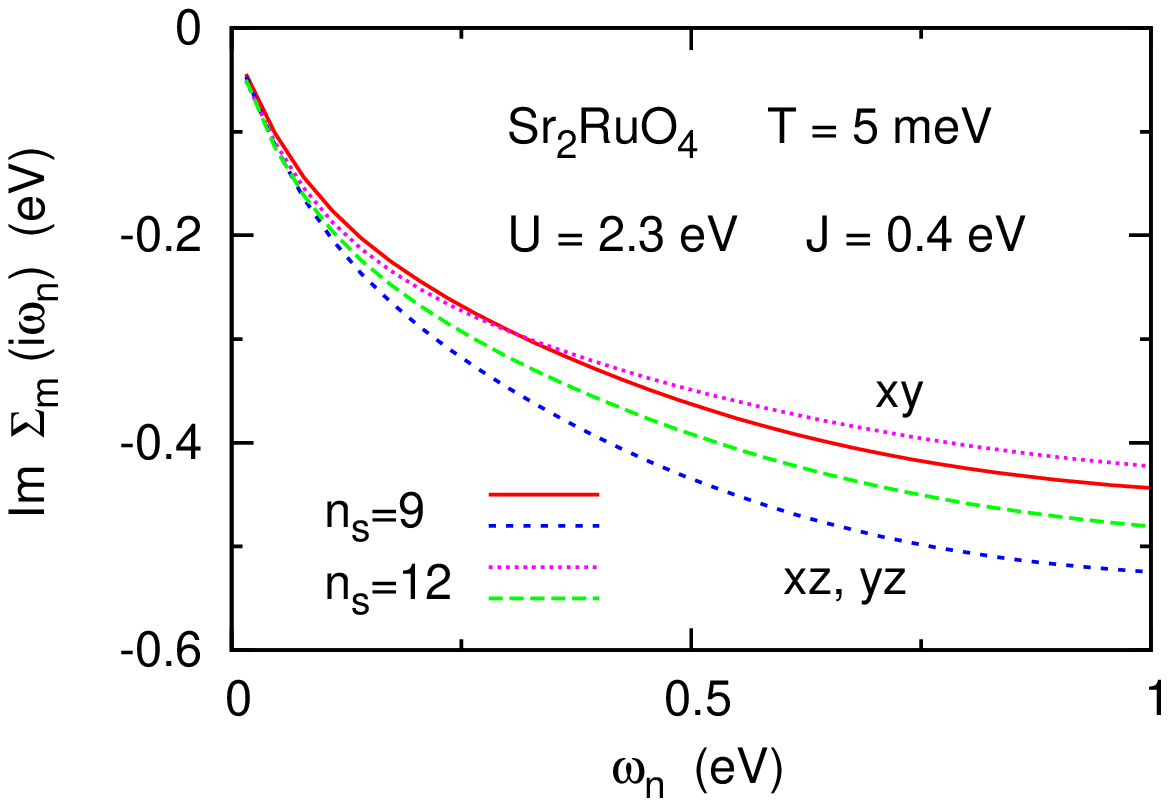} 
\includegraphics[width=4.5cm,height=6.5cm,angle=-90]{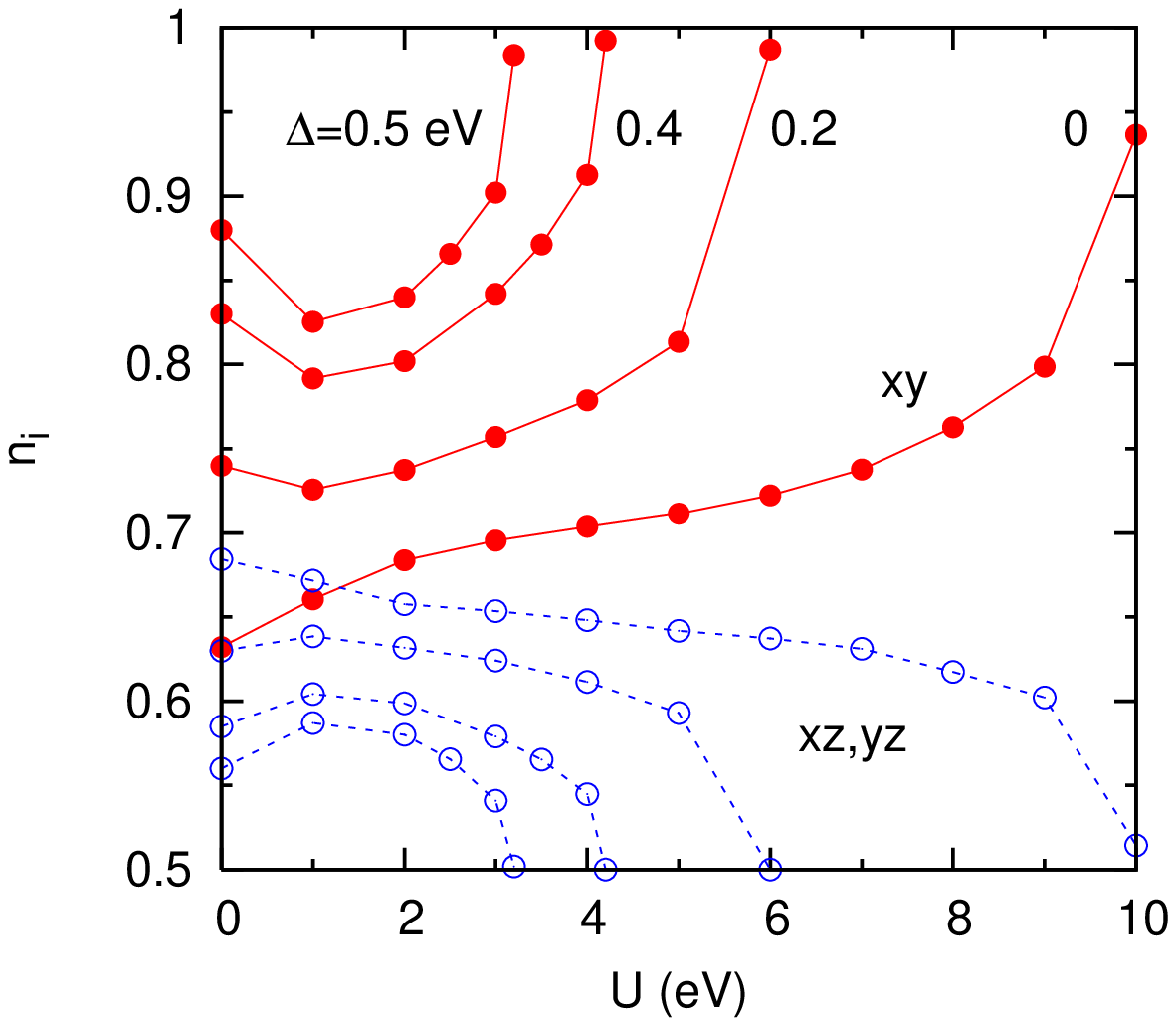} 
\end{center}
\caption{(Color online)
Upper panel: Imaginary part of Sr$_2$RuO$_4$ $t_{2g}$ self-energy components
for cluster sizes $n_s=9$ and $n_s=12$;  $U=2.3$~eV, $J=0.4$~eV, $T=5$~meV. 
Lower panel:
Orbital occupancies (per spin) 
of Ca$_{2-x}$Sr$_x$RuO$_4$  $t_{2g}$ subbands as functions 
of Coulomb energy for various crystal field splittings $\Delta$, derived within
ED DMFT ($n_s=9$) \cite{prl2007}. Solid (red) dots: $n_{xy}$, empty (blue) dots: 
$n_{xz,yz}$. Hund exchange $J=U/4$; $T=20$~meV.  
}\label{SRO2}\end{figure}

The upper panel of Figure~\ref{SRO2} illustrates the convergence of the ED 
self-energy components with cluster size for a tight-binding Hamiltonian 
including up to second-neighbor hopping interactions \cite{prl2000}.  
The results are for full Hund exchange and the values of $U$ and $J$ 
correspond to those derived in recent constrained RPA calculations
\cite{mravlje}. There is good agreement for two and three bath levels per 
impurity orbital, especially at low energies. The small differences are 
mainly caused by slightly different orbital occupancies obtained for these 
two bath sizes. For $n_s=12$, we find $n_{xy}=0.66$ and  $n_{xz,yz}=0.67$.
The uncorrelated subband occupancies are $n_{xy}=0.63$ and $n_{xz,yz}=0.68$.
Thus, Coulomb correlations in the pure Sr compound lead to decreasing orbital 
polarization.  The effective mass enhancement is approximately 
$m^*/m_{\rm LDA}\approx 4$, in agreement with continuous-time QMC 
calculations \cite{mravlje} 
and with experiment \cite{maeno}. In this sense, it is appropriate to 
regard Sr$_2$RuO$_4$ as a strongly correlated material.     

For a tight-binding model similar to the one used in Ref.~\cite{prl2000}, 
continuous-time QMC results \cite{mrav} for the self-energy agree very well 
with those in Figure~\ref{SRO2}. In particular, they show little difference 
between the $d_{xy}$ and $d_{xz,yz}$ effective mass enhancements.
Stronger $t_{2g}$ anisotropy is obtained for a more accurate single-particle 
Hamiltonian, for which the $d_{xy}$ van Hove singularity lies closer to $E_F$ 
and the difference in $d_{xy}$ and $d_{xz,yz}$ band widths is less pronounced 
than in the tight-binding model \cite{mravlje}.   

\begin{figure}  [t!] 
\begin{center}
\includegraphics[width=4.5cm,height=6.5cm,angle=-90]{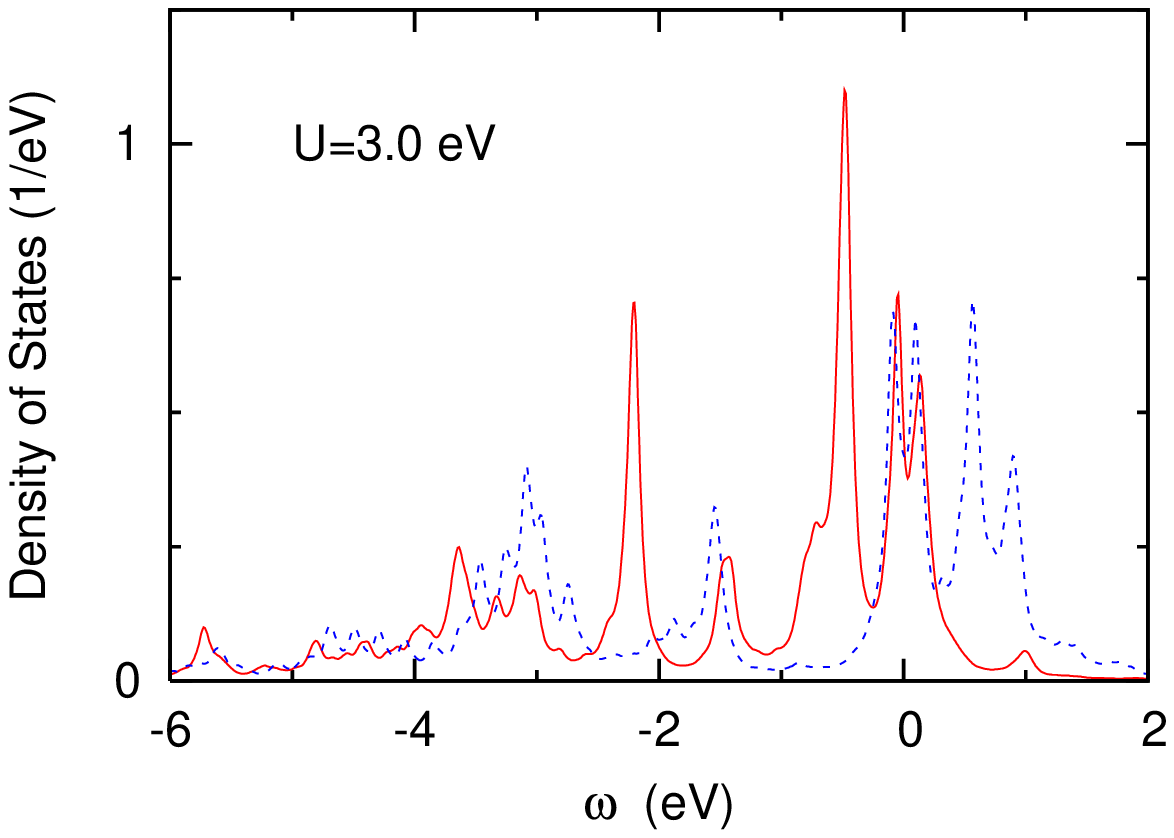} 
\includegraphics[width=4.5cm,height=6.5cm,angle=-90]{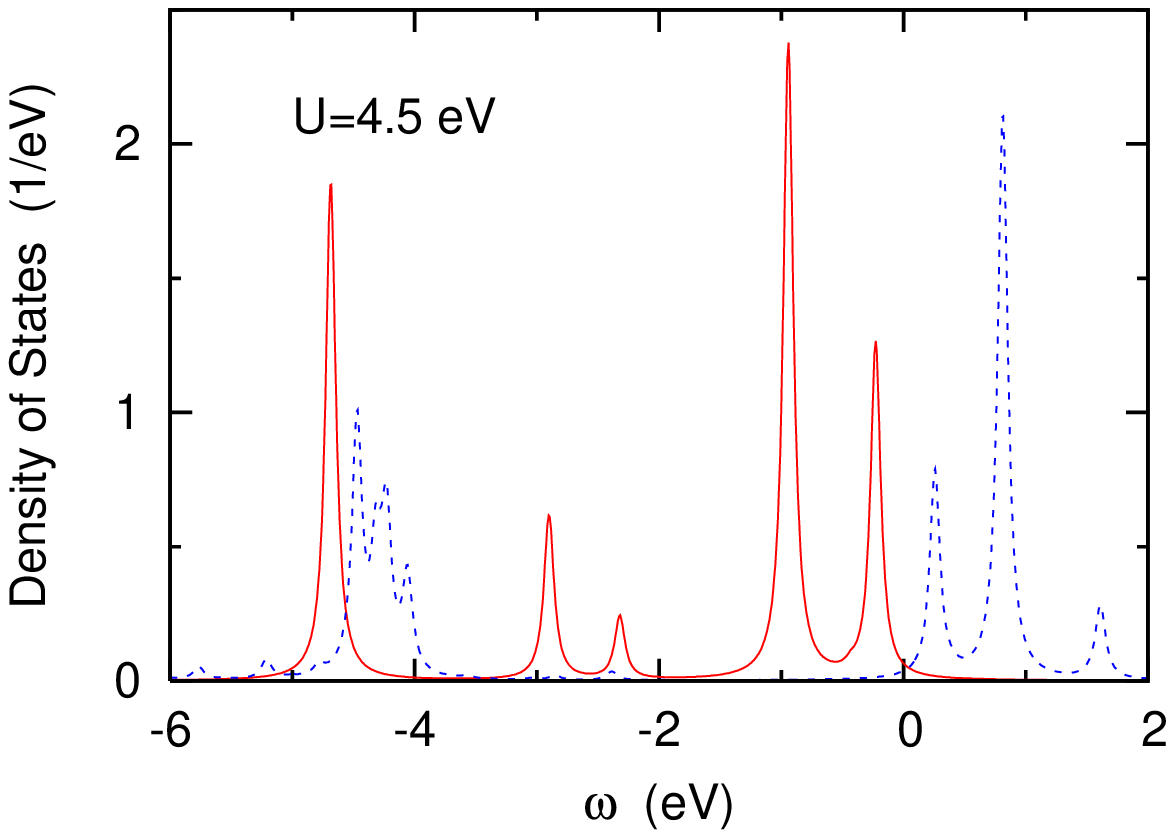} 
\end{center}
\caption{(Color online)
Quasiparticle spectra of Ca$_2$RuO$_4$ for crystal-field splitting $\Delta=0.4$~eV
at $U=3.0$~eV (upper panel) and $U=4.5$~eV (lower panel) for $J=U/4$, $T=20$~meV. 
Solid (red) curves:
$d_{xy}$ band, dashed (blue) curves: $d_{xz,yz}$ bands. The spectra are derived
from the interacting cluster Green's function for $n_s=9$, with broadening 
$\gamma=50$~meV \cite{prl2007}. 
}\label{CRO}\end{figure}

Substitution of Sr via Ca leads to rotation, tilting and flattening of 
Oxygen octahedra and, as a result, to a lowering of the $d_{xy}$ band
with respect to the $d_{xx,yz}$ bands \cite{fang,gorelov}. 
To investigate the influence of Coulomb correlations in the presence of this 
structurally induced interorbital charge transfer we have performed ED DMFT 
calculations, with an additional crystal field $\Delta=\epsilon_{xz,yz}-\epsilon_{xy}$ 
to account for the downward shift of the $d_{xy}$ bands \cite{prl2007}. 
As shown in the lower panel of Figure~\ref{SRO2}, correlations greatly enhance 
the $d_{xx,yz}\rightarrow d_{xy}$ charge transfer (except at small values of $U$).
In fact, for realistic values of $\Delta$, $U$, and $J$, the $d_{xy}$ band 
gets completely filled and the half-filled $d_{xz,yz}$ bands become Mott 
insulating. This orbital polarization is also evident in the cluster spectra
which are depicted in Figure~\ref{CRO} for $\Delta=0.4$~eV. At $U=3.0$~eV, 
the system is metallic, and all three $t_{2g}$ orbitals contribute to the
density of states at $E_F$. At $U=4.5$~eV, however, the $d_{xy}$ band is 
pushed below $E_F$ and the doubly degenerate $d_{xz,yz}$ bands are split 
into lower and upper Hubbard bands. Within the accuracy of the ED DMFT 
calculations, the filling and Mott-splitting of subbands takes place at the 
same critical $U_c\approx 4.2$~eV. Thus, there are no orbital-selective, 
successive Mott transitions for the narrow $d_{xz,yz}$ and wide $d_{xy}$ subbands, 
in contrast to the mechanism proposed by Anisimov {\it et al.}~\cite{anisimov}. 
Recent continuous-time QMC DMFT results for Ca$_2$RuO$_4$ by Gorelov 
{\it et al.}~\cite{gorelov} based on detailed LDA calculations also do 
not find successive transitions. Orbital selective Mott phases can appear, 
however, if the $t_{2g}$ crystal field splitting is assumed to have the 
opposite sign \cite{medici}.  

We point out that the large critical Coulomb energies seen in the lower 
panel of Figure~\ref{SRO2} for small values of $\Delta$ are a consequence
of the assumption $J=U/4$. If $J$ is held fixed at a realistic value, $U_c$ 
becomes accordingly smaller. For $U=3.1$~eV and $J=0.7$~eV, as in 
Ref.~\cite{gorelov}, we have $J \approx U/4$ so that the orbital 
polarization obtained for $\Delta\approx 0.4,\ldots,0.5$~eV is 
qualitatively correct.

\bigskip
\centerline{\bf {3.2.2. \  LaTiO$_3$}}
\bigskip

\begin{figure}  [t!] 
\begin{center}
\includegraphics[width=4.5cm,height=6.5cm,angle=-90]{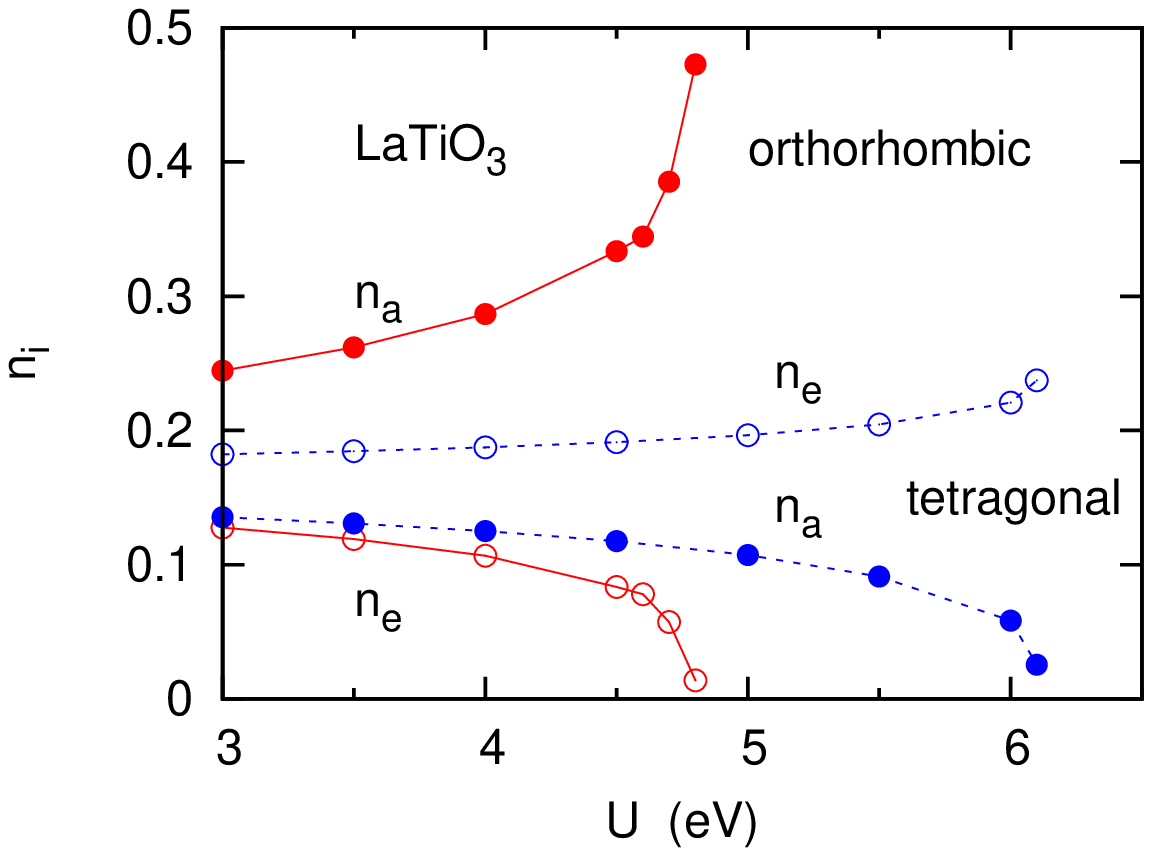} 
\includegraphics[width=4.5cm,height=6.5cm,angle=-90]{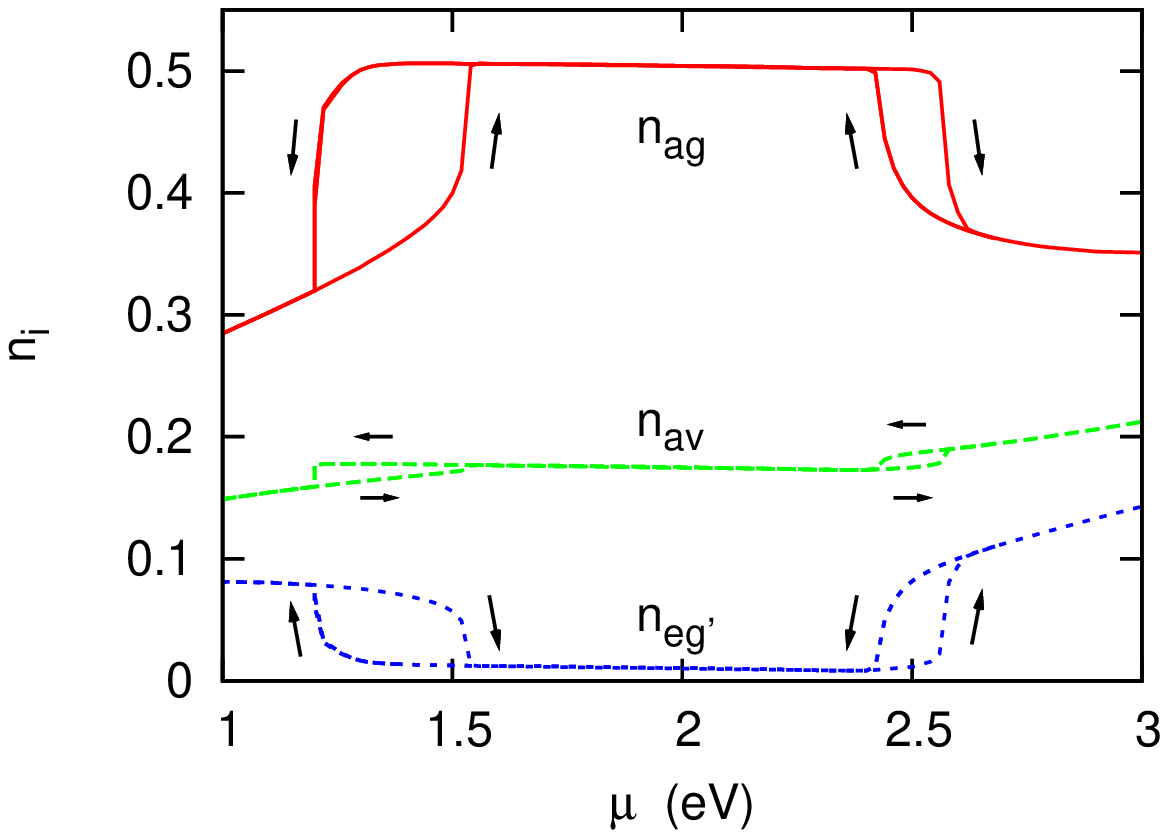} 
\end{center}
\caption{(Color online)
Upper panel: 
LaTiO$_3$ subband occupancies for orthorhombic and tretragonal bulk geometry. 
Since the crystal field splitting has the opposite sign in these structures,
the orbital polarization between the singly degenerate $a_g$ ($d_{xy}$) and 
doubly degenerate  $e_{g'}$ ($d_{xz,yz}$) states is reversed \cite{LTO2:prb77}.   
Lower panel:
LaTiO$_3$ $t_{2g}$ subband occupancies per spin as functions of 
chemical potential. $U=5$~eV, Hund exchange $J=0.65$~eV, $T=20$~meV.
Also shown is the average occupancy $n_{av}=(n_{ag}+2n_{eg'})/3$. The arrows
indicate the hystereses for increasing and decreasing $\mu$ \cite{LTO1:prb77}.
}\label{LTO}\end{figure}

The nature of the Mott transition in Ca$_2$RuO$_4$, where the correlation 
induced crystal field enhancement leads to strong orbital polarization, 
is similar to the one discussed by Pavarini {\it et al.}~\cite{pavarini} 
for LaTiO$_3$ ($3d^1$). In contrast to cubic SrVO$_3$, the 
orthorhombic structure of LaTiO$_3$ gives rise to a splitting of about 0.2~eV 
between the singly degenerate  $a_g$ and doubly degenerate $e_{g'}$ states, 
implying appreciably different subband occupancies: $n_{a_g}\approx 0.23$
and $n_{e_{g'}}\approx 0.135$ per spin. This orbital polarization is strongly 
enhanced by the local Coulomb interaction, leading to virtually empty $e_{g'}$
subbands and a Mott phase in the half-filled $a_g$ band. Finite-temperature ED 
DMFT results \cite{LTO2:prb77} which are shown in upper panel of Figure~\ref{LTO} 
are consistent with this picture.

Interestingly, when thin layers of LaTiO$_3$ are deposited on cubic SrTiO$_3$, 
as in  LaTiO$_3$/SrTiO$_3$ heterostructures, the initial growth appears to 
be tetragonal \cite{suzuki}. 
This substrate-enforced reconstruction gives rise to an increased
$t_{2g}$ band width and to a reversal of the $t_{2g}$ crystal field.  As also 
shown in the upper panel of Figure~\ref{LTO}, the result of these effects is an 
increase of the critical Coulomb energy for the Mott transition, and qualitatively
different subband occupancies. Now the singly-degenerate $d_{xy}$ band is 
emptied, and the quarter-filled $d_{xz,yz}$ bands are split into lower and 
upper Hubbard bands. These results are relevant for the interpretation 
of the metallicity observed LaTiO$_3$/SrTiO$_3$ heterostructures \cite{hwang}. 
If realistic values of $U$ for LaTiO$_3$ are about $5.0,\ldots,5.5$~eV, the
results in Figure~\ref{LTO} imply that thin LaTiO$_3$ layers are strongly correlated
metals rather than Mott insulators. Thus, the experimentally observed metallic 
behavior is not associated with the atomic interface Ti layers with nominal
$3d^{0.5}$ occupancy, but with the entire LaTiO$_3$ slabs. 

The lower panel of Figure~\ref{LTO} demonstrates that the $t_{2g}$ orbital
polarization is also highly sensitive to electron and hole doping \cite{LTO1:prb77}. 
Since the subband compressibilities $\kappa_m=\partial n_m/\partial\mu$ diverge 
at the Mott transition, small changes away from the nominal $n=1$ occupancy 
cause significantly larger changes in subband occupancies. In fact, weak overall 
hole doping gives rise to much larger hole doping in the half-filled $a_g$
band, accompanied by electron doping in the $e_{g'}$ bands. Conversely,
weak overall electron doping of the system gives rise to much larger electron 
doping in the empty $e_{g'}$ bands, accompanied by hole doping of the ${a_g}$ 
band. Thus, near the Mott transition, pronounced internal charge rearrangement 
takes place among the relevant subbands. 

ED DMFT cluster spectra illustrating the doping dependence in LaTiO$_3$, and 
cluster spectra corresponding to the orthorhombic and tetragonal bulk phases, 
can be found in Refs.~\cite{LTO1:prb77} and \cite{LTO2:prb77}, respectively.

\bigskip
\centerline{\bf {3.2.3. \ V$_2$O$_3$ }}
\bigskip
 
\begin{figure}  [t!] 
\begin{center}
\includegraphics[width=4.5cm,height=6.5cm,angle=-90]{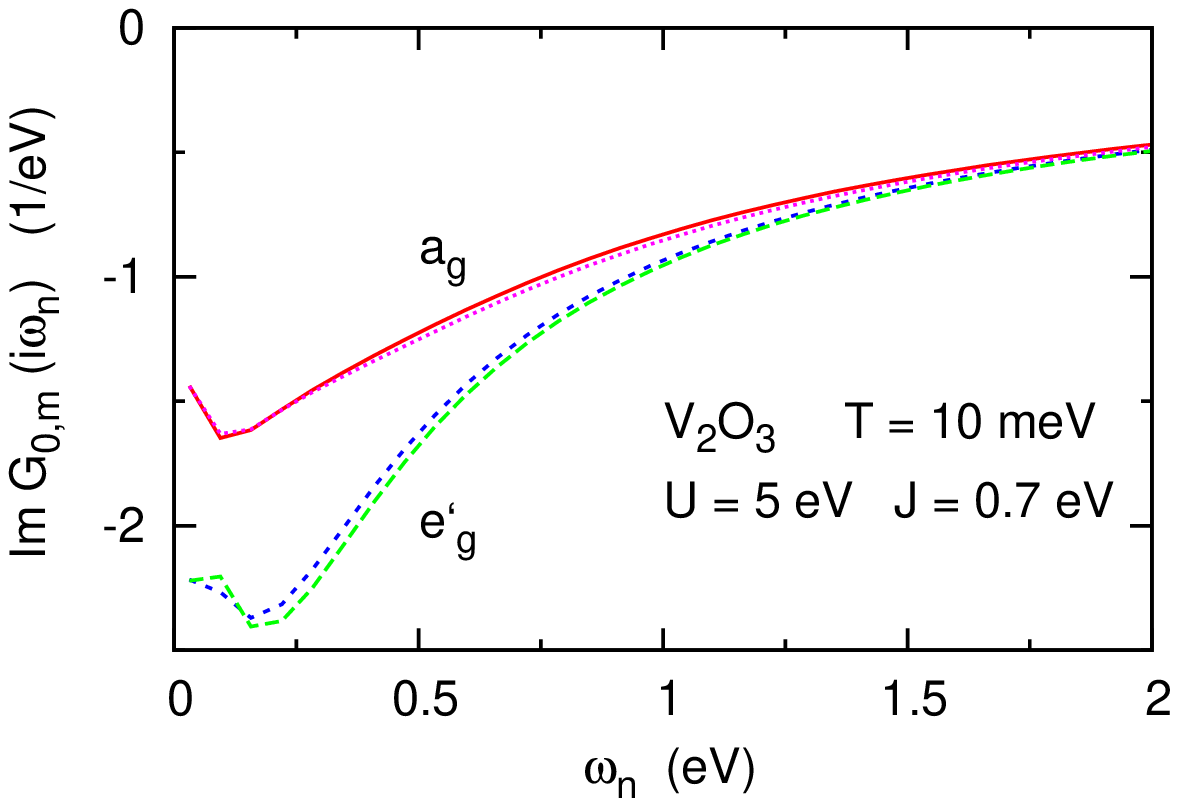} 
\includegraphics[width=4.5cm,height=6.5cm,angle=-90]{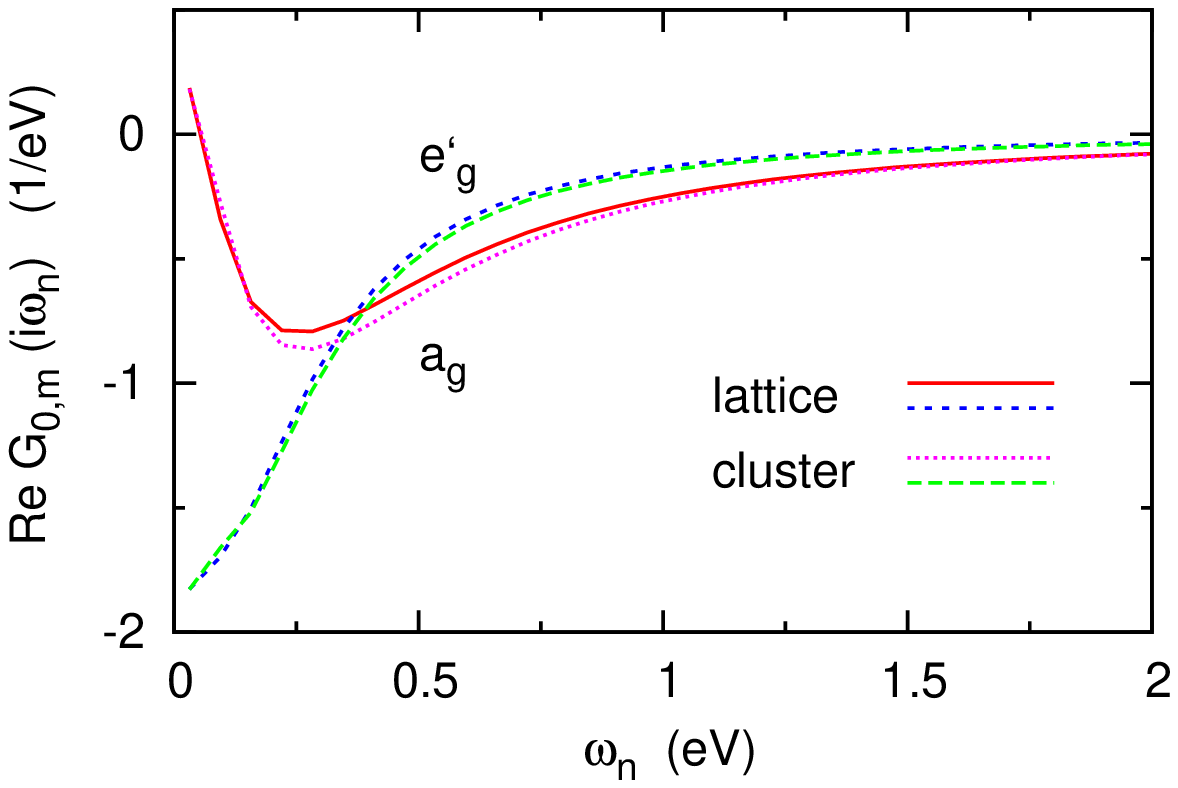} 
\end{center}
\caption{(Color online)
Discretization of V$_2$O$_3$ bath Green's function components, using 
two bath levels per impurity orbital; $U=5$~eV, $J=0.7$, 
$T=10$~meV. Upper (lower) panel: imaginary (real) part.
}\label{V2O3.g0}\end{figure}

Another material that exhibits enhanced orbital polarization near the Mott 
transition is V$_2$O$_3$ ($3d^2$) \cite{keller}. 
In this case, the corundum lattice geometry ensures that the doubly-degenerate 
$e_{g'}$ bands have a slightly larger binding energy than the singly-degenerate 
$a_g$ band. In Ref.~\cite{PRB78}, we have used ED DMFT to analyze this orbital
polarization for $n_s=9$, i.e., using two bath levels per impurity orbital.   
Figure~\ref{V2O3.g0} illustrates the quality of the discretization of the 
bath Green's function $G_{0,m}(i\omega_n)$ components. 
As pointed out above, cluster spectra are always gapped, so that 
Im\,$G^{cl}_{0,m}(i\omega_n)\rightarrow 0$ at very low frequencies. 
This limiting behavior is discernible, however, only below 
$\omega_0=\pi T=0.031$~eV, where the lattice and cluster Green's functions
strongly deviate from one another. Thus, by choosing a Matsubara grid with 
a temperature not much lower than $T\approx 10$~meV, these discrepancies are 
almost completely avoided and the finite cluster can serve as an adequate 
representation of the lattice continuum.

\begin{figure}  [t!] 
\begin{center}
\includegraphics[width=4.5cm,height=6.5cm,angle=-90]{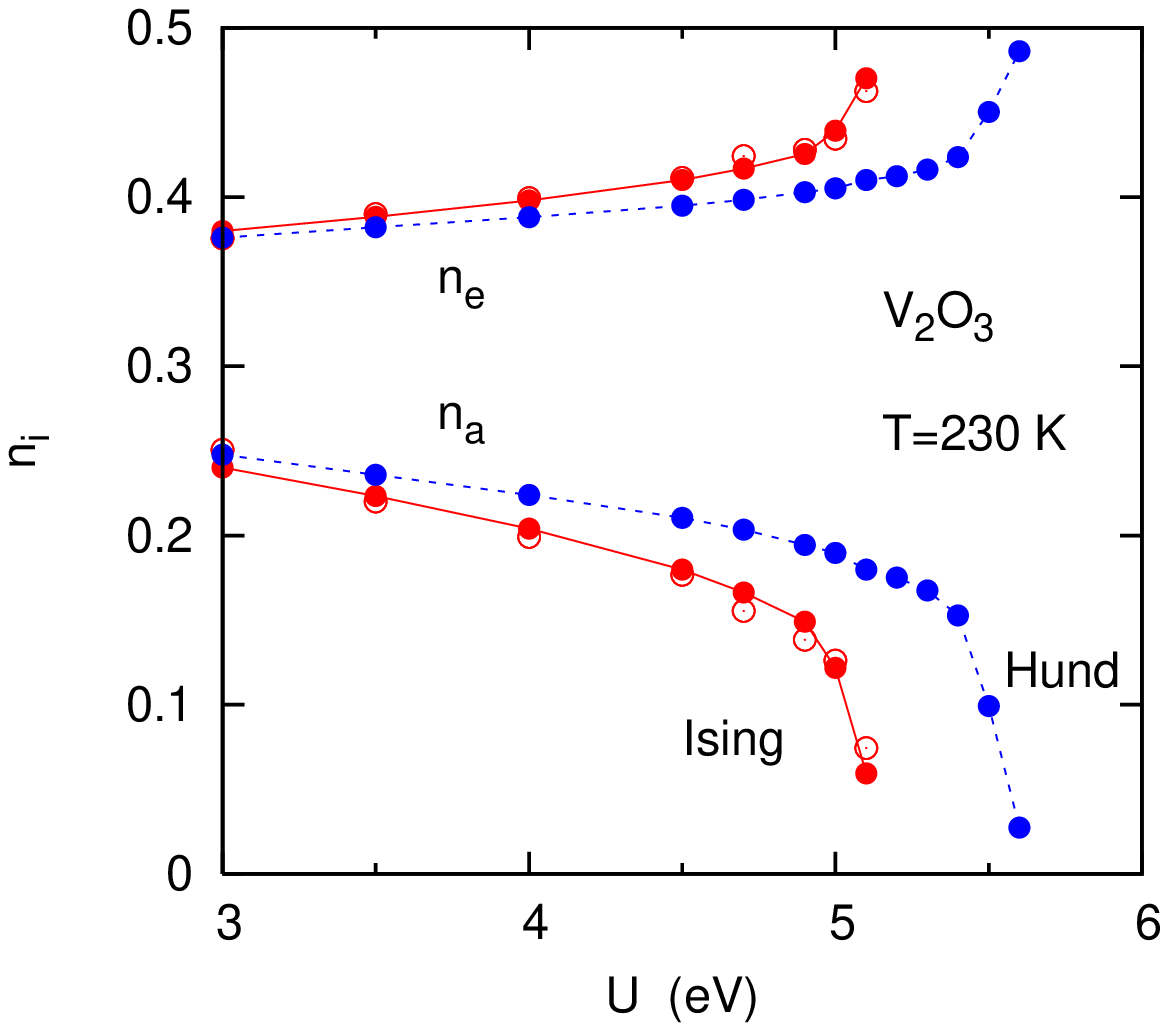} 
\includegraphics[width=4.5cm,height=6.5cm,angle=-90]{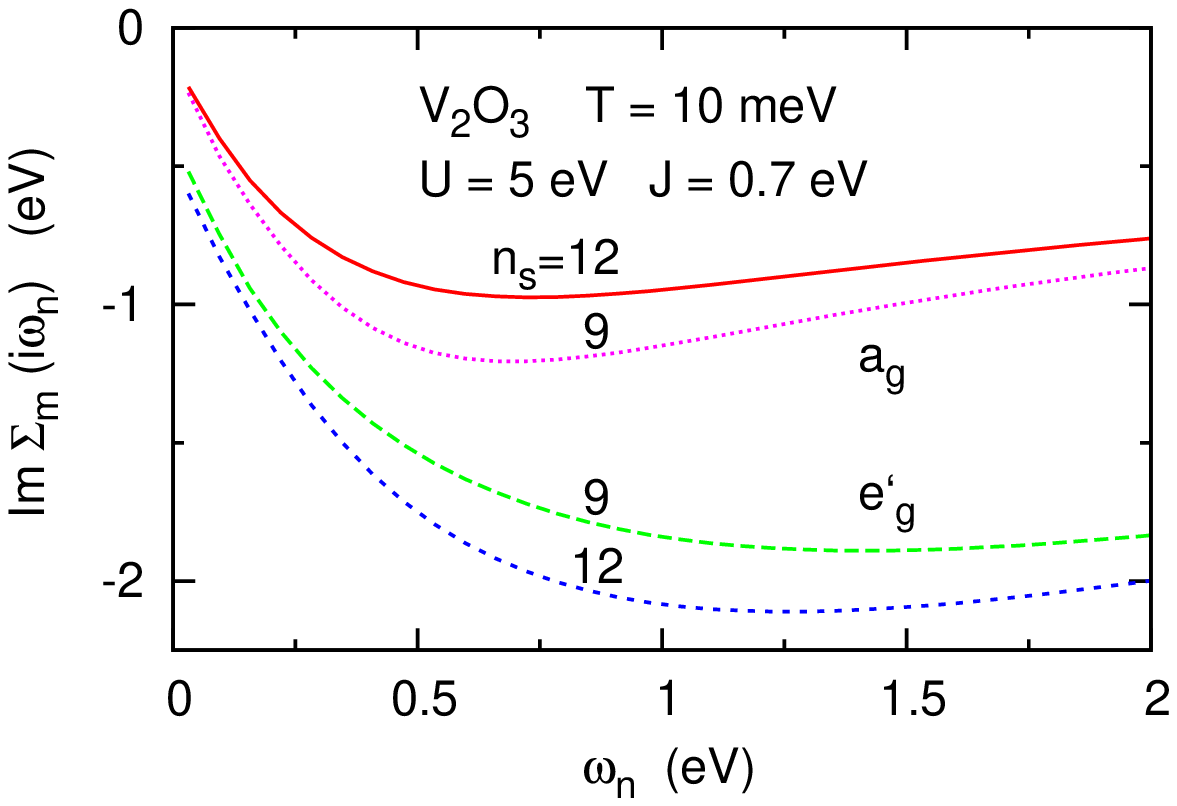} 
\end{center}
\caption{(Color online)
Upper panel: 
V$_2$O$_3$ subband occupancies per spin as functions of Coulomb energy.
Red symbols: Ising exchange, blue symbols: full Hund exchange.
Solid dots: ED DMFT \cite{PRB78}, empty: dots: QMC DMFT \cite{keller}. 
Lower panel: imaginary part of self-energy components for two and three 
bath levels per impurity orbital ($n_s=9$ and $n_2=12$, respectively) 
at $U=5$~eV, $J=0.7$, $T=10$~meV. 
}\label{V2O3}\end{figure}

Figure~\ref{V2O3} shows the Coulomb driven enhancement of the V$_2$O$_3$ orbital 
polarization. Near $U\approx 5$~eV, the $a_g$ band is pushed above the Fermi 
level and the half-filled $e_{g'}$ bands are split into lower and upper Hubbard 
peaks.
Note that the precise value of the critical Coulomb energy depends sensitively
on the exchange terms included among the many-body interactions. In the case
of only density-density terms (Ising-like exchange), we find $U_c\approx 5.1$~eV,
in agreement with Hirsch-Fye QMC DMFT calculations by Keller 
{\it et al.}~\cite{keller}. 
Full Hund exchange, on the other hand, shifts $U_c$ upward to about $5.6$~eV, 
a trend that is consistent with results obtained in ED and NRG studies of 
simpler two-band models \cite{prl2005,pruschke}.         
       
The lower panel of Figure~\ref{V2O3} shows the convergence of the self-energy 
components in the metallic phase close to the Mott transition with bath size. 
The nearly half-filled $e_{g'}$ subbands are seen to exhibit a significant 
low-energy scattering rate which is reminiscent of the spin freezing transition 
in the degenerate three-band model discussed in Section 3.1. The nearly empty 
$a_{g}$ band, on the other hand, is much more itinerant. Nevertheless, it exhibits 
a large effective mass enhancement of about $5$, and a small but finite scattering 
rate caused by the interaction with the almost Mott-localized $e_{g'}$ states.  
The differences for $n_s=9$ and $n_s=12$ are presumably caused by slightly 
different orbital occupancies. Close to a transition, the self-energy is 
particularly sensitive to small parameter changes. A similar effect was found
close to the spin-freezing transition in the degenerate three-band model 
discussed in Section 3.1.    


\bigskip
\centerline{\bf {3.2.4. \ Na$_{0.3}$CoO$_2$} }
\bigskip

Despite several years of experimental and theoretical investigations, basic 
features of the electronic properties of the intercalated layer compound 
Na$_{x}$CoO$_2$ are still controversial. The doping concentration $x=0.3$ 
is of particular interest since this material becomes superconducting 
when small amounts of water are added \cite{takada}.

As a result of the octahedral crystal field, the Co $3d$ bands are split
into $t_{2g}$ and $e_g$ subbands. With Na doping the occupancy of the $t_{2g}$ 
sector can be continuously tuned between $n=5$ and $n=6$. The rhombohedral 
symmetry at Co sites yields a further splitting of $t_{2g}$ orbitals into
singly-degenerate $a_g$ and doubly-degenerate $e_{g'}$ states. Because of 
the nearly two-dimensional hexagonal structure, the Fermi surface as 
predicted by LDA calculations \cite{singh.NCO} exhibits a large $a_g$ 
hole pocket centered at $\Gamma$, and six small $e_{g'}$ hole pockets 
along the $\Gamma K$ directions of the Brillouin Zone. 
These small pockets have so far not been observed in ARPES measurements
\cite{hasan,yang}.
As discussed above, local Coulomb interactions can give rise to charge 
transfer between subbands, implying increasing or decreasing orbital
polarization, depending on the details of the single-particle Hamiltonian.
 Several groups have carried out DMFT calculations in order to explore whether 
the small pockets might disappear because of $a_g$ to $e_{g'}$ electron transfer 
\cite{NCO.prl2005,perroni,NCO.EPJB,lechermann,marianetti}.

\begin{figure}  [t!] 
\begin{center}
\includegraphics[width=4.5cm,height=6.5cm,angle=-90]{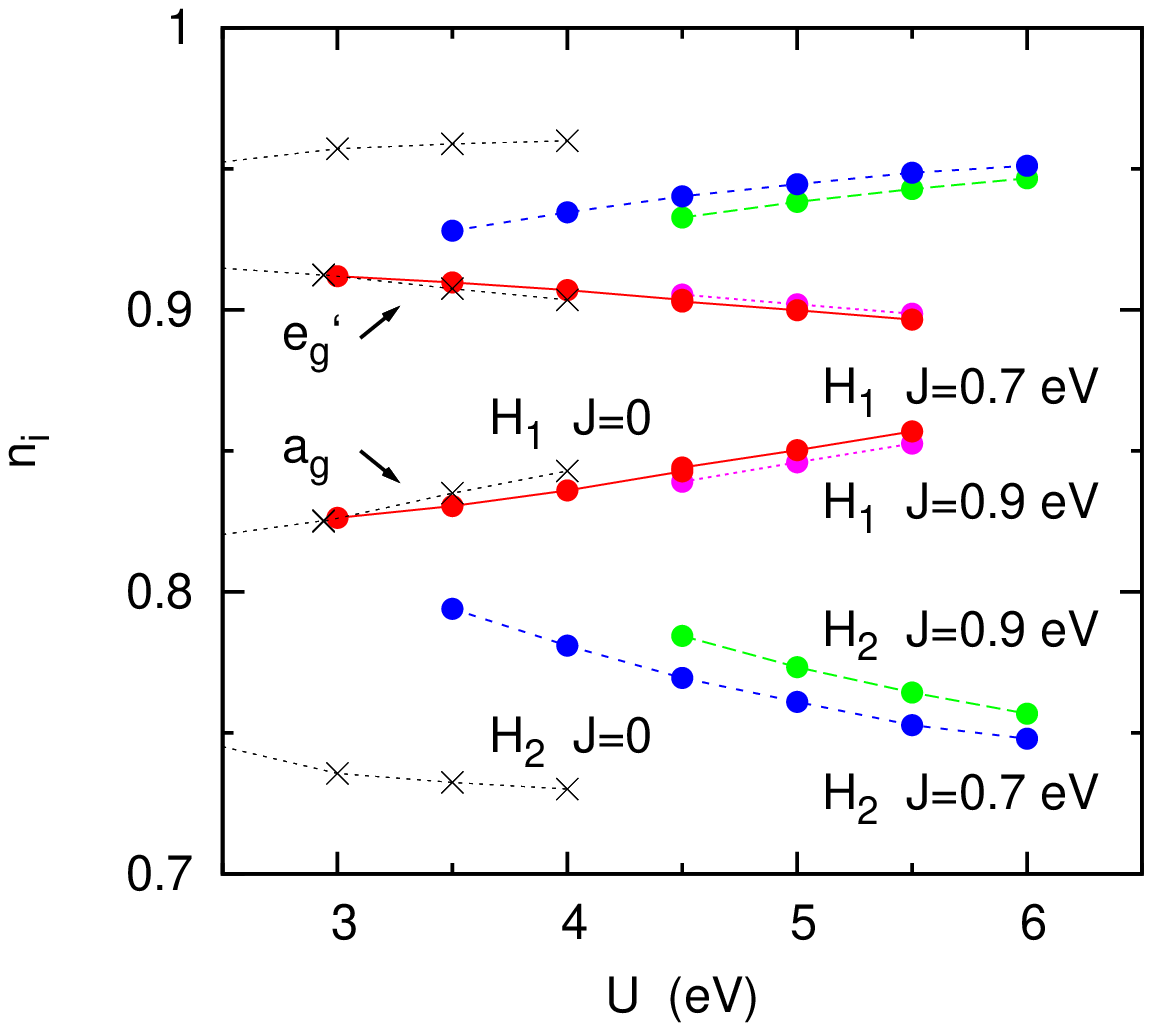} 
\includegraphics[width=4.5cm,height=6.5cm,angle=-90]{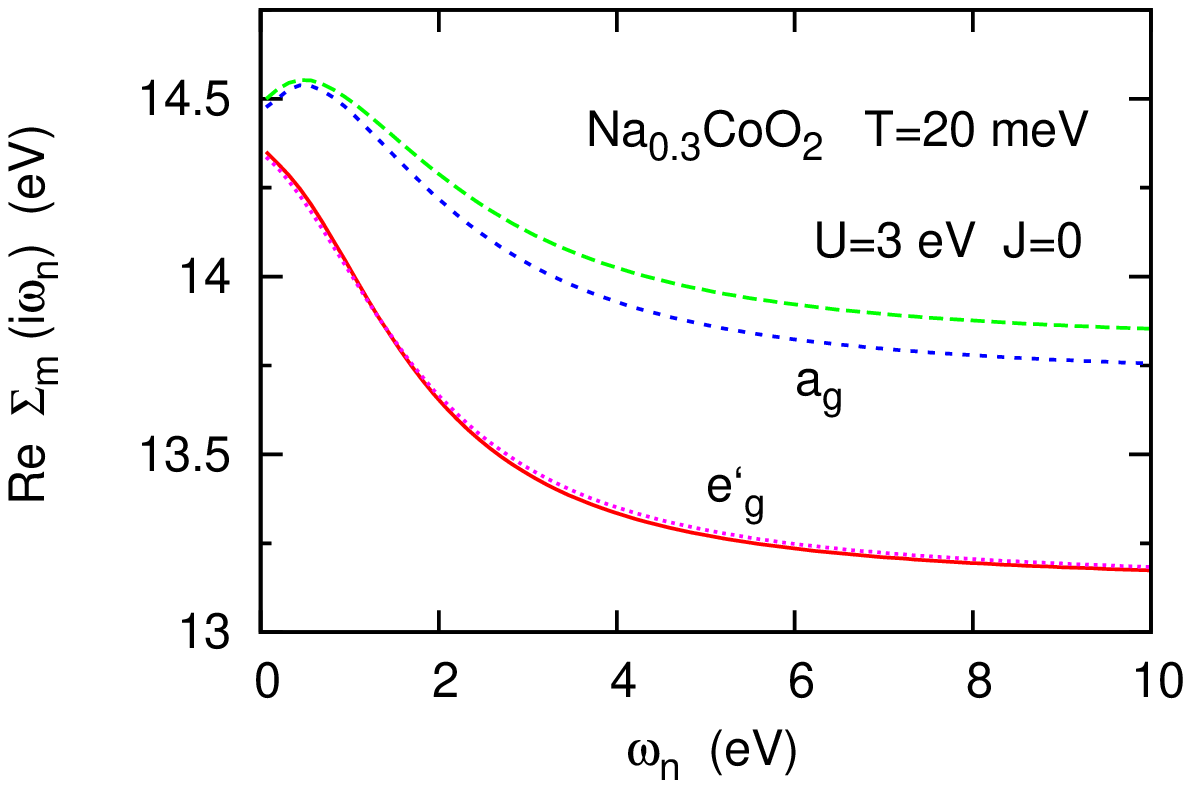} 
\includegraphics[width=4.5cm,height=6.5cm,angle=-90]{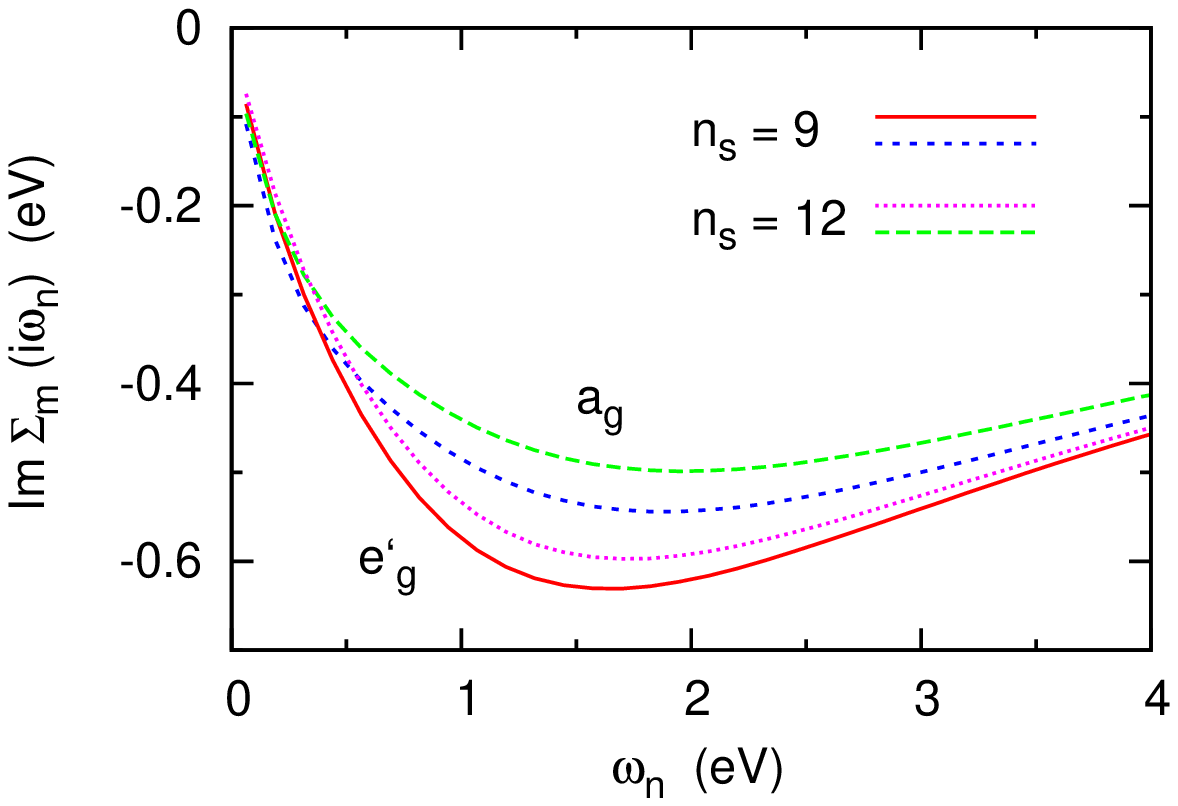} 
\end{center}
\caption{(Color online)
Upper panel:
Na$_{0.3}$CoO$_2$ subband occupancies as functions of Coulomb energy 
for two single-particle Hamiltonians $H_1$ and $H_2$ (see text) and 
different Hund exchange energies $J$ \cite{NCO.EPJB}.    
Lower two panels:
Real and imaginary components of Na$_{0.3}$CoO$_2$ self-energy calculated 
within ED DMFT for two and three bath levels per impurity $t_{2g}$ orbital;
Hamiltonian $H_2$, $U=3$~eV, $J=0$, and $T=20$~meV.
}\label{NaCoO}\end{figure}

The upper panel of Figure~\ref{NaCoO} summarizes the degree of orbital polarization 
obtained for slightly different single-particle Hamiltonians $H_1$ and $H_2$,
as well as different Hund couplings $J$ \cite{NCO.EPJB}.   The important 
difference between these Hamiltonians is that, because of different 
band crossings in the tight-binding fits, the effective crystal field 
$\Delta=\epsilon_{a_g} - \epsilon_{e_{g'}}$ is about $-130$~meV for $H_1$, 
while it is $-10$~meV for $H_2$. The band crossings obtained for $H_1$
correspond to those of high-quality LAPW LDA calculations.
As shown in the figure, these two tight-binding fits yield opposite trends
of the orbital occupancy with increasing local Coulomb energy. $H_1$ gives
rise to decreasing orbital polarization, whereas $H_2$ leads to increasing 
orbital polarization. These different trends are maintained even if the
Coulomb and energies are varied over wide ranges, $U\approx 3,\ldots, 6$~eV
and $J=0,\ldots,0.9$~eV. Evidently, correlation effects are not strong enough
to fill the $e_{g'}$ pockets.

The lower panels of Figure~\ref{NaCoO} illustrate the convergence of the 
ED self-energy components of Na$_{x}$CoO$_2$ with cluster size, using 
Hamiltonian $H_2$. The small differences obtained for $n_s=9$ and $n_s=12$ are 
partly related to the slightly different occupancies: $n_{a_g} = 0.753\ (0.740)$ 
and $n_{e_g'}= 0.946\  (0.959)$ for  $n_s=9\ (n_s= 12)$.
The important low-frequency features are seen to be in very good agreement 
for these cluster sizes, in particular, the small upward shift of $a_g$ states 
relative to $e_{g'}$ by about $0.15$~eV. We also point out that these ED 
results for $H_2$ are fully consistent with the ones obtained within QMC
for the same Hamiltonian \cite{marianetti}. In fact, the detailed comparison 
of the respective self-energy components shows, that, if identical input 
parameters are employed, nearly quantitative agreement is found for these 
impurity solvers \cite{marianetti.private}.

\begin{figure}  [t!] 
\begin{center}
\includegraphics[width=4.5cm,height=6.5cm,angle=-90]{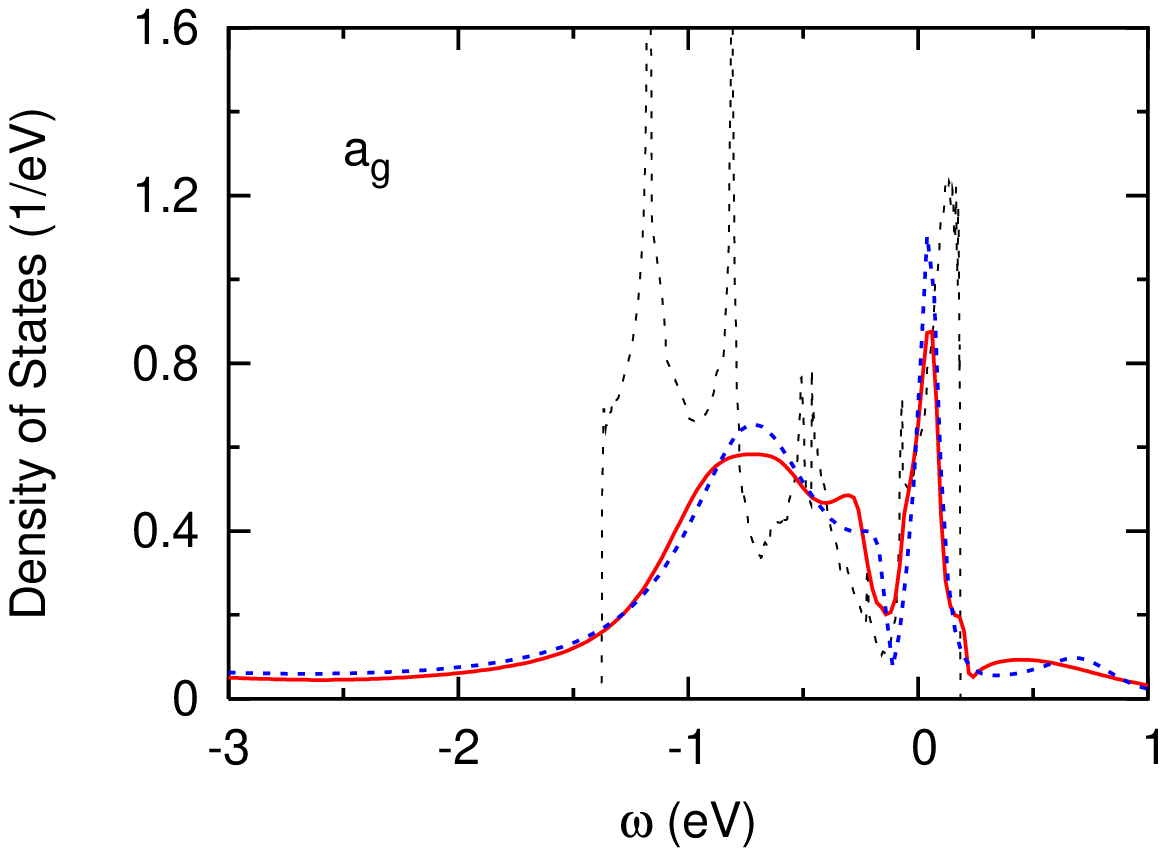} 
\includegraphics[width=4.5cm,height=6.5cm,angle=-90]{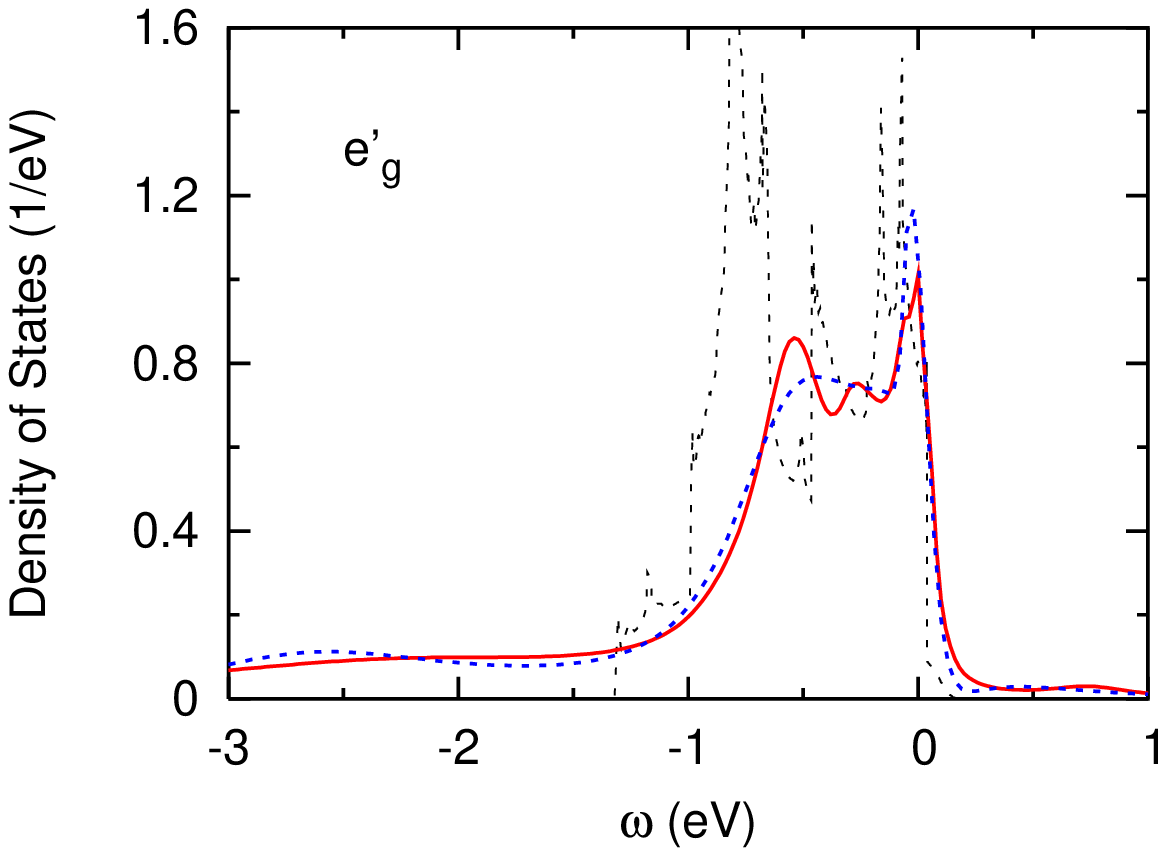} 
\end{center}
\caption{(Color online)
Quasiparticle spectra for $a_g$ (upper panel) and $e'_g$ (lower panel) subbands 
of Na$_{0.3}$CoO$_2$ at $U=3$~eV, $J=0.75$~eV, $T=10$~meV.
Solid (red) curves: spectra derived from lattice Green's function $G_m(\omega)$ 
after extrapolation of $\Sigma_m(i\omega_n)$ to real $\omega$; 
dashed (blue) curves: spectra derived via extrapolation of $G_m(i\omega_n)$; 
dotted (black) curves: single-particle density of states 
\cite{perroni}.    
}\label{NaCoO.A}\end{figure}

The $e_{g'}$ pockets can be made to vanish by adjusting the $t_{2g}$ 
crystal field to $\Delta=40,\ldots,90$~meV \cite{marianetti}. 
As a result, the $e_{g'}$ bands are pushed down, and Coulomb correlations 
can fill them completely. The physical origin of such a modification of the 
crystal field is presently unknown. A possible source of the discrepancy 
between theory and experiment is that ARPES is surface sensitive. As shown 
by Pillay {\it et al.}~\cite{pillay}, the surface layer of Na ions can give 
rise to an effective $t_{2g}$ crystal field, which substantially lowers the 
$e_{g'}$ bands with respect to the $a_g$ bands. 
  
The results shown in Figure~\ref{NaCoO} underline the complex interplay between 
kinetic energy and Coulomb repulsion in strongly correlated materials. 
While in some systems, such as LaTiO$_3$ and V$_2$O$_3$, orbital polarization 
tends to be enhanced by onsite Coulomb interactions, other systems, such as 
Ca$_{2-x}$Sr$_x$RuO$_4$ and Na$_{x}$CoO$_2$, exhibit more complex patterns. 
Because of their two-dimensional geometry, these oxides comprise subbands
of very different widths and spectral shapes. Orbital fluctuations
can then be enhanced or reduced as a result of Coulomb correlations.

As pointed out before, quasi-particle spectra at real $\omega$ can be derived by 
analytic continuation of the lattice Green's function $G_m(i\omega_n)$, or by first 
extrapolating $\Sigma_m(i\omega_n)$ to real $\omega$ and then evaluating $G_m(\omega)$. 
The comparison shown in Figure~\ref{NaCoO.A} demonstrates that both methods are 
consistent. Nonetheless, the latter method is seen to retain finer spectral details 
originating from the single-particle Hamiltonian. For instance, the $e_g$ spectrum 
shows two peaks below $E_F$ which evidently are the shifted and broadened density 
of states features near $0.4$ eV and $0.8$ eV below the Fermi level. Also, the 
peak close to $E_F$ exhibits some of the fine structure of the single-particle 
density of states. These details are lost if the spectrum is instead derived
via analytic continuation of the lattice Green's function.

\subsection{3.3. \ Degenerate Five-Band Model}

The discovery of superconductivity in iron pnictides and chalcogenides has 
stimulated a variety of investigations concerning the role of Coulomb 
correlations in these materials. 
In contrast to high-$T_c$ cuprates, which have antiferromagnetic 
Mott insulators as parent compounds, pnictides are correlated magnetic metals 
showing significant deviations from Fermi-liquid behavior. Moreover, 
as a result of the multi-band nature of pnictides, the interplay of Coulomb 
and exchange interactions give rise to phenomena not found in cuprates.

As there is no clear distinction between $t_{2g}$ and $e_g$ subbands in compounds 
such as FeAsLaO and FeSe, we have recently extended ED DMFT to five bands, where 
each of the impurity orbitals hybridizes with two bath levels ($n_s=15$) \cite{FeAsLaO}.
Because of the enormous size of the resulting Hilbert space, the cluster
Green's function is evaluated at $T=0$, i.e., for $\beta\rightarrow\infty$ 
in Eq.~(\ref{Gclm}), while the Matsubara grid corresponds to $T\approx 0.01$. 
 
The main result of this study is the identification of a spin freezing 
transition near the nominal $n=6$ occupancy of the Fe $3d$ states, i.e., 
approximately one electron away from the Mott phase at half-filling.       
Towards larger occupancies (electron doping), ordinary Fermi liquid behavior
is restored, whereas occupancies lower than $n\approx 6$ (hole doping) give
rise to more pronounced non-Fermi-liquid-like deviations caused by a non-vanishing 
scattering rate. Accordingly, the spin-spin correlation function   
$C_{m}(\tau)=\langle S_{mz}(\tau)S_{mz}(0)\rangle$ changes from Pauli-like
at large doping to Curie-Weiss-like at smaller doping. These findings 
are closely related to the spin-freezing transition \cite{prlwerner} 
in the degenerate three-band model near $n=2$ occupancy, i.e., about one 
electron away from the Mott phase at half-filling (see Section 3.1). 
   
Making use of an accurate single-particle Hamiltonian $t({\bf k})$, derived 
from maximally localized Wannier functions, and constrained RPA interaction 
parameters calculated within the same basis \cite{miyake}, it was also 
shown that, because of larger Coulomb interactions, FeSe lies on the 
non-Fermi-liquid side of the spin freezing transition, while FeAsLaO 
corresponds to a moderately correlated Fermi liquid \cite{FeSe}. 
Deviations from Fermi-liquid behavior were recently also shown to occur 
in the paramagnetic phase of $\alpha$-Fe \cite{Fe}.

\begin{figure}  [t!] 
\begin{center}
\includegraphics[width=4.5cm,height=6.5cm,angle=-90]{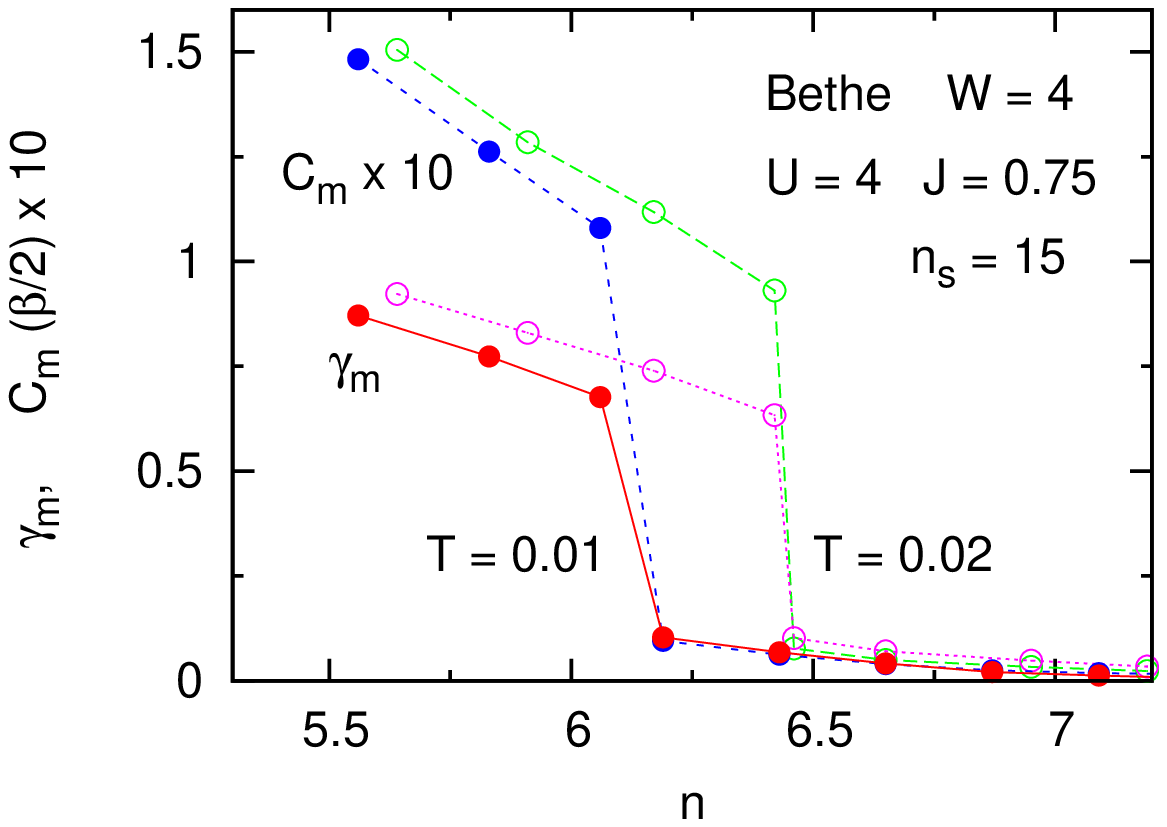} 
\includegraphics[width=4.5cm,height=6.5cm,angle=-90]{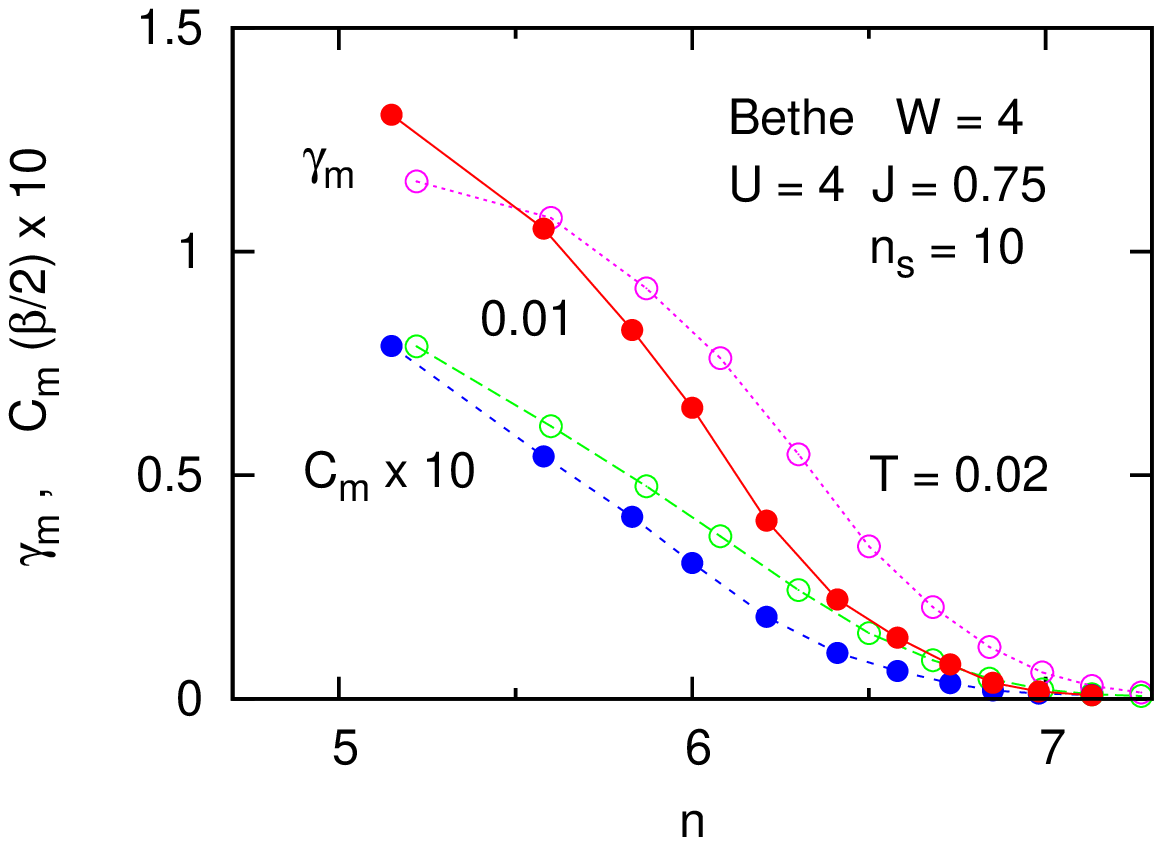} 
\end{center}
\caption{(Color online)
Low-energy scattering rates and spin-spin correlations derived within 
ED DMFT for a degenerate five-band model with Bethe lattice density of 
states ($W=4$); $U=4$, $J=0.75$. Solid dots: $T=0.01$, empty dots: $T=0.02$.
Upper panel: $n_s=15$; lower panel $n_s=10$. 
\label{five}
}\end{figure}

To elucidate the extent to which these results are related to crystal field effects 
and orbital polarization, we consider a fully degenerate five-band model, with 
semi-elliptical density of states components of width $W=4$ ($W=4$~eV corresponds 
to typical pnictide compounds). Using interaction parameters similar to those
in FeSe, we find a spin-freezing transition at about $n=6.2$ occupancy, 
implying non-Fermi-liquid properties for $n=6$. The upper panel of Figure
\ref{five} shows the variation of the low-energy scattering rate,
$\gamma_m = \vert {\rm Im}\, \Sigma_m(i\omega_n\rightarrow0)\vert $, with band 
occupancy for $T=0.01$ and $T=0.02$, assuming two bath levels per impurity orbital. 
Both Matsubara grids indicate a Fermi-liquid to non-Fermi-liquid transition 
near $n=6.2,\ldots,6.4$. The spin correlation function $C_m(\beta/2)$ changes 
from Pauli to Curie-Weiss behavior in the same occupancy range. 

Because of the importance of strongly correlated five-band transition metals and
transition metal oxides, it is of great interest to investigate whether ED DMFT 
calculations with only one bath level per impurity orbital can provide useful 
qualitative information concerning crucial dynamical properties of the system.
Calculations for $n_s=10$ (the dimension of largest sector of the Hilbert space
is $N=63504$) are extremely fast, requiring only few minutes per iteration 
at typical temperatures. Thus, for a qualitative exploration of the phase
diagram they would be highly valuable, before more accurate but 
time-consuming calculations for $n_s=15$ are carried out.  
 
\begin{figure}  [t!] 
\begin{center}
\includegraphics[width=4.5cm,height=6.5cm,angle=-90]{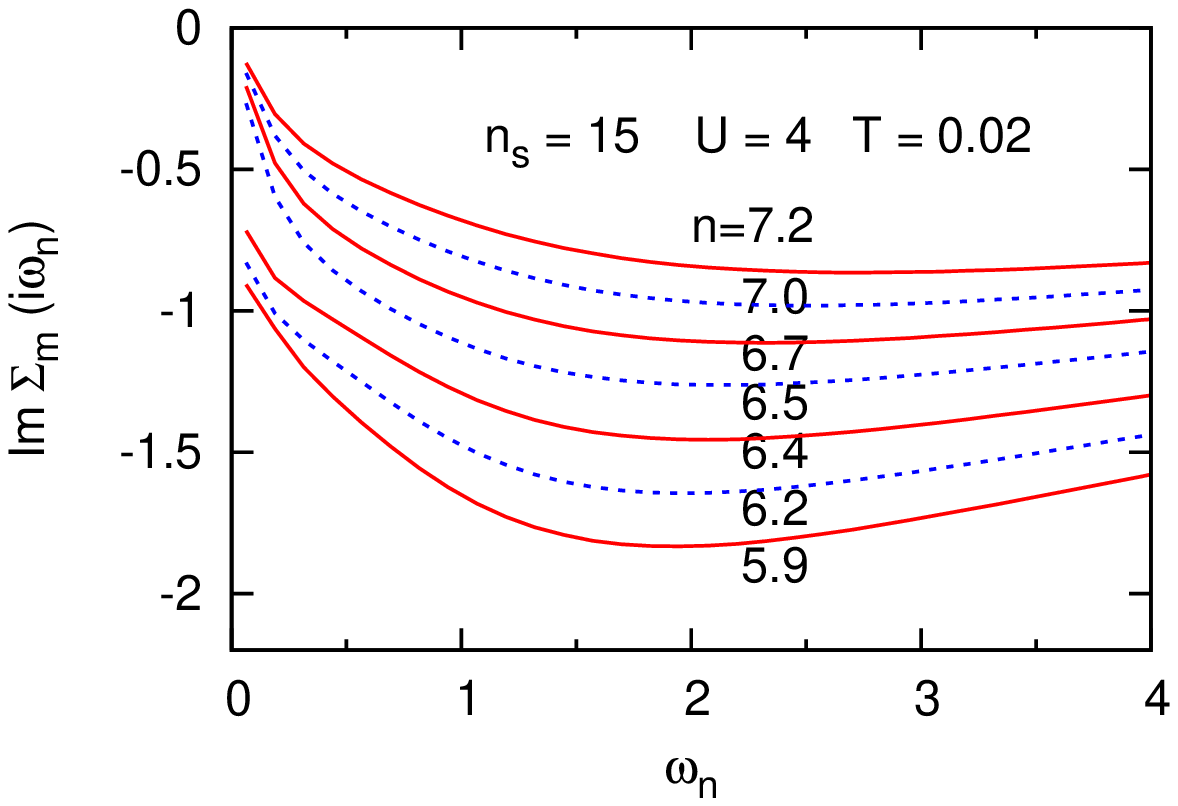} 
\includegraphics[width=4.5cm,height=6.5cm,angle=-90]{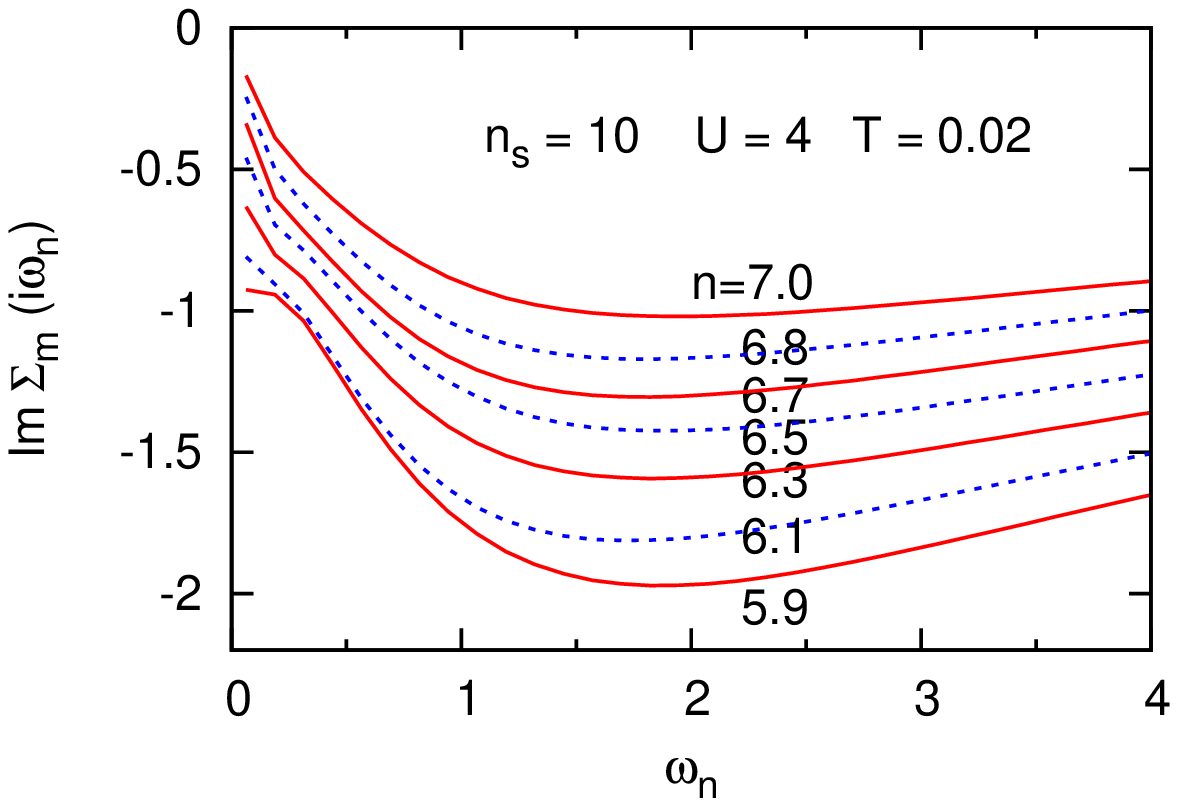} 
\end{center}
\caption{(Color online)
Self-energy of degenerate five-band model for several occupancies near the 
spin freezing transition; $U=4$, $T=0.02$. Upper panel: $n_s=15$, two bath
levels per impurity orbital. Lower panel: $n_s=10$, one bath level per 
impurity orbital. Note the onsets and low-energy kinks. 
\label{five.sigma}
}\end{figure}

The lower panel of Figure~\ref{five} shows the scattering rate and spin 
correlation function as functions of band occupancy for $n_s=10$, i.e., 
for only one bath level per impurity orbital. Both quantities now exhibit a 
less abrupt change from Fermi-liquid behavior at large $n$ to non-Fermi-liquid 
behavior at small $n$ than for $n_s=15$. Nevertheless, qualitatively, 
these results also suggest the existence of a spin-freezing transition 
near $n=6.5$ occupancy. Evidently, this transition is not caused 
by orbital polarization.
As in the case of the three-band model \cite{prlwerner}, it is purely
related to many-electron Coulomb and exchange interactions at about 
one electron away from half-filling.

Figure~\ref{five.sigma} illustrates the influence of the bath size on the 
self-energy. While uncertainties at low frequencies are larger for $n_s=10$ 
than for $n_s=15$, the qualitative picture is the same in both cases. At large 
occupancies, Im\,$\Sigma_m(i\omega_n)\sim\omega_n$ for $\omega_n\rightarrow0$, 
consistent with Fermi-liquid properties. At about $n\approx 6.5$,
Im\,$\Sigma_m$ begins to exhibit an onset, indicating the breakdown of 
Fermi-liquid behavior due to the presence of frozen moments. 
It would be very interesting to compare the ED results shown in Figs.~\ref{five} 
and \ref{five.sigma} with corresponding data obtained via continuous-time QMC DMFT.    

The overall consistency of the five-band results for $n_s=10$ with the more
accurate ones for $n_s=15$ might be very useful for future applications 
involving five-band materials. The computationally rather straightforward 
ED DMFT approach for $n_s=10$ could then be used to scan a wide range of 
system parameters, such as temperature, pressure (band width), crystal field 
splitting, doping concentration, Coulomb and exchange interactions, etc.       
For greater accuracy, more demanding calculations with $n_s=15$ could then 
be performed near critical values of some of these parameters.
One could also consider extending finite-temperature ED DMFT to $f$-electron 
systems with $n_c=7$ orbitals, using $7$ bath levels, i.e. $n_s=14$.        

\begin{figure} [t!] 
\begin{center}
\includegraphics[width=4.5cm,height=6.5cm,angle=-90]{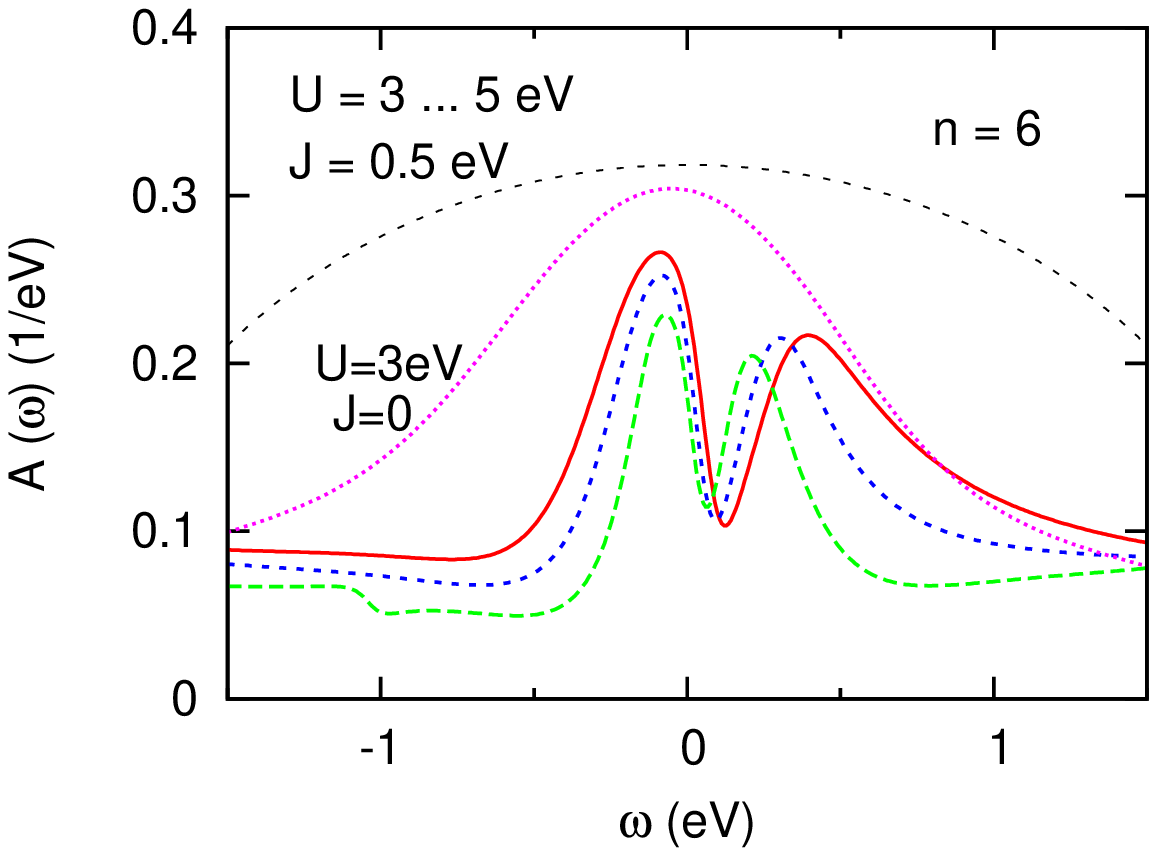} 
\includegraphics[width=4.5cm,height=6.5cm,angle=-90]{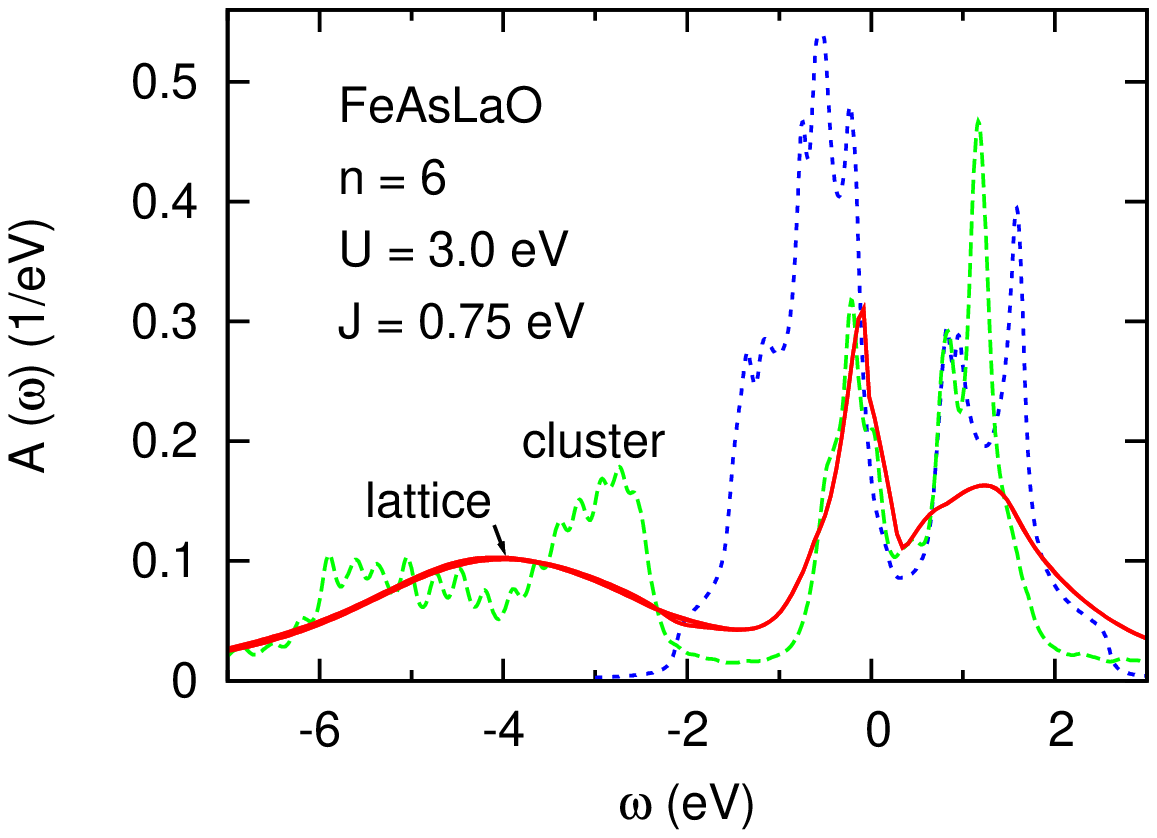} 
\end{center}
\caption{(Color online)
Upper panel:
Low-energy region of spectral distribution of degenerate five-band model ($W=4$ eV)
at occupancy $n=6$, calculated within ED DMFT ($n_s=10$) for several Coulomb 
energies: $U=3$~eV (solid red curve), $U=4$~eV (short-dashed blue curve), 
and $U=5$~eV (long-dashed green curve); $J=0.5$~eV. Spectra for $J=0.75$~eV
and $J=1$~eV are similar. Also shown is the result for $U=3$~eV, $J=0$ (dotted 
magenta curve). The bare density of states is indicated by the black dashed curve.
Lower panel:
Fe $3d$ quasiparticle spectra of FeAsLaO, calculated within ED DMFT ($n_s=15$) 
for $U=3$~eV, $J=0.75$~eV \cite{ELL5J}. Solid (red) curve: lattice spectrum;
long-dashed (green) curve: cluster spectrum. The bare $3d$ density of states 
is indicated by the short-dashed blue curve \cite{miyake}. 
\label{ELL5J}
}\end{figure}

We close this subsection by pointing out that the self-energies shown in 
Figure~\ref{five.sigma} do not only exhibit a low-energy onset towards 
half-filling, but also a kink near $\omega_n\approx0.1$~eV. Extrapolation 
of $\Sigma_m(i\omega_n)$ to real $\omega$ reveals that this kink is associated 
with a resonance in Im\,$\Sigma_m(\omega)$ which gives rise to a pseudogap in 
the density of states slightly above the Fermi level. In fact, the non-vanishing 
low-energy scattering rate may be viewed as a result of this resonance in the
self-energy. As shown in Ref.~\cite{ELL5J}, the pseudogap is remarkably stable 
over a wide range of Coulomb and exchange energies, as long as $J$ is not too 
small. The upper panel of Figure~\ref{ELL5J} illustrates some of these
quasi-particle spectra, which are derived by analytic continuation of the
lattice Green's function. Moreover, the local spin correlation function exhibits 
a pronounced maximum at low energies, suggesting that the peak in the self-energy 
corresponds to a collective mode induced by spin fluctuations. These features 
disappear at small $J$ (see the spectrum for $U=3$~eV, $J=0$). 
This behavior is intimately related to the fact that, 
at half-filling, the system is a Mott insulator for $J=0.5,\ldots,1.0$~eV, 
while it becomes metallic for small $J$. Thus, the Mott phase in this multi-band 
system, and the collective mode in  Im\,$\Sigma_m(\omega)$ away from half-filling, 
are a consequence of Hund coupling. Qualitatively similar behavior is also
found in the degenerate three-band Hubbard model discussed in Section 3.1.    

The pseudogap induced via Hund coupling found in the degenerate five-band model 
might be relevant also for the understanding of the electronic properties of 
iron-based pnictides. As shown in Figure~\ref{ELL5J} (lower panel) for FeAsLaO, 
the bare density of states as well as the quasi-particle spectrum (obtained
via analytic continuation of the lattice Green's function) reveal a pronounced
minimum slightly above $E_F$. (Also shown is the corresponding cluster spectrum, 
which agrees well with the lattice spectrum in the vicinity of $E_F$.) Similar 
features have been obtained for a variety of pnictides (see, for instance, 
Ref.~\cite{yin}). The results for 
the degenerate five-band suggest that the pseudogap above $E_F$ is a generic 
feature caused by multi-band Coulomb correlations within the $3d$ shell 
and that its existence depends crucially on realistic values of Hund coupling. 
The paramagnetic quasi-particle distribution of actual pnictides should 
therefore consist of a combination of correlation features associated with 
$J$ and signatures related to the bare density of states. Evidence for this
behavior seems to have been observed in recent experimental optical data for
 BaFe$_2$Se$_2$ \cite{wang2,schafgans}.


\subsection{3.4. \ Triangular Lattice}

The role of spatial fluctuations is of great importance for a variety of 
strongly correlated systems, in particular, in the high-$T_c$ cuprates,
in organic crystals, and in graphene. In the next three subsections we 
use the cluster extension of finite-temperature ED DMFT to elucidate the 
non-local properties of the self-energy for various two-dimensional
single-band materials.    
 
The layered charge transfer salts $\kappa$-(BEDT-TTF)$_2 X$, where $X$ denotes 
inorganic monovalent anions such as Cu[N(CN)$_2$]Cl or Cu$_2$(CN)$_3$, exhibit 
a remarkably rich series of phases as function of hydrostatic pressure, 
including Fermi-liquid and bad-metallic behavior, superconductivity, 
as well as paramagnetic and antiferromagnetic insulating phases 
\cite{lefebre,kurosaki}.
Since these compounds have triangular lattice geometries, with equivalent or 
inequivalent nearest neighbor hopping interactions, a feature of special 
interest is magnetic frustration. Thus, long-range magnetic ordering is 
increasingly frustrated if the lattice becomes more isotropic.

\begin{figure}  [t!] 
\begin{center}
\includegraphics[width=4.5cm,height=6.5cm,angle=-90]{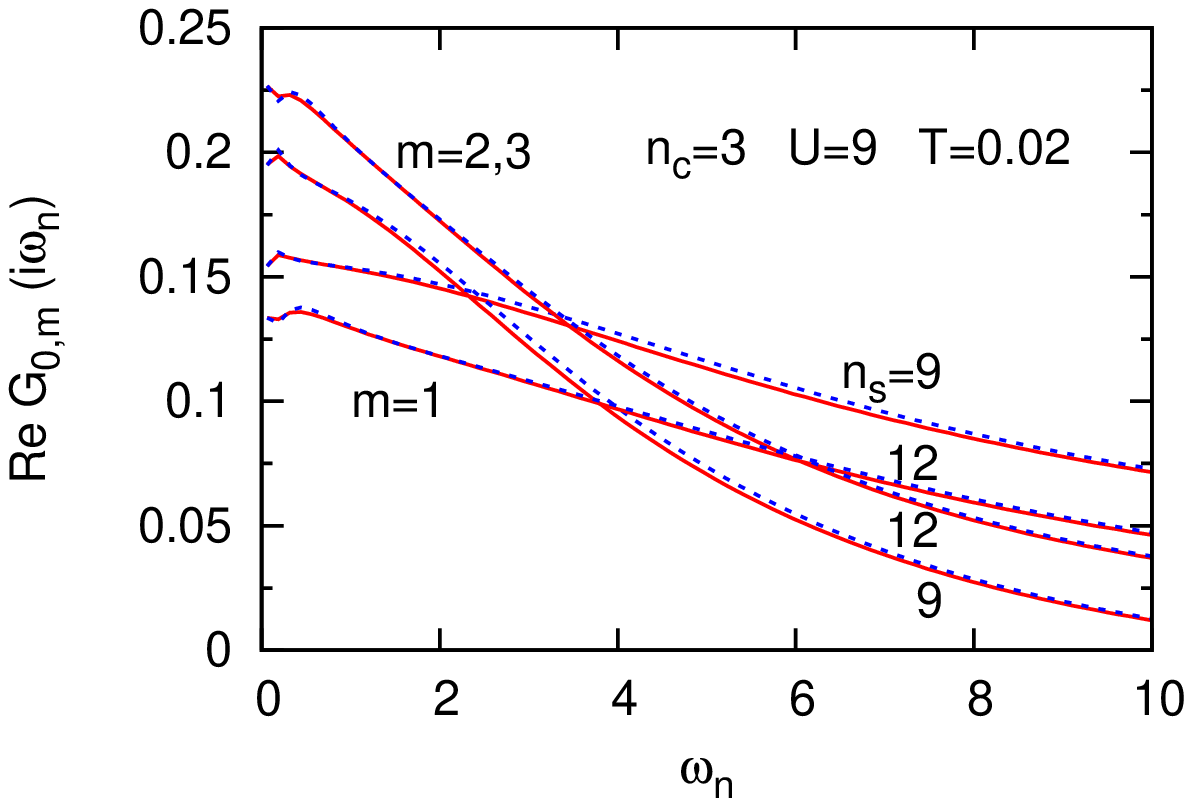} 
\includegraphics[width=4.5cm,height=6.5cm,angle=-90]{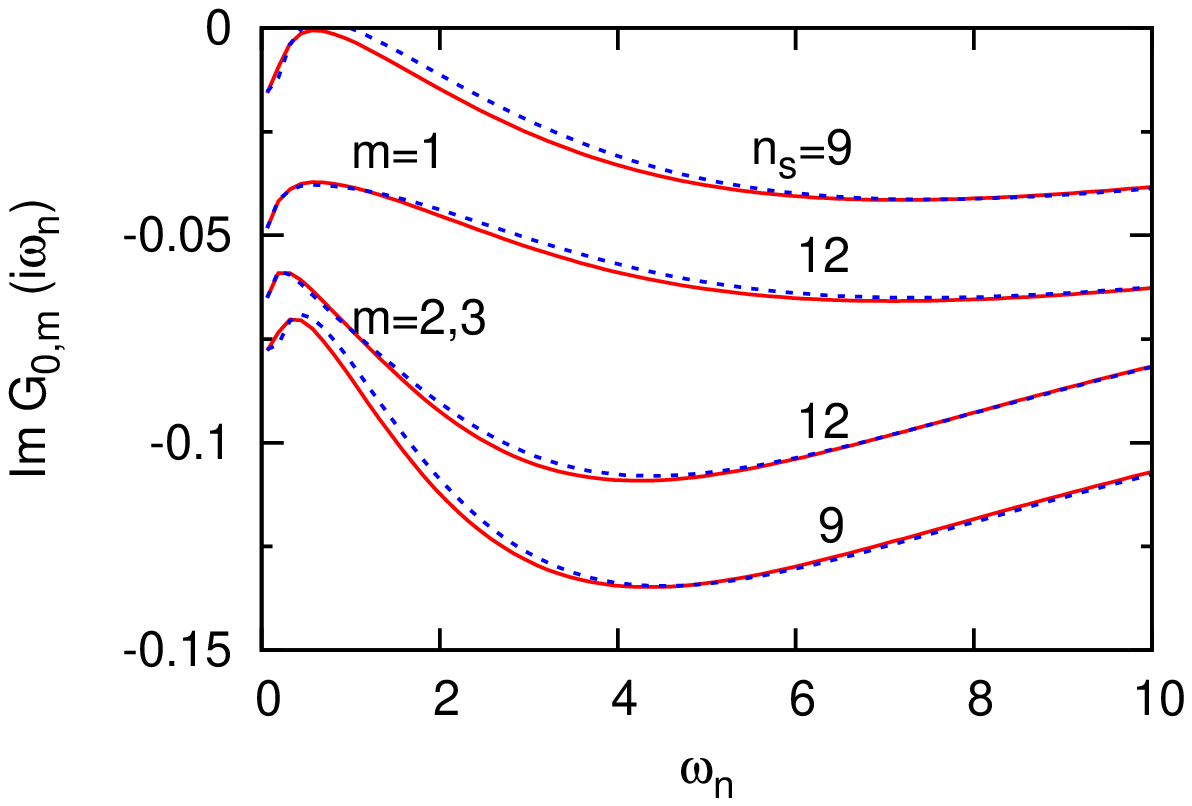} 
\end{center}
\caption{(Color online)
Discretization of bath Green's function components $G_{m,0}(i\omega_n)$
for isotropic triangular lattice using two ($n_s=9$) and three ($n_s=12$) bath 
levels per molecular orbital; $t=1$, $U=9$, $T=0.02$. 
For clarity, the results for $n_s=9$ are shifted up ($m=1$) or down ($m=2,3$)
by $0.025$. Upper panel: real part; lower panel: imaginary part.
}\label{tre.g0}\end{figure}

\begin{figure}  [t!] 
\begin{center}
\includegraphics[width=4.5cm,height=6.5cm,angle=-90]{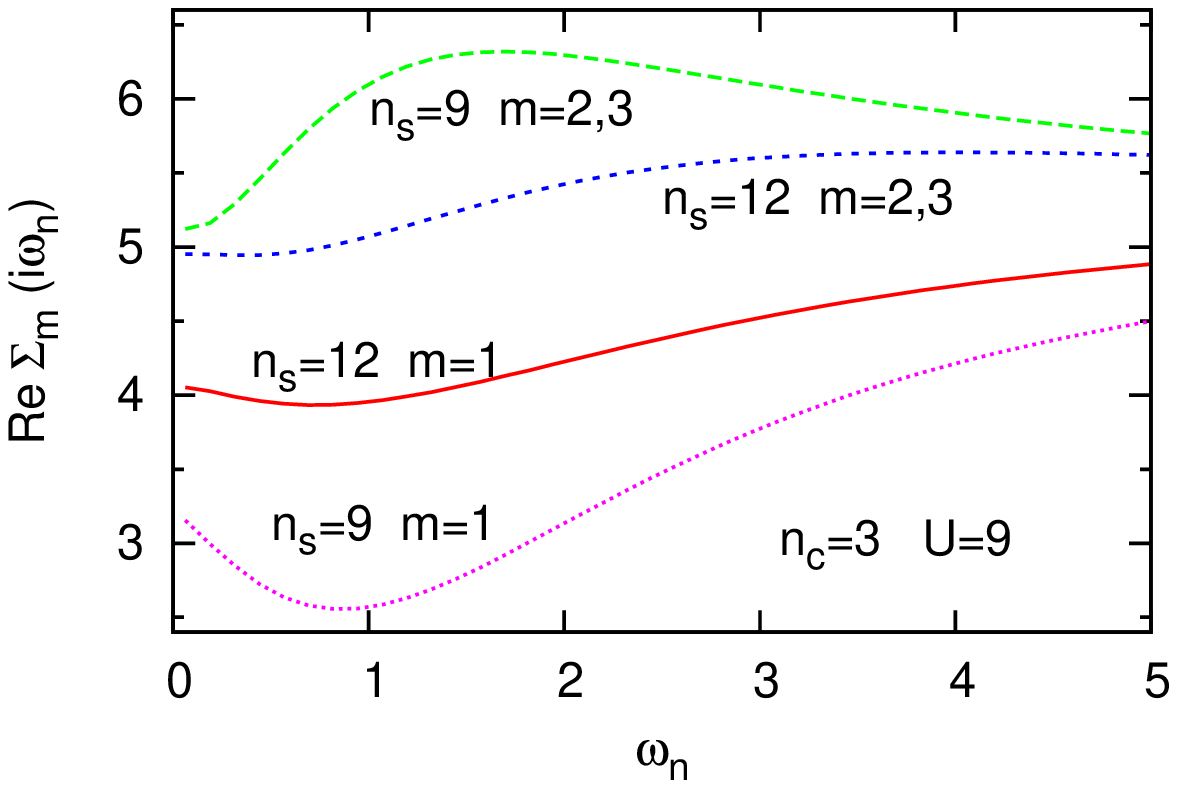} 
\includegraphics[width=4.5cm,height=6.5cm,angle=-90]{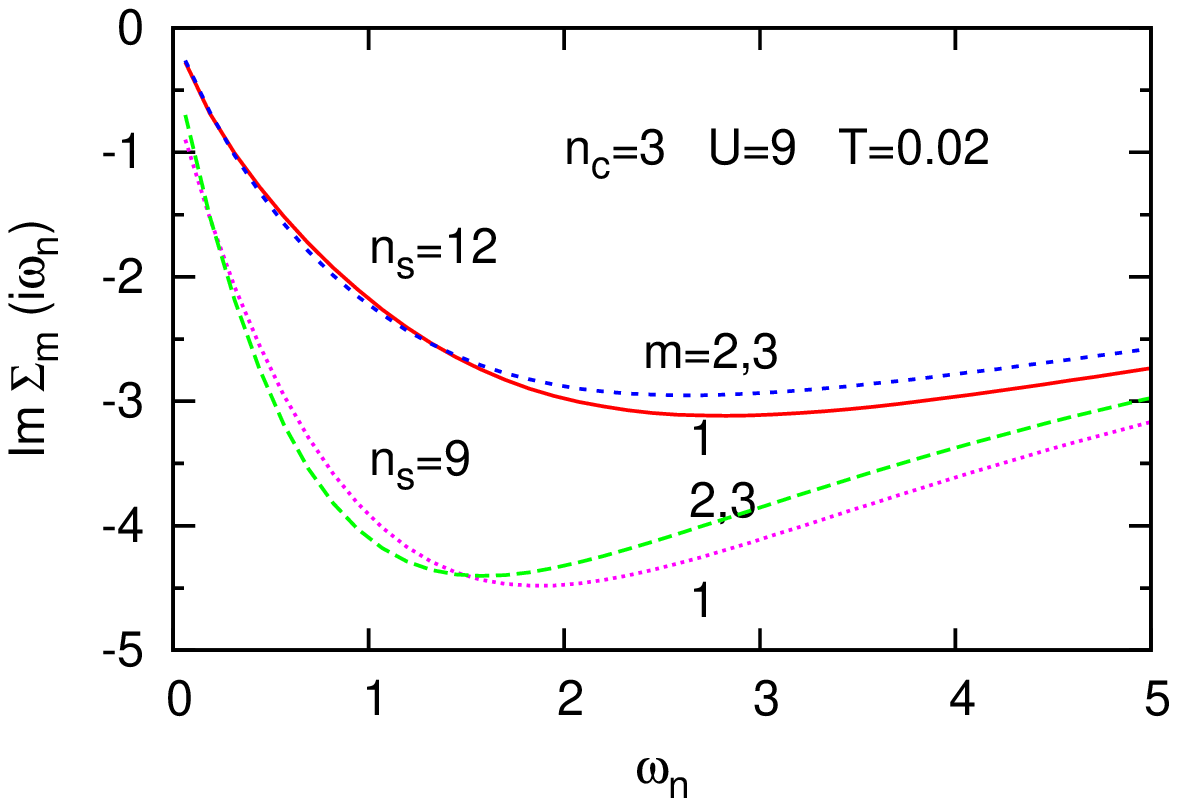} 
\end{center}
\caption{(Color online)
Self-energy components for isotropic triangular lattice using two ($n_s=9$) 
and three ($n_s=12$) bath levels per molecular orbital; $t=1$, $U=9$, $T=0.02$.   
Upper panel: real part; lower panel: imaginary part.
}\label{tre.s}\end{figure}

In Ref.~\cite{PRB79} a cluster extension of DMFT combined with finite-temperature 
ED was used to study the Mott transition in two-dimensional frustrated systems, 
in particular, the variation of the $T/U$ phase diagram with the degree of 
anisotropy. In the isotropic, fully frustrated limit ($t'=t$), the phase diagram 
is qualitatively similar to the one obtained within the single-site DMFT. 
On the other hand, for intermediate anisotropy $t'\approx 0.8t$, re-entrant 
behavior was found. Analogous QMC CDMFT results \cite{ohashi} also exhibit 
re-entrant behavior, albeit at higher temperatures. Also, only the phase 
boundary $U_{1c}(T)$ was determined. More recent continuous-time QMC DMFT
calculations for $t'=0.44t$ yield $T_c=0.13t$ \cite{sentef}, in approximate 
agreement with the ED result, $T_c\approx 0.11t$ \cite{PRB79}.

The unit cells employed in Ref.~\cite{PRB79} consist of three sites, 
and the bath was assumed to contain six levels ($n_s=9$). 
In Ref.~\cite{PRB78} it was shown that, despite this small bath size, 
for $T=0.02$ a rather accurate discretization of the bath Green's 
function is achieved ($t=1$, band width $W=9$). Figure~\ref{tre.g0} 
illustrates how the quality of this  discretization improves further if three
bath levels per site are included ($n_s=12$). We consider here only the
isotropic case $t'=t$, so that in the molecular-orbital basis the Green's
function is diagonal, as indicated in Eq.~(\ref{G3}). The components 
$G_{0,m}(i\omega_n)$ can then be discretized independently. Although the 
fit for $n_s=9$ is already quite good, several more detailed low-energy 
features are reproduced better for $n_s=12$.    
 
Figure~\ref{tre.s} shows the self-energy components for both bath sizes.
The differences in this case are caused by the fact that the Coulomb energy 
$U=9$ is very close to the lower boundary $U_{c1}$ of the coexistence 
region \cite{PRB79}. Moreover, the value of $U_{c1}$ is slightly smaller 
for $n_s=9$ than for $n_s=12$. Similar differences close to a phase boundary  
were also found in the degenerate three-band model discussed in Section 3.1, 
for instance, in the upper panel of Figure~4. 
The chemical potential $\mu=3$ nearly coincides with the spin freezing 
transition, so that small changes due to different 
bath sizes can lead to larger changes in scattering rates and other 
dynamical properties. Thus, near phase boundaries the size of the bath 
plays a more subtle role. The important point, however, is that this merely
affects the precise values of certain quantities, such as the critical Coulomb 
energies, and not the qualitative aspects of the phase diagram.           
      
ED DMFT cluster spectra for isotropic and anisotropic triangular lattices
can be found in Refs.~\cite{PRB78} and \cite{PRB79}, respectively. 

\subsection{3.5. \ Square Lattice}

One of the most widely studied applications of cluster DMFT is the two-dimensional
Hubbard model for the square lattice which is believed to capture some of the 
essential ingredients of the physics of cuprate high-temperature superconductors. 
The parent compounds at half-filling are Mott insulators while in the overdoped 
region at large electron or hole concentrations ordinary Fermi-liquid behavior
is observed. In the underdoped region, significant deviations from Fermi-liquid
properties are found, including pseudogap formation and a strongly momentum
dependent scattering rate along the Fermi surface. Evidently, such effects
cannot be described within a single-site or local version of DMFT. The most 
natural unit cell for taking into account short-range fluctuations are 
$2\times2$ clusters as shown in the center diagram of Figure 2.  
Several groups have applied $T=0$ ED CDMFT to investigate the non-local 
properties of the single-band Hubbard model for a square lattice  
\cite{capone2004,civelli,kyung,merino,koch,sakai}.
     
Here we focus on the Hamiltonian defined in Eq.~(\ref{Hcl}), with $t=0.25$ and 
$t'=-0.075$ as first and second neighbor hopping interactions, respectively
(band width $W=2$).
In Ref.~\cite{PRB80} we have used finite-temperature ED CDMFT to study 
various aspects of this model, in particular, the collective mode in the 
$(\pi,0)$ component of the self-energy which gives rise to the opening of 
the pseudogap at low doping, the increase of the scattering rate with
decreasing doping, and the striking asymmetry between electron and hole 
doping caused by the second-neighbor hopping $t'$. For instance, at $U=10t=2.5$ 
the Mott transition driven by electron doping is first-order, with characteristic 
hysteresis behavior of the orbital occupancies and double occupancy. 
In contrast, the transition induced by hole doping is continuous. 
In the latter case, upon doping the van Hove singularity at $(\pi,0)$ shifts 
toward the Fermi energy, whereas for electron doping it moves away from $E_F$.

\begin{figure}  [t!] 
\begin{center}
\includegraphics[width=4.0cm,height=6.5cm,angle=-90]{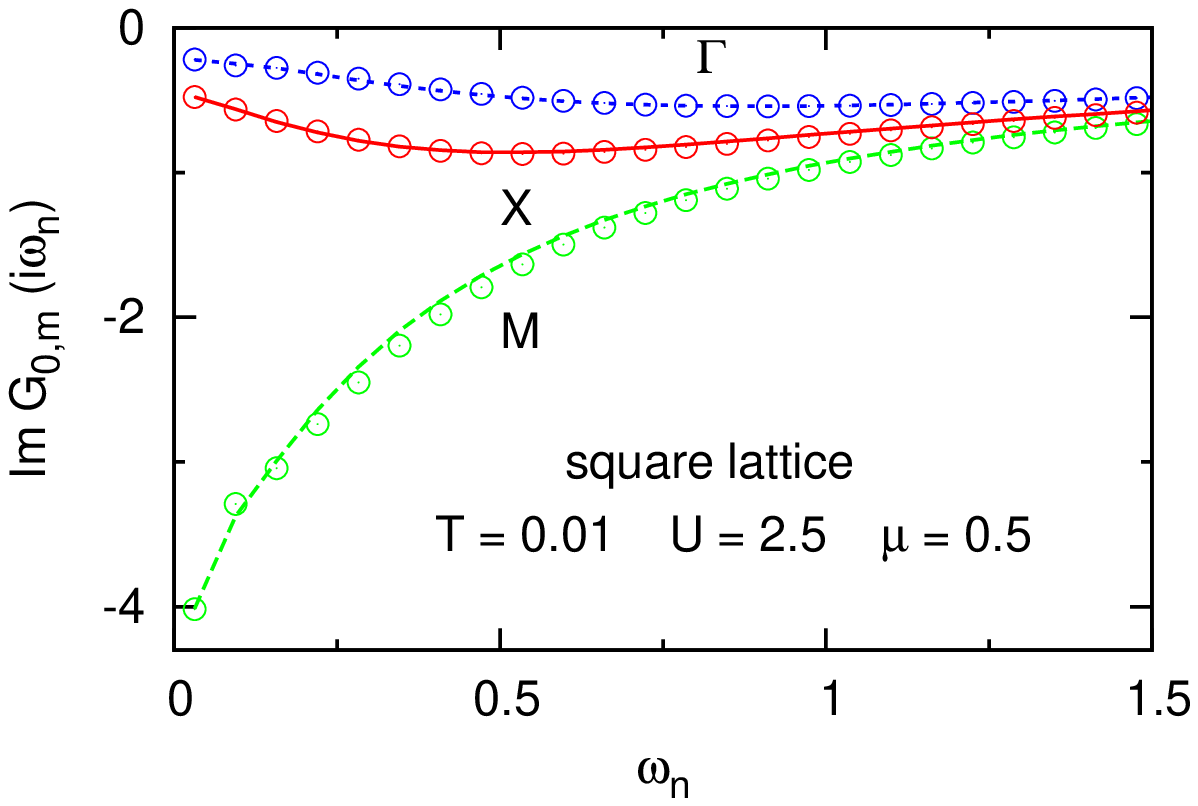} 
\includegraphics[width=4.0cm,height=6.5cm,angle=-90]{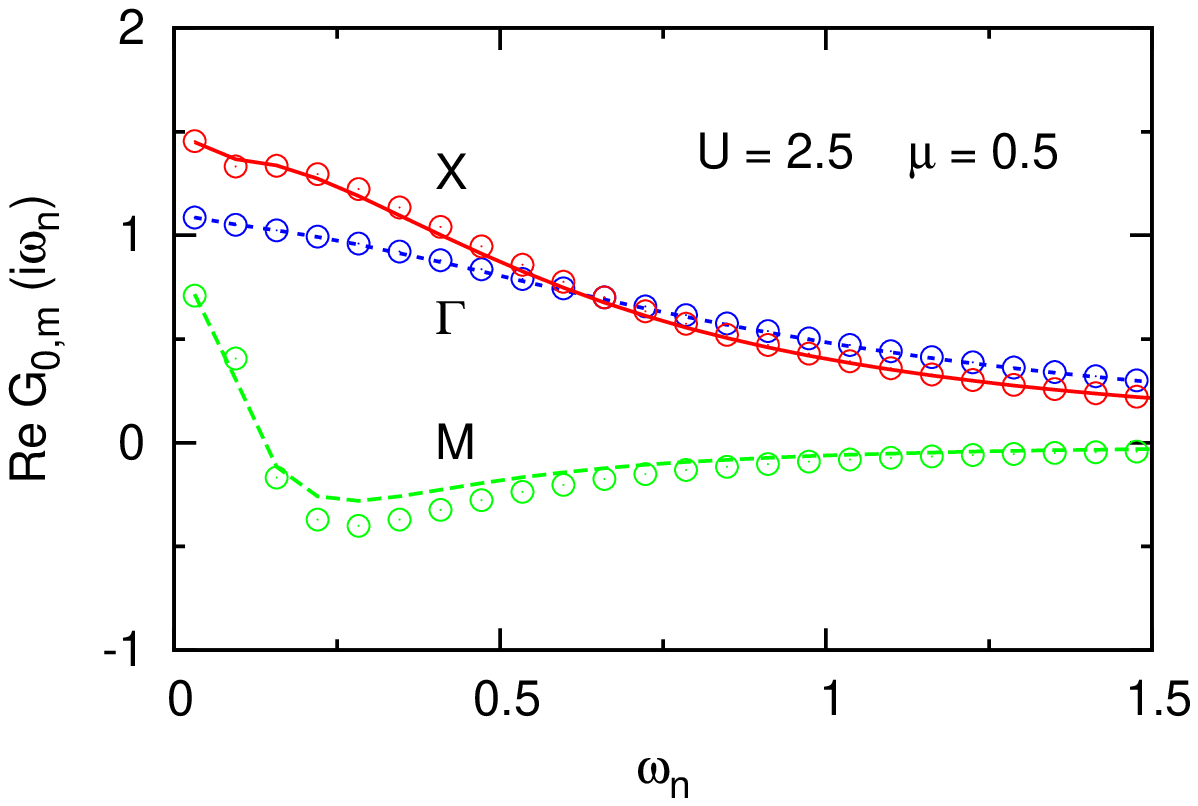} 
\includegraphics[width=4.0cm,height=6.5cm,angle=-90]{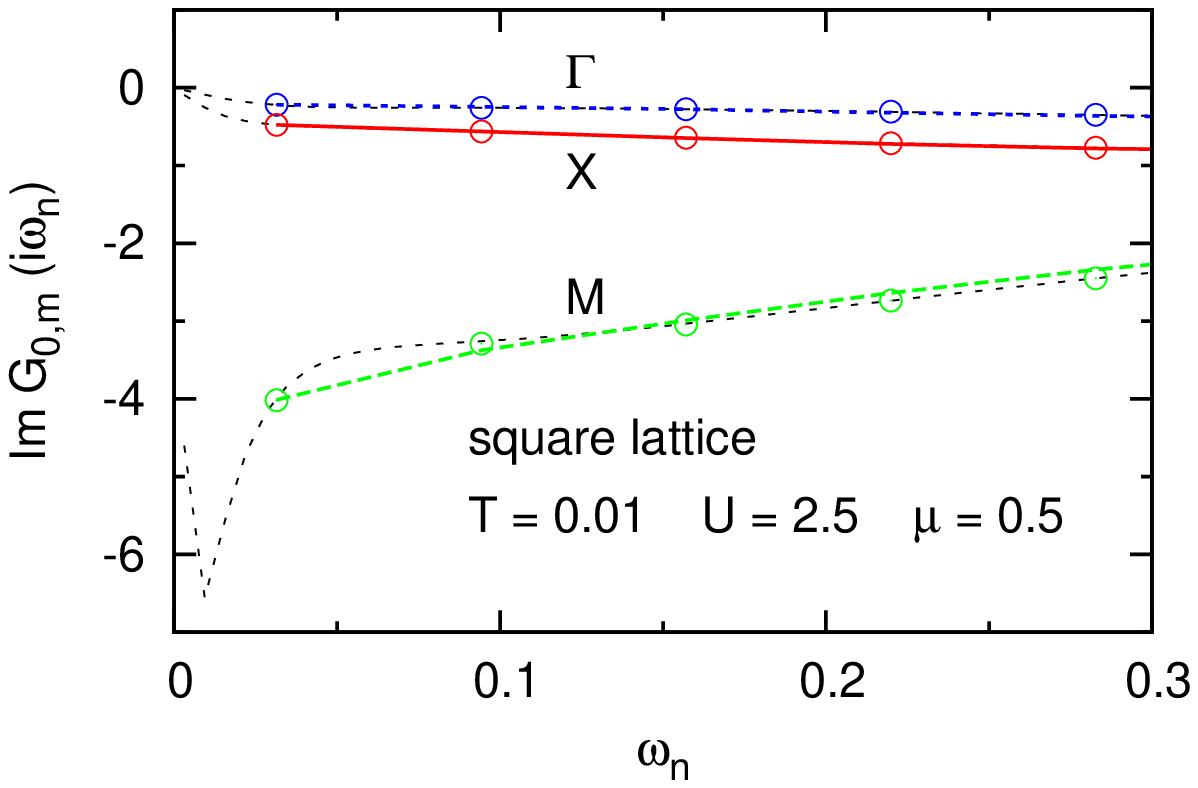} 
\includegraphics[width=4.0cm,height=6.5cm,angle=-90]{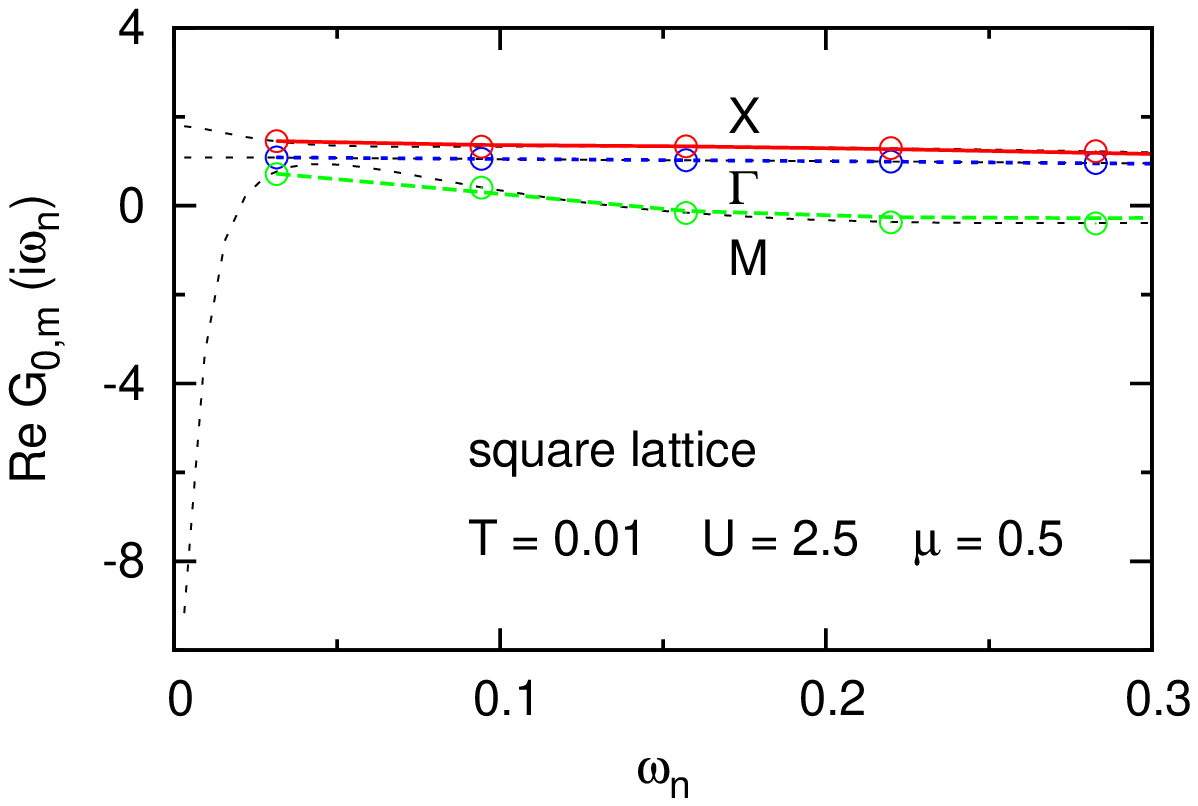} 
\end{center}
\caption{(Color online)
Upper two panels: Comparison of diagonal components of lattice bath  
Green's function $G_{0,m}(i\omega_n)$ (solid and dashed colored curves) 
and cluster Green's function $G_{0,m}^{cl}(i\omega_n)$ (symbols) for square 
lattice, using two bath levels per impurity molecular orbital ($n_s=12$); 
$U=2.5$, $T=0.01$, $\mu=0.5$ \cite{PRB80}.  
Lower two panels: Amplified low-energy region. Black dotted curves: 
$G_{0,m}^{cl}(i\omega_n)$ on a Matsubara grid corresponding to $T=0.001$. 
}\label{DOPE4.Fig2}\end{figure}

To implement ED CDMFT for the square lattice, we use the diagonal molecular-orbital 
or plaquette \cite{jarrell2001,civelli} basis, as indicated in Eq.~(\ref{G4}), 
which yields three 
independent Green's function  and self-energy components ($m=1,\ldots,3$). 
We denote them as $\Gamma=(0,0)$, $M=(\pi,\pi)$, and $X=(\pi,0)$, where the latter 
is degenerate with $Y=(0,\pi)$. Each of the impurity levels hybridizes only with 
its own bath consisting of two levels, as indicated in the left-hand diagram 
of Figure 1. The upper panels of Figure~\ref{DOPE4.Fig2} show the discretization 
of the bath Green's function components via the supercluster consisting of impurity 
cluster plus bath. The Coulomb energy $U=2.5$ is chosen so that at half-filling 
the system is a  Mott insulator, where $U_c\approx 5.6t=1.4$ \cite{park}. 
The chemical potential
$\mu=0.5$ corresponds to about 8 \% hole doping. At $T=0.01$, each component 
$G_{0,m}(i\omega_n)$ is seen to be fitted very well with only five parameters:
the impurity level, two bath levels, and two hopping terms.  For completeness,
the values are provided in Table I.

\begin{table}[b!]
\caption{Impurity levels $\varepsilon_m$, bath levels $\varepsilon_k$, 
and impurity bath hopping matrix elements $V_{m,k}$ for square lattice; 
$n_s=12$, $U=2.5$, $T=0.01$, $\mu=0.5$. The corresponding Green's function 
components are shown in Figure~\ref{DOPE4.Fig2}.}
\bigskip
\begin{tabular}{ |r|r|r|r|r|r| } 
\hline
\,m \ &$\varepsilon_m$ \ &$\varepsilon_{k=m+4}$ \ &$\varepsilon_{k'=m+8}$ \ 
    &$V_{k=m+4}$ \  &$V_{k'=m+8}$ \ \\
\hline
1 \ &$-0.42339$ \  &$ 0.03516$ \ &$-0.05844$ \ &$0.07602$ \ &$0.09822$ \ \\
2 \ &$ 0.59461$ \  &$-0.02611$ \ &$ 0.10462$ \ &$0.08436$ \ &$0.17005$ \ \\  
3 \ &$-0.07018$ \  &$ 0.08829$ \ &$-0.03705$ \ &$0.13478$ \ &$0.09028$ \ \\
\hline
\end{tabular}
\end{table}

The results shown in Figure~\ref{DOPE4.Fig2} are derived via minimization of 
$\vert G_0 - G_0^{cl}\vert$ as in Eq.~(\ref{diff}), for $W_n=1/\omega_n^N$ with 
$N=1$. Appendix A.2 demonstrates that discretizations of similar accuracy are achieved 
for $N=0$ and $N=2$. Also, minimization of $\vert 1/G_0 - 1/G_0^{cl}\vert$, as 
in Eq.~(\ref{diff'}), for $W_n=1/\omega_n^N$, $N=0,1,2$, is of nearly the same   
quality.

Transforming the molecular-orbital parameters back to the non-diagonal site 
basis, the values given in Table I yield the onsite level 
$\varepsilon=(\varepsilon_1+ \varepsilon_2+ 2\varepsilon_3)/4= 0.0078$, 
and the hopping energies   
$\tau =-(\varepsilon_1- \varepsilon_2)/4 = 0.2545\approx t $, and 
$\tau'=-(\varepsilon_1+ \varepsilon_2- 2\varepsilon_3)/4 =-0.078 \approx t'$. 
Thus, the impurity levels in the diagonal 
molecular-orbital basis are well determined by the asymptotic behavior 
of $G_{0}(i\omega_n)$. The small deviations from $\varepsilon=0$, $\tau=t$, and 
$\tau'=t'$ suggest that at low $\omega_n$ a slightly more accurate discretization 
of $G_0$ can be achieved than by enforcing these conditions.

The two lower panels of 
Figure~\ref{DOPE4.Fig2} show the low-energy region of $G_0(i\omega_n)$ in more 
detail. The dotted black curves represent the cluster Green's function components 
$G_{0,m}^{cl}(i\omega_n)$ on a much finer Matsubara grid, corresponding to 
$T=0.001$. Since cluster spectra are always gapped, large oscillations occur at 
low frequencies $\omega_n< 0.01\pi$, in order to accommodate the limit 
Im\,$G_{0,m}^{cl}(i\omega_n)\rightarrow 0$ for $\omega_n\rightarrow 0$. 
Evidently, more bath levels would be required to achieve a satisfactory 
representation of extremely low temperature and low frequency properties.
[For the hopping $t=1/4$ used here, the temperature in Fig.~14 of 
 Ref.~\cite{koch} is $T=t/256=1/1024$.] 
On the other hand, as long as the discussion is limited to 
temperatures of about $T=0.005$ or higher, and to frequencies larger than 
about $\omega_0=\pi T$, two bath levels per impurity orbital are adequate.

\begin{figure}  [t!] 
\begin{center}
\includegraphics[width=4.5cm,height=6.5cm,angle=-90]{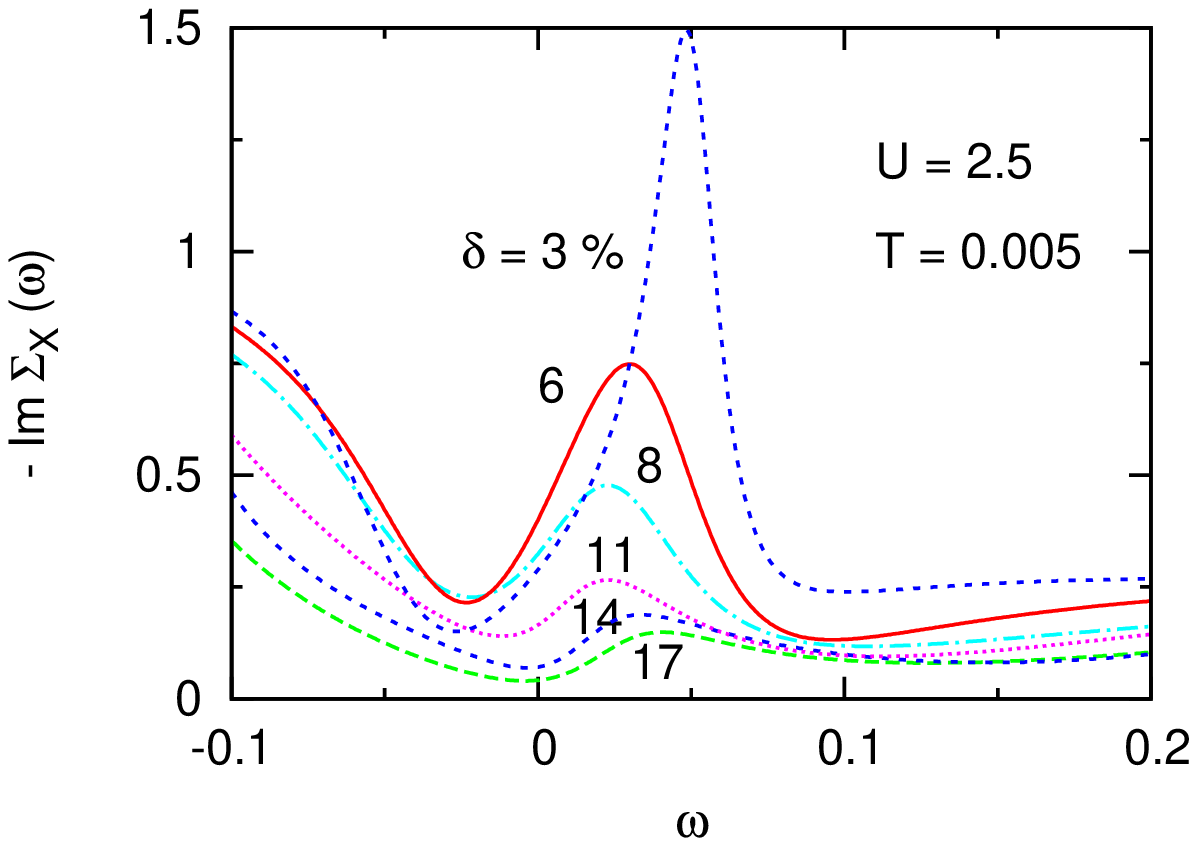} 
\includegraphics[width=4.5cm,height=6.5cm,angle=-90]{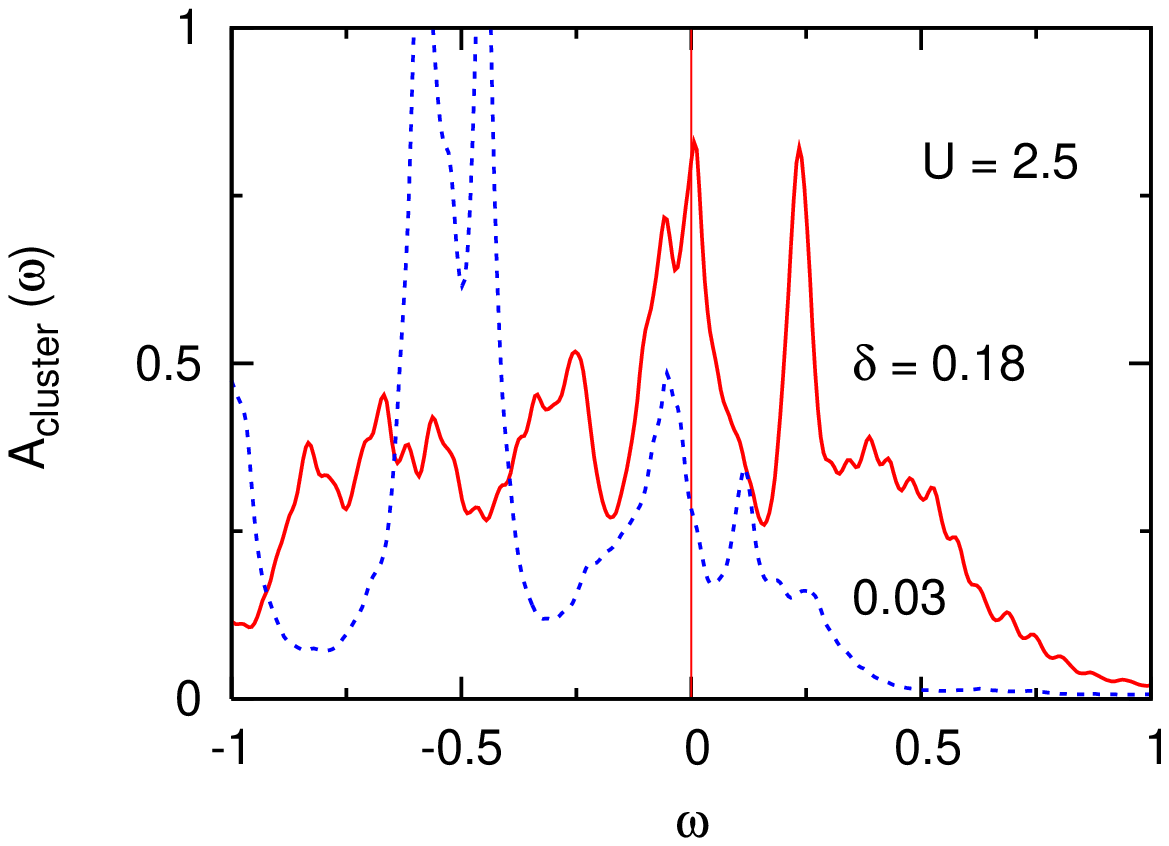} 
\includegraphics[width=4.5cm,height=6.5cm,angle=-90]{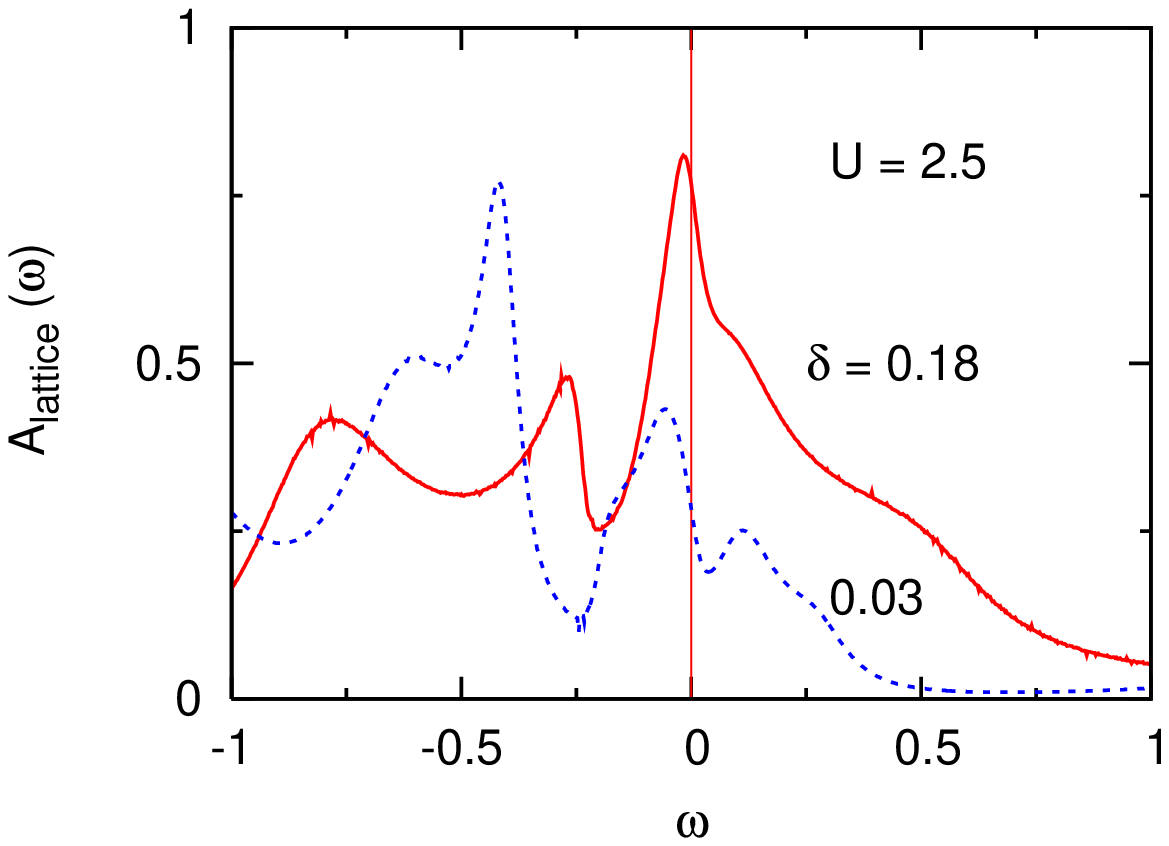} 
\end{center} 
\caption{(Color online)
Upper panel:
Low-energy region of imaginary part of self-energy $\Sigma_{X=(\pi,0)}(\omega)$ 
obtained via extrapolation to real frequencies for several hole doping 
concentrations (broadening $\gamma=0.005$); $U=2.5$, $T=0.005$. 
Middle panel:
Cluster spectral distributions. 
Solid red curve: hole doping $\delta=0.18$ with peak at $E_F=0$;  
dashed blue curve: $\delta=0.03$ exhibiting a pseudogap above $E_F$.
Lower panel: analogous spectra obtained via analytic continuation of lattice 
Green's function \cite{PRB80}.
}\label{DOPE4.Fig12}\end{figure}

As shown in Ref.~\cite{PRB80}, temperatures in the range $T=0.005,\ldots,0.010$ 
are low enough to provide evidence of a collective mode in the 
$(\pi,0)$ component of the self-energy, once it is extrapolated from the 
Matsubara axis to real $\omega$. This is illustrated in the upper panel of 
Figure~\ref{DOPE4.Fig12}, 
which reveals a transition from Fermi liquid properties in the overdoped 
region ($\delta>15$\ \%) to pronounced non-Fermi-liquid behavior in the 
underdoped region  ($\delta<15$\ \%). These self-energies are derived via
analytic continuation from the Matsubara axis to $\omega+i\gamma$ with 
$\gamma=0.005$.  
The collective mode in Im\,$\Sigma_{(\pi,0)}(\omega)$ gives rise to strong 
damping of electronic states near $E_F$, consistent with a pseudogap in the 
quasi-particle density of states \cite{jarrell2001,kyung}. This is also shown
in Figure~\ref{DOPE4.Fig12}, where the middle panel represents 
cluster spectra at hole dopings $\delta=0.03$ and $\delta=0.18$ (broadening 
$\gamma=0.02$), while the lower panel shows the corresponding spectra obtained 
via analytic continuation of the lattice Green's function to real $\omega$. 
About $400\ldots600$ Matsubara points are taken into account and the same 
broadening is assumed as in the cluster spectra. The low-frequency behavior
of both spectral distributions, in particular, the pseudogap slightly above 
$E_F$ and the concomitant particle-hole asymmetry in the underdoped region, 
is seen to be in good agreement. 
Farther below or above $E_F$, the lattice and cluster spectra exhibit large 
differences. The lattice spectra can also be calculated by first extrapolating 
the self-energy to real frequencies and then evaluating Eq.~(\ref{Gm}) at real 
$\omega$. The results are fully consistent with the spectra derived
via direct extrapolation of the lattice Green's function. 

Additional cluster spectra as a function of hole doping can be found in 
Ref.~\cite{swt}, which discusses the correlation-induced transfer of spectral 
weight between low and high energies in the two-dimensional Hubbard model.

\begin{figure}  [t!] 
\begin{center}
\includegraphics[width=4.5cm,height=6.5cm,angle=-90]{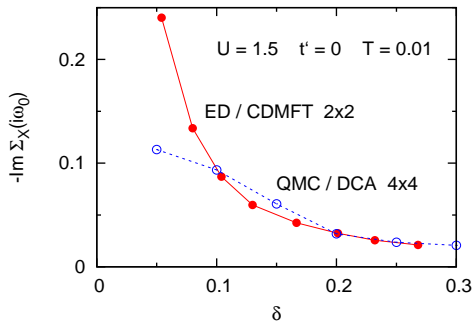} 
\end{center} 
\caption{(Color online)
Imaginary part of self-energy component $\Sigma_{X=(\pi,0)}(i\omega_0)$ at 
first Matsubara frequency as a function of doping for $U=1.5$, $t'=0$.
Solid (red) symbols: ED CDMFT ($n_s=12$) results for $2\times2$ clusters at 
$T=0.01$ \cite{PRB80};  
open (blue) symbols: QMC DCA results for  $4\times4$ clusters at $T=0.014$
\cite{vidhya}.     
}\label{sigmavid}\end{figure}

A dynamical cluster calculation (DCA) for the square lattice was recently 
carried out for $4\times4$ clusters, using Hirsch-Fye QMC as an impurity 
solver \cite{vidhya}. Figure~\ref{sigmavid} shows the comparison of the 
$(\pi,0)$ component of the self-energy at the lowest Matsubara frequency 
with the corresponding ED CDMFT results obtained in Ref.~\cite{PRB80} for
$2\times2$ clusters. Both approaches are in good agreement for doping 
$\delta\ge0.1$. The difference at $\delta=0.05$ is most likely related to the 
singular nature of Im\,$\Sigma_{(\pi,0)}(i\omega_n)$ close to the Mott transition. 
Of course, the larger $4\times4$ cluster provides a much better momentum 
resolution, in particular, a better distinction between scattering properties
along the nodal and anti-nodal directions of the Brillouin Zone. Nonetheless,
the crossover from Fermi-liquid to non-Fermi-liquid behavior is driven by
$\Sigma_{(\pi,0)}$ which is seen to be well represented within finite-$T$
ED CDMFT. For a systematic study of the variation of the self-energy
components with cluster size, see Ref.~\cite{gull.prb82}. 
As also shown in Ref.~\cite{PRB80}, at lower temperature the Fermi-liquid 
to non-Fermi-liquid transition as a function of doping becomes sharper, in 
agreement with the QMC results in \cite{vidhya,gull.prb82}. 

The doping variation of the self-energy shown in Figure~\ref{sigmavid}, with 
Fermi-liquid behavior in the overdoped region ($\delta>0.2$), non-Fermi-liquid
properties at lower doping ($0.05< \delta<0.15$), and a Mott phase at 
half-filling ($\delta=0$), is strikingly similar to the one discussed 
above for the degenerate three-band and five-band models in sections 
3.1 and 3.3, respectively. In the latter cases, Mott insulating phases 
may occur also at integer occupancies away from half-filling, but these
tend to have considerably larger critical Coulomb energies
\cite{prlwerner,FeAsLaO,FeSe}.
On the other hand, at not too large values of $U$, these systems exhibit 
non-Fermi-liquid properties due to the formation of local moments, including 
a finite low-energy scattering rate, up to about one electron or hole away 
from half-filling, i.e., $\vert\delta\vert < 0.2,\ldots,0.3$. 
Only at larger doping, ordinary Fermi-liquid behavior is restored. 
Since multi-site correlations in the single-band Hubbard model may be 
formulated within a diagonal molecular-orbital basis, where the on-site 
repulsion is converted into intra-orbital and inter-orbital Coulomb and 
exchange interactions, the fact that common physical phenomena appear in 
both types of systems is plausible. The parallel treatment
of these materials within DMFT, and the close relation between key physical
properties, suggests that the existence of inter-orbital or inter-site 
interaction channels gives rise to fundamentally new physics that is absent 
in the single-band, single-site limit.

\subsection{3.6. \ Honeycomb Lattice}
  
\begin{figure}  [t!] 
\begin{center}
\includegraphics[width=4.5cm,height=6.5cm,angle=-90]{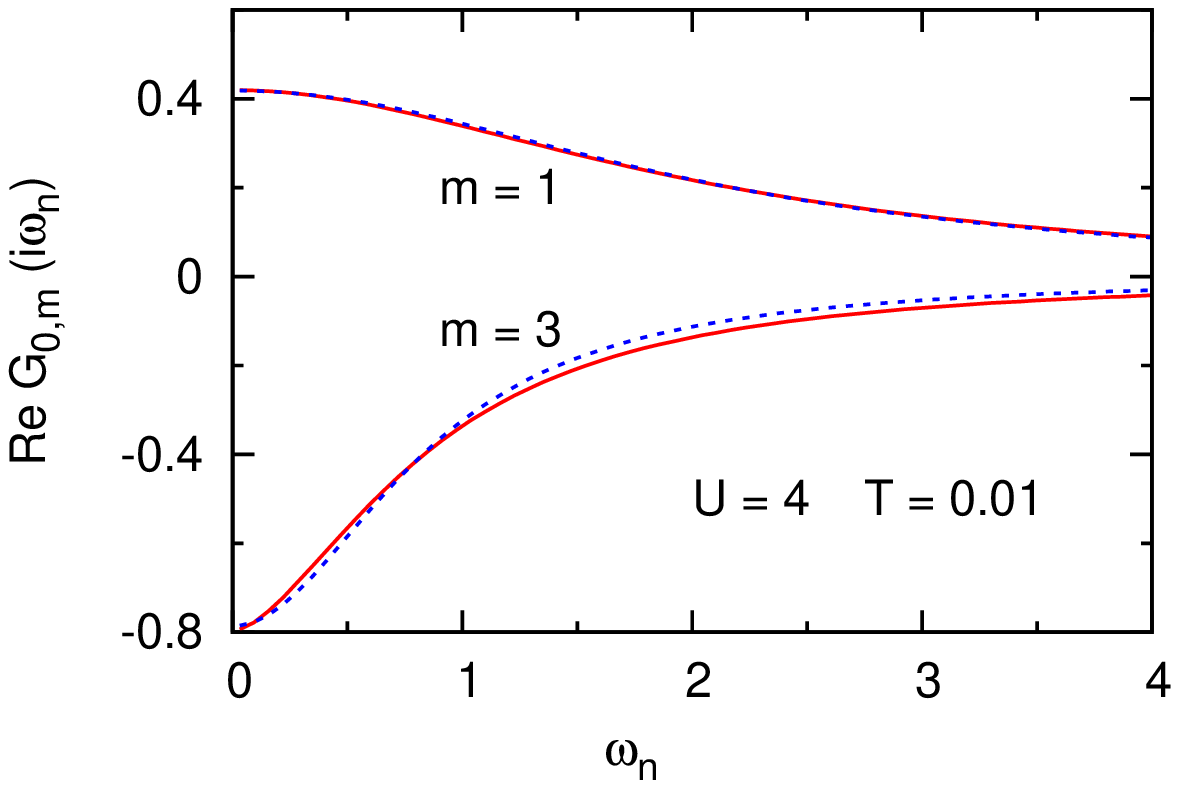} 
\includegraphics[width=4.5cm,height=6.5cm,angle=-90]{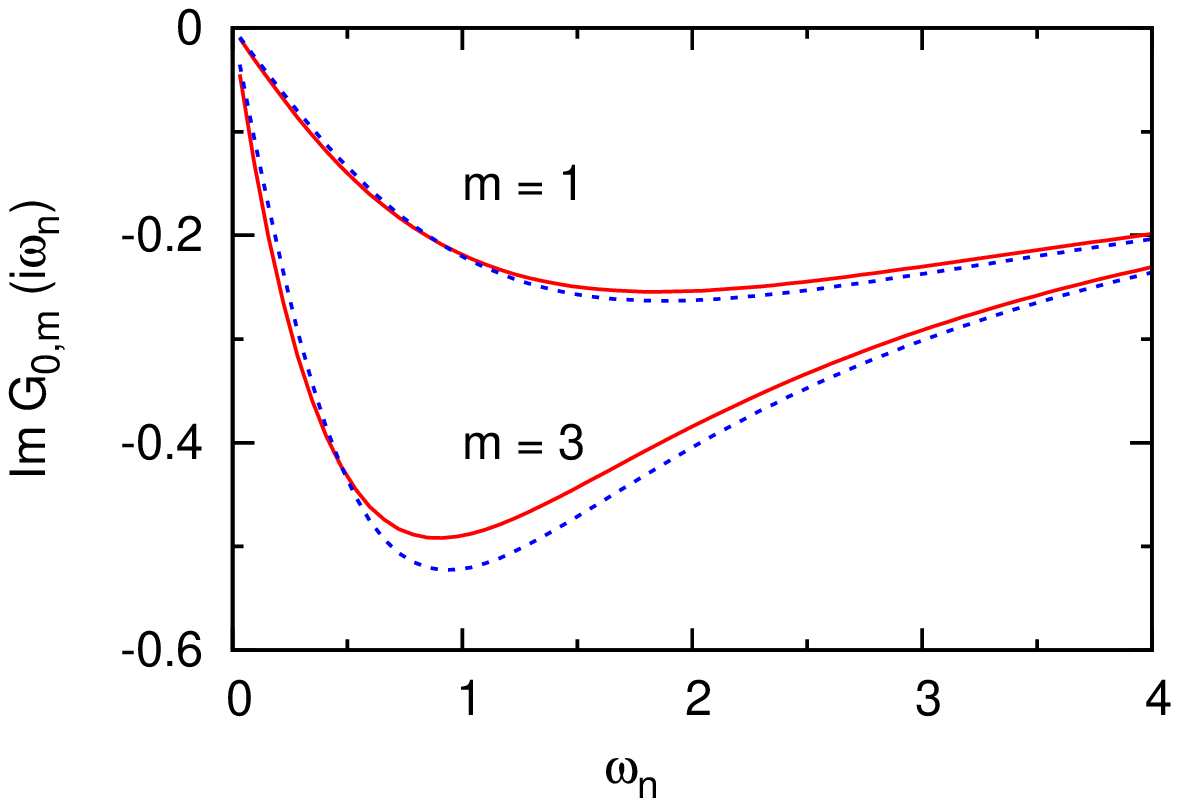} 
\end{center} 
\caption{(Color online)
Comparison of lattice bath Green's function $G_{0,m}(i\omega_n)$ 
(solid red curves) and cluster Green's function $G_{0,m}^{cl}(i\omega_n)$ 
(dashed blue curves) for honeycomb lattice using only one bath level per 
impurity orbital ($n_s=12$); $U=4$, $T=0.01$.
Upper panel: real part; lower panel: imaginary part.
}\label{hong0}\end{figure}

\begin{figure} [t!] 
\begin{center}
\includegraphics[width=4.0cm,height=6.5cm,angle=-90]{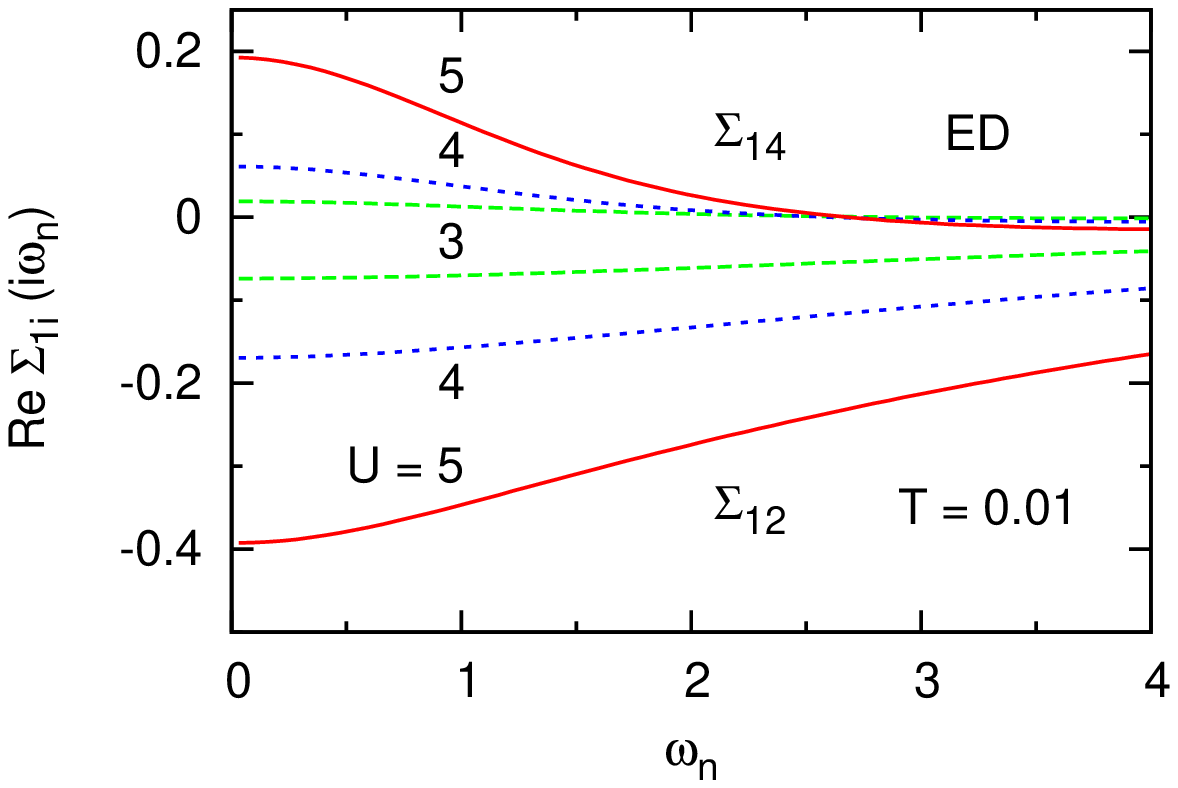} 
\includegraphics[width=4.0cm,height=6.5cm,angle=-90]{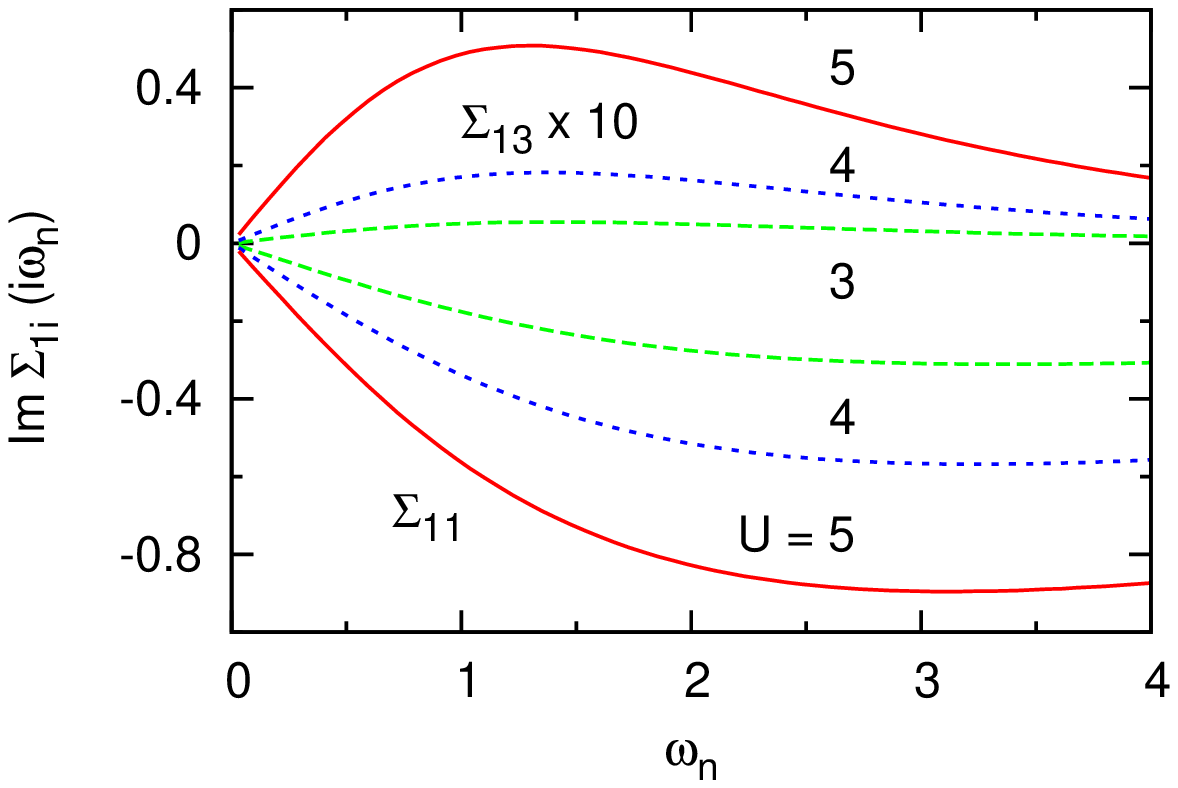} 
\includegraphics[width=4.0cm,height=6.5cm,angle=-90]{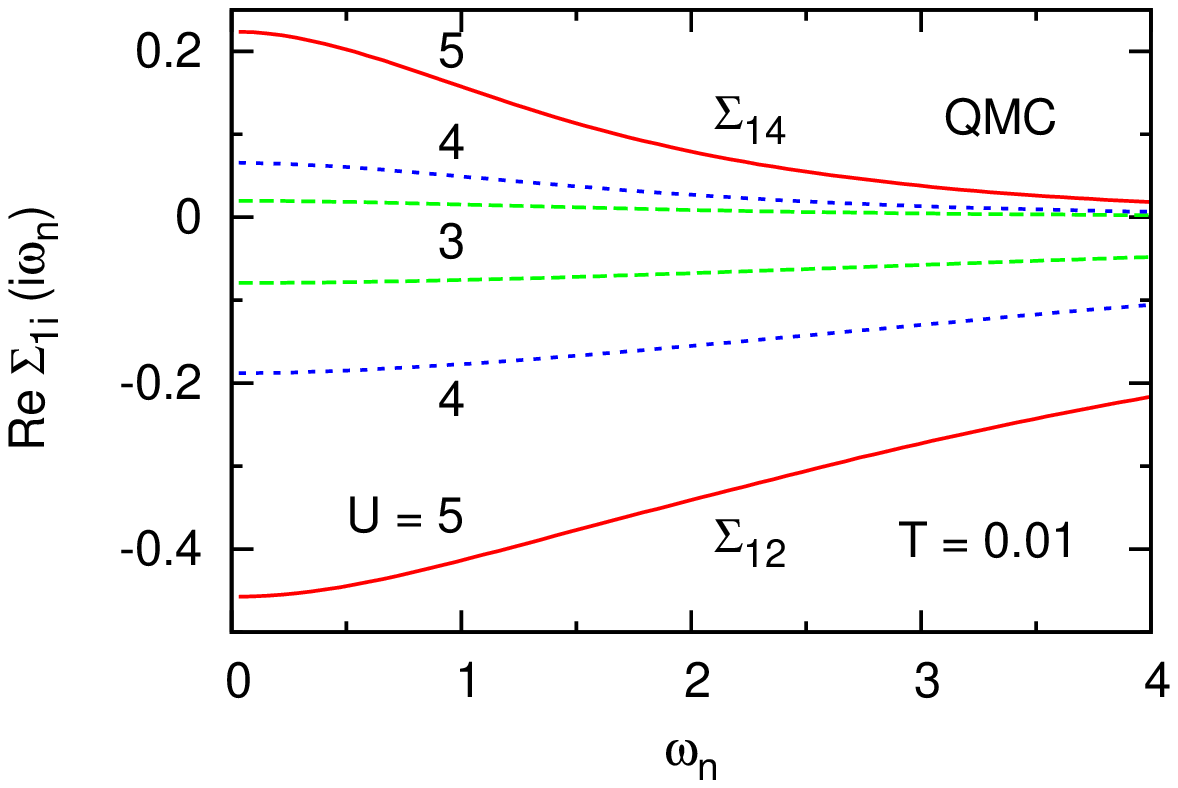} 
\includegraphics[width=4.0cm,height=6.5cm,angle=-90]{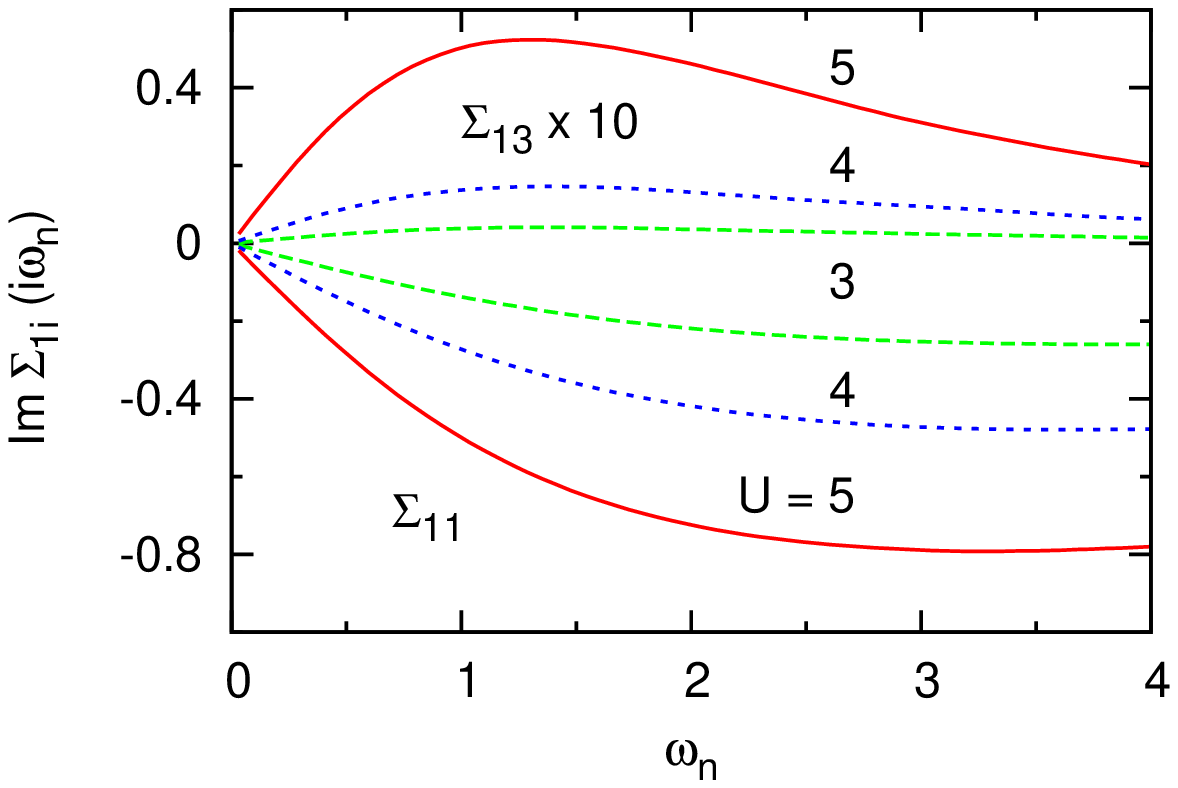} 
\end{center} 
\caption{(Color online)
Self-energy components $\Sigma_{1i}(i\omega_n)$ ($i=1,\ldots,4$) of honeycomb 
lattice obtained within  cluster DMFT for three Coulomb energies at $T=0.01$.
For symmetry reasons, $\Sigma_{11}$ and $\Sigma_{13}$ are imaginary, while 
$\Sigma_{12}$ and $\Sigma_{14}$ are real. Upper two panels: ED \cite{honey}; 
lower two panels: continuous-time QMC \cite{wu}. 
}\label{honsig.ed}\end{figure}
 
We have recently used finite-temperature ED in combination with CDMFT to study 
the semi-metal to insulator transition in the honeycomb lattice at half-filling
\cite{honey}.
The Hubbard Hamiltonian is given by Eq.~(\ref{Hcl}), with hopping matrix
elements $t_{ij}=t=1$ for nearest neighbors and $t_{ij}=0$ otherwise. 
The impurity cluster in which spatial fluctuations are accounted for 
explicitly consists of six sites, as illustrated in the right-hand diagram 
of Figure 2. Because of the symmetry of this cluster, the lattice Green's
function can be diagonalized as indicated in Eq.~(\ref{G6}). In the site 
basis $G_{11}$ and $G_{13}=G_{15}$ are imaginary, while $G_{12}=G_{16}$ 
and $G_{14}$ are real. Thus, the diagonal components satisfy $G_2=-G^*_1$ 
and $G_4=-G^*_3$, giving two independent complex functions. Analogous 
relations hold for the diagonal components of the cluster Green's function 
and self-energy. Although a $T=0$ cluster calculation with two bath levels 
per impurity molecular orbital ($n_s=18$) is feasible \cite{Jaime.ns=18}, 
the aim of this section is to demonstrate that, for the honeycomb 
lattice at half-filling, adequate results can be obtained even if only one 
bath level per impurity orbital is employed. Since for $n_s=12$ the Hilbert 
space is still fairly large, the level spacing of interacting states is rather 
dense so that a reasonable representation of the low-energy behavior of the 
self-energy can be achieved.

As discussed in Section 2.4, one criterion for the accuracy of ED DMFT 
is the discretization of the lattice Green's function 
$G_0$ via the finite bath.  In the present case, the $k$-sum on the right-hand
side of Eq.~(\ref{G0cl}) consists of only one term, so that each diagonal
element $G_{0,m}(i\omega_n)$ is fitted with three parameters, the impurity 
level $\varepsilon_m$, the  bath level $\varepsilon_{k}$, and the hopping term 
$V_{k}$, where $k=m+6$. Figure~\ref{hong0} shows the comparison of the 
lattice and cluster Green's functions for $U=4$, i.e., close to the Mott 
transition near $U_c\approx 3.5$. Comparisons of similar quality are found 
at $U=3$ and $U=5$, i.e., on the semi-metallic and insulating sides of the 
transition. Undoubtedly, one reason for the surprisingly good agreement is 
the semi-metallic nature at half-filling, which ensures that the imaginary 
part of the lattice Green's function vanishes in the limit of small $\omega_n$ 
even for $U<U_c$. Thus, there is no disparity here of representing a metallic 
lattice Green's function via a finite-size cluster Green's function. 

The role of Coulomb correlations in the honeycomb lattice has recently also 
been studied within CDMFT by Wu {\it et al.}~\cite{wu} who used continuous-time 
QMC as impurity solver. Figure~\ref{honsig.ed} shows the ED and QMC self-energy 
components in the site basis, where $\Sigma_{11}$ and $\Sigma_{13}$ are 
imaginary, while $\Sigma_{12}$ and $\Sigma_{14}$ are real. The comparison 
at three different Coulomb energies ranging from the semi-metallic to the Mott 
insulating phases indicates that there is very good agreement between the ED 
and continuous-time QMC results. As can be seen in Refs.~\cite{honey} and 
\cite{wu}, similar agreement is found for the double occupancies and for the 
quasi-particle spectra.

\begin{figure}  [t!!] 
\begin{center}
\includegraphics[width=4.5cm,height=6.5cm,angle=-90]{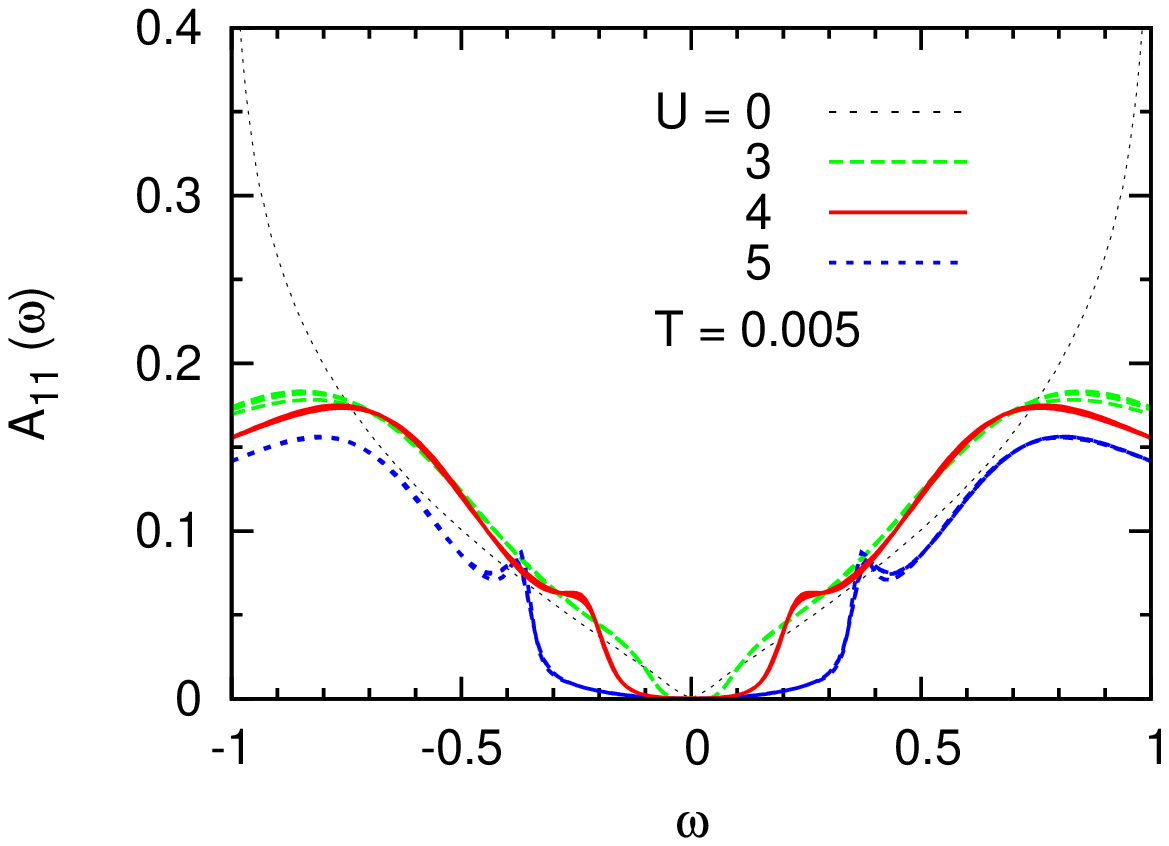} 
\includegraphics[width=4.5cm,height=6.5cm,angle=-90]{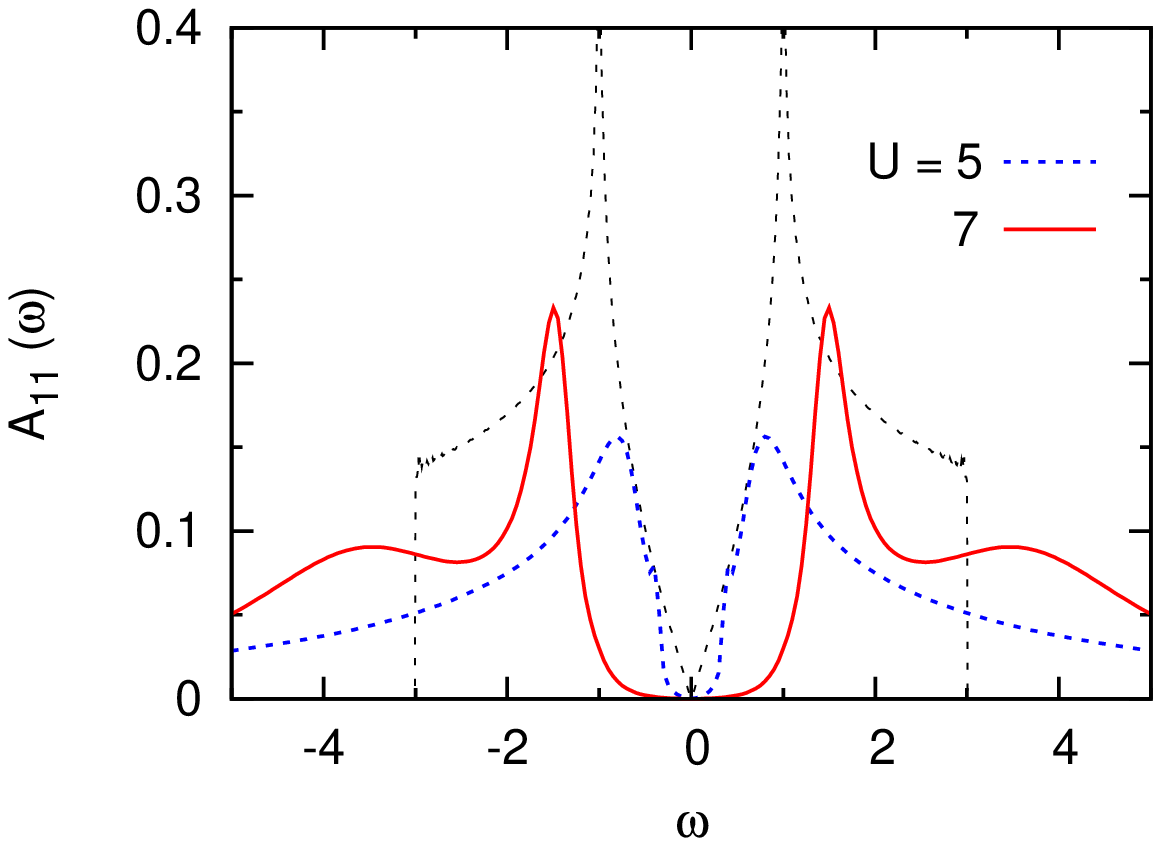} 
\end{center}
\caption{(Color online)
Upper panel: low-energy region of quasiparticle distribution of honeycomb lattice 
for several Coulomb energies at $T=0.005$ \cite{honey}. 
The noninteracting density of states is indicated by the black dotted curve.  
Between 50 and 200 Matsubara points are used to extrapolate the lattice
Green's function to real $\omega$.  
Lower panel: density of states over wider energy range for $U=5$ and $U=7$.
}\label{A11}\end{figure}

Figure~\ref{A11} (upper panel) shows the low-energy region of the interacting 
density of states for several Coulomb energies. These spectra are derived from 
an extrapolation of the local lattice Green's function $G_{11}(i\omega_n)$
to real $\omega$. To illustrate the stability of this extrapolation, 
at each value of $U$ several curves are plotted for 50 to 200 Matsubara 
points, with an additional small energy broadening of the order of 
$0.1\omega^2$. (For $\vert\omega\vert>1$ the broadening is kept constant
at $0.1$.) At $U=3$, a tiny gap or pseudogap is seen which is near the 
limit of what can be resolved within ED/DMFT. At $U=4$, a full gap of 
width $\Delta\approx 0.25$ has opened. Its width increases approximately 
to $\Delta\approx 0.6$ when the Coulomb energy is increased to $U=5$.
This trend is consistent with the one found in Refs.~\cite{wu,meng}.
The variation of the gap at larger $U$, and the appearance of Hubbard bands,
are indicated in Figure~\ref{A11} (lower panel). 
These results suggest that non-local correlations in the honeycomb lattice 
induce a paramagnetic semi-metal to insulator Mott transition in the range 
$U=3\ldots4$, in striking contrast to single-site DMFT which yields 
$U_c\approx 10\ldots 13$ \cite{jafari,tran}. 

\section{4.  \ Summary and Outlook} 

The accuracy of DMFT based on finite-temperature exact diagonalization 
has been discussed for a variety of strongly correlated materials. 
The cases include three-band and five-band single-site applications of 
DMFT, appropriate to transition metal compounds such as Ca$_2$RuO$_4$, 
V$_2$O$_3$, LaTiO$_4$, Na$_x$CoO$_2$, FeAsLaO, etc. 
In addition, short-range correlations in two-dimensional single-band Hubbard models 
are discussed for triangular, square, and honeycomb lattices, which represent
key aspects of organic molecular crystals, high-$T_c$ cuprates, and graphene.         
The main focus of this work is on the role of the temperature and size
of the bath which is used to discretize the host lattice surrounding the
correlated impurity. In contrast to the simpler single-band, single-site
model, which requires typically 3 to 5 bath levels to achieve convergence,
the multi-band and multi-site materials discussed here are usually well 
represented using two bath levels per orbital or site, as long as the 
temperature is not too low.

The accuracy of the results is illustrated using three different criteria:
(i) the convergence of the self-energy with bath size, (ii) the quality of
the discretization of the bath Green's function in terms of a finite cluster,
and (iii) the comparison with analogous results obtained within continuous-time
QMC DMFT. For a total of 10 to 15 impurity and bath levels, the 
Hilbert space for correlated multi-band or multi-site systems is rather large.
Thus, because of the indirect coupling among the baths pertaining to different
impurity levels, the spectrum of excited states is very dense, even if only 
two bath levels per impurity orbital or site are included. Moreover, at 
temperatures larger than 5 to 10~meV, the bath discretization is usually 
very accurate. At lower temperature, accordingly more bath levels per orbital 
or site should be employed. 
For five correlated orbitals within local DMFT only one bath level per orbital 
was found to give qualitatively correct results, including the transition from 
a Mott phase at half-filling to non-Fermi-liquid and Fermi-liquid behavior for 
increasing doping. Also, results for the self-energy of the single-band Hubbard 
model on the honeycomb lattice at half-filling with only one bath level per site 
were found to be in excellent agreement with analogous QMC CDMFT results.  

Since ED DMFT calculations, like those within QMC DMFT, are performed along  
the Matsubara axis, an extra step is needed to evaluate the Green's function 
and self-energy to real energies. An attractive feature of ED is that the 
cluster quantities can be calculated directly close to the real axis. Moreover,
the corresponding lattice quantities can be derived via analytic continuation. 
As we have shown for several systems, this extrapolation is usually rather 
stable in the low-energy region, where the results agree well with the 
(broadened) cluster spectra.        

The examples discussed in this paper suggest that finite-temperature ED DMFT is 
a versatile and accurate scheme for the description of electronic properties 
of a wide range of strongly correlated materials. It is applicable at large
Coulomb energies and for full Hund exchange. Moreover, it is free of sign 
problems and statistical errors. Since it is most suitable at low temperatures 
ED DMFT can be regarded as complementary to continuous-time QMC DMFT. Studies
at higher temperatures pose no problem in principle, but computational times
increase due to the larger number of excited states that come into play.
In the future it would be interesting to explore the correspondence between 
ED and QMC at higher temperatures than the ones discussed in the present work,
and to establish the range of accuracy of both schemes at very low temperatures.    
It would also be very interesting to study more $d$-electron and possibly 
$f$-electron materials in order to investigate further the accuracy of ED DMFT
when only one bath level per impurity orbital is included.  
\bigskip

{\bf Acknowledgments.} We like to thank M. Aichhorn, E. Gull, E. Koch, 
J. Merino, G. Sangiovanni, and Ph.~Werner for useful discussions. 
Comments on the manuscript by E. Gull, D. S\'en\'echal, and A.-M. S. Tremblay 
are gratefully acknowledged. We also thank Philipp Werner for the QMC data in 
Figure~3 and Wei Wu for those in Figure~\ref{honsig.ed}.  
The numerical work was mainly carried out on the Juropa Computer of the
Research Center J\"ulich. \\
a.liebsch@fz-juelich.de,  ishida@chs.nihon-u.ac.jp

\vskip8mm
\centerline {\bf Appendix: \ Bath Discretization}
\vskip5mm

One of the key approximations in ED DMFT is the discretization of the bath
Green's function $G_0$ in terms of a finite cluster. Whereas $G_0$ for the 
infinite lattice is continuous at real $\omega$, the analogous cluster version
$G_0^{cl}$ consists of a small number of discrete lines. Nonetheless, at finite 
temperatures both quantities are smooth functions along the imaginary Matsubara
axis. To find the bath levels $\varepsilon_k$ and impurity--bath hopping matrix
elements $V_{mk}$, the distance functions in Eq.~(\ref{diff}) or (\ref{diff'}) 
are minimized. Evidently, this minimization implies some uncertainty due to the
different weight functions $W_n$ that may be used to place more or less weight
on the important low-energy region. In this Appendix we discuss in more detail
the  discretization of the bath Green's function in the case of the three-band 
model (see Section 3.1) and the square lattice (Section 3.5).

\begin{figure}  [t!] 
\begin{center}
\includegraphics[width=4.0cm,height=6.5cm,angle=-90]{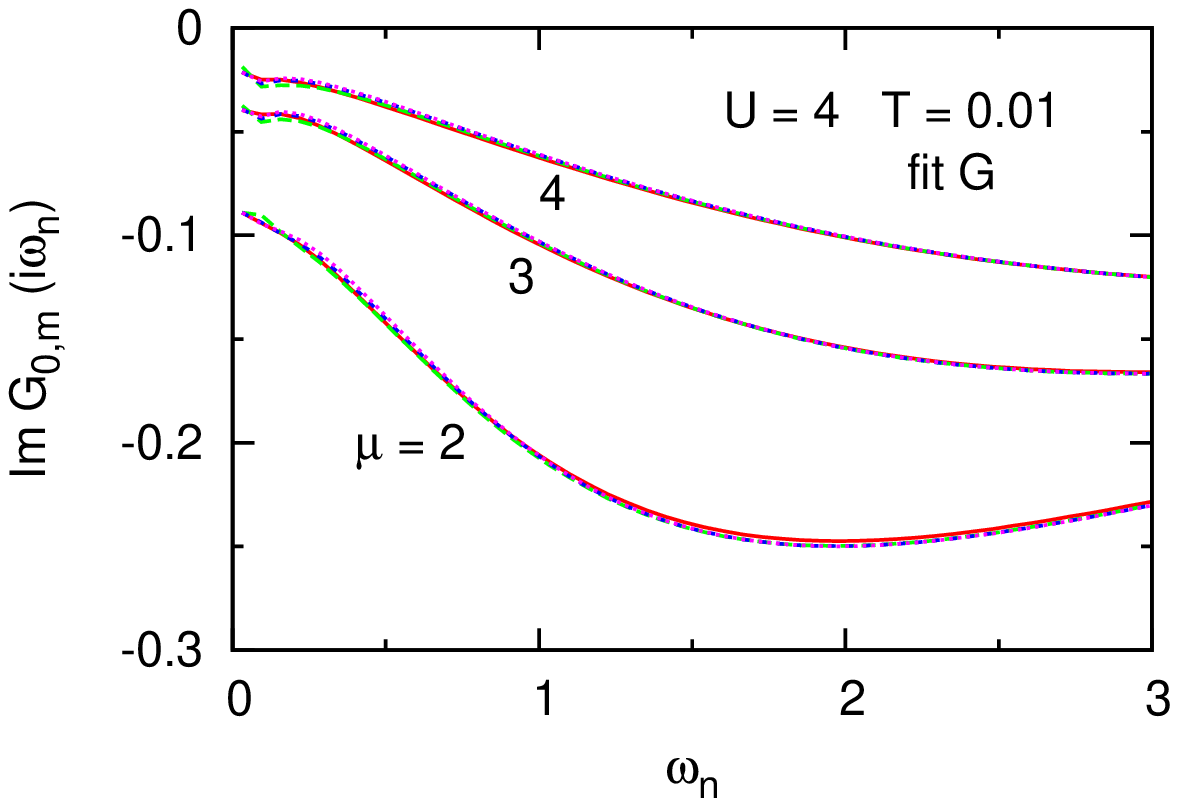} 
\includegraphics[width=4.0cm,height=6.5cm,angle=-90]{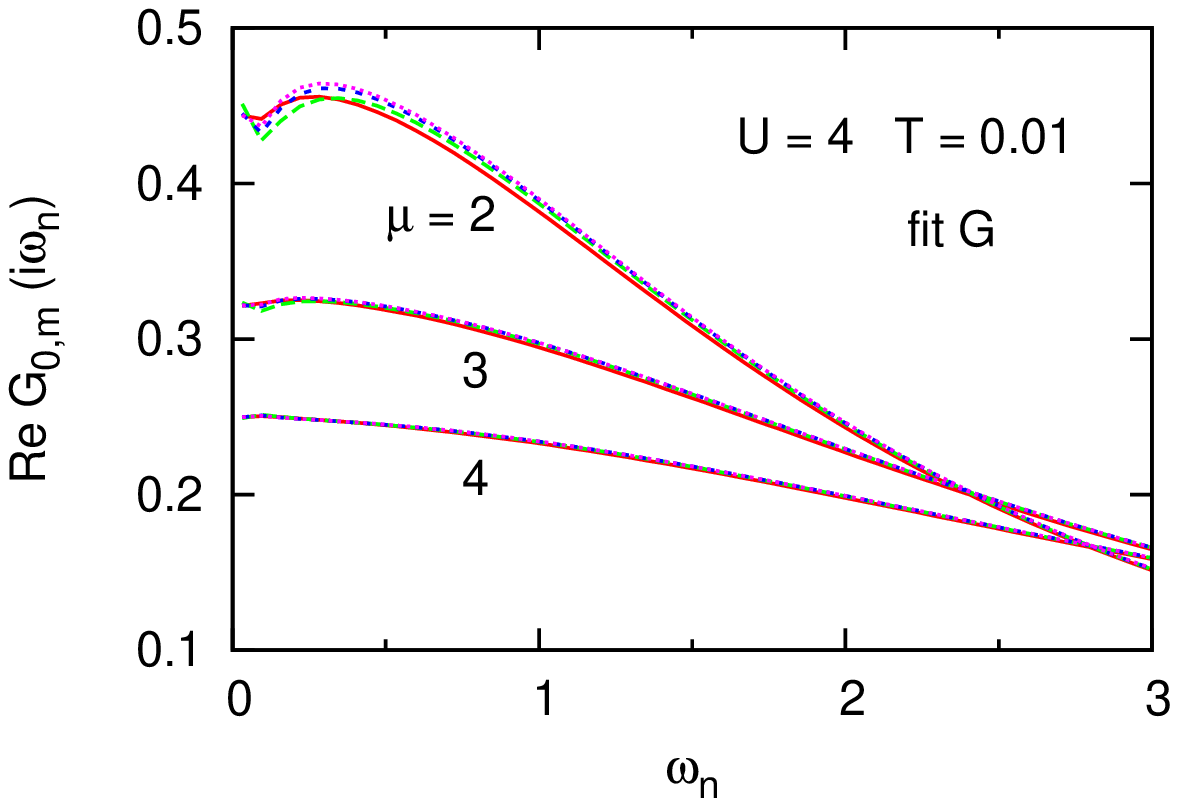} 
\includegraphics[width=4.0cm,height=6.5cm,angle=-90]{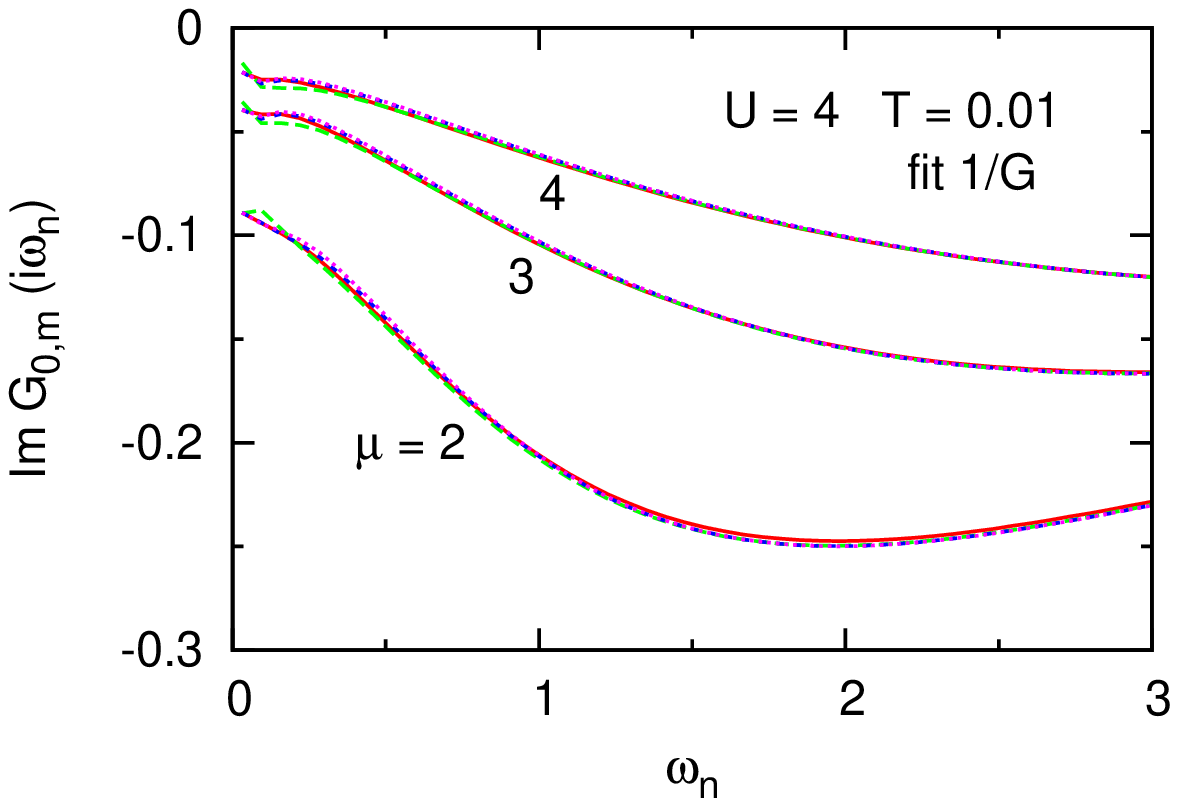} 
\includegraphics[width=4.0cm,height=6.5cm,angle=-90]{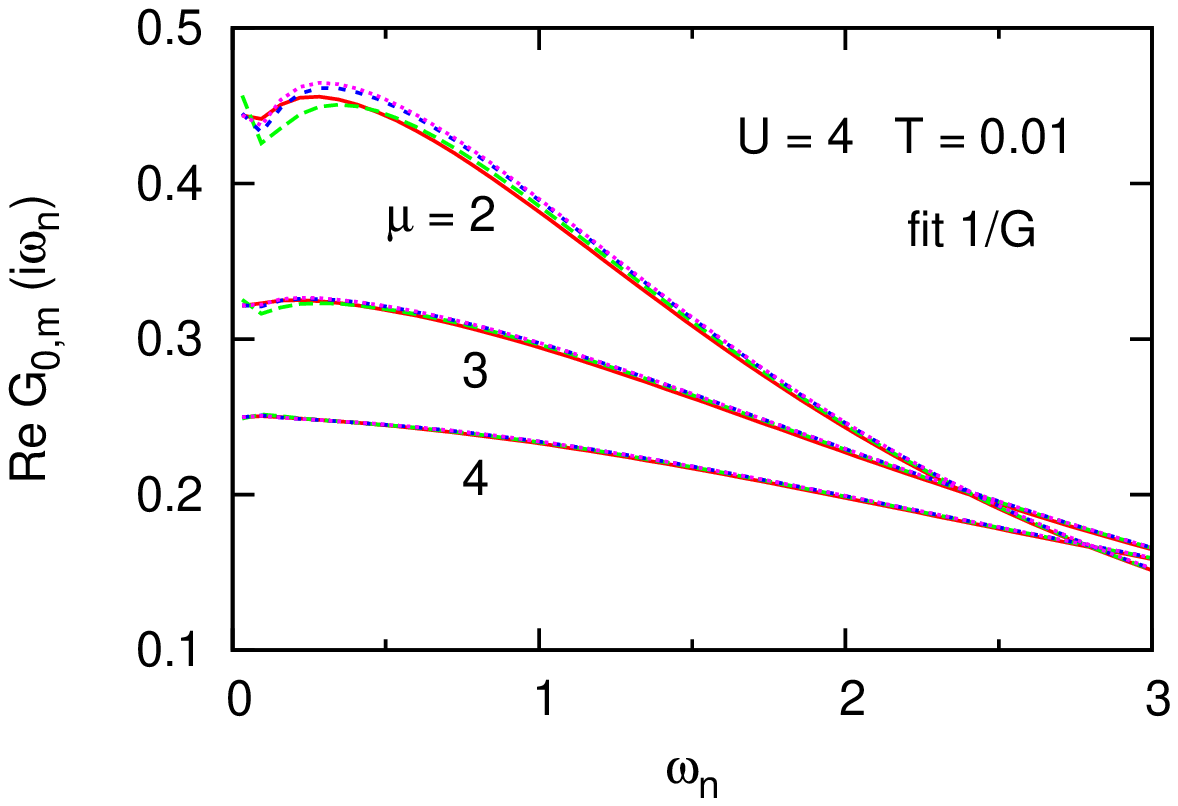} 
\end{center}\vskip-3mm
\caption{(Color online)
Discretization of bath Green's function for degenerate three-band model as in 
Section 3.1 using two bath levels per impurity orbital ($n_s=9$). 
Upper two panels: minimization of $\vert G_0-G_0^{cl}\vert$, see Eq.~(\ref{diff}); 
lower two panels: minimization of $\vert 1/G_0 - 1/G_0^{cl}\vert$, 
see Eq.~(\ref{diff'}).  
Solid red curves: lattice bath Green's function $G_{0,m}(i\omega_n)$ (same as in 
Figure~5); remaining curves: cluster Green's functions derived for weight functions 
$W_n=1/\omega_n^N$ with 
$N=0$ (dashed green curves),
$N=1$ (dashed blue curves), and 
$N=2$ (dotted magenta curves).  
}\label{app.1}\end{figure}

\vskip8mm 
\centerline {\bf A.1. Degenerate Three-Band Model} 
\vskip5mm

To illustrate the quality of the bath discretization more systematically, we show in 
Figure~\ref{app.1} the fits obtained for the three-band model discussed in Section 
3.1 for weight functions $W_n=1/\omega_n^N$ with $N=0,1,2$. In all cases, the 
impurity levels are held fixed at $\varepsilon_{m=1,2,3}=0$, in order to focus on the 
variation of the bath levels. Thus, there are four fit parameters per impurity orbital. 
The two upper panels are for the distance function specified in Eq.~(\ref{diff}),
while the lower panels are for minimization according to Eq.~(\ref{diff'}). 
The range of chemical potentials is chosen to cover the Fermi-liquid region 
($\mu=2$, $n\approx1.5$) as well as the non-Fermi-liquid region close to the Mott
phase ($\mu=4$, $n\approx2.5$). The intermediate case ($\mu=3$, $n\approx2$) 
is near the spin-freezing transition.

\begin{figure}  [t!] 
\begin{center}
\includegraphics[width=4.5cm,height=6.5cm,angle=-90]{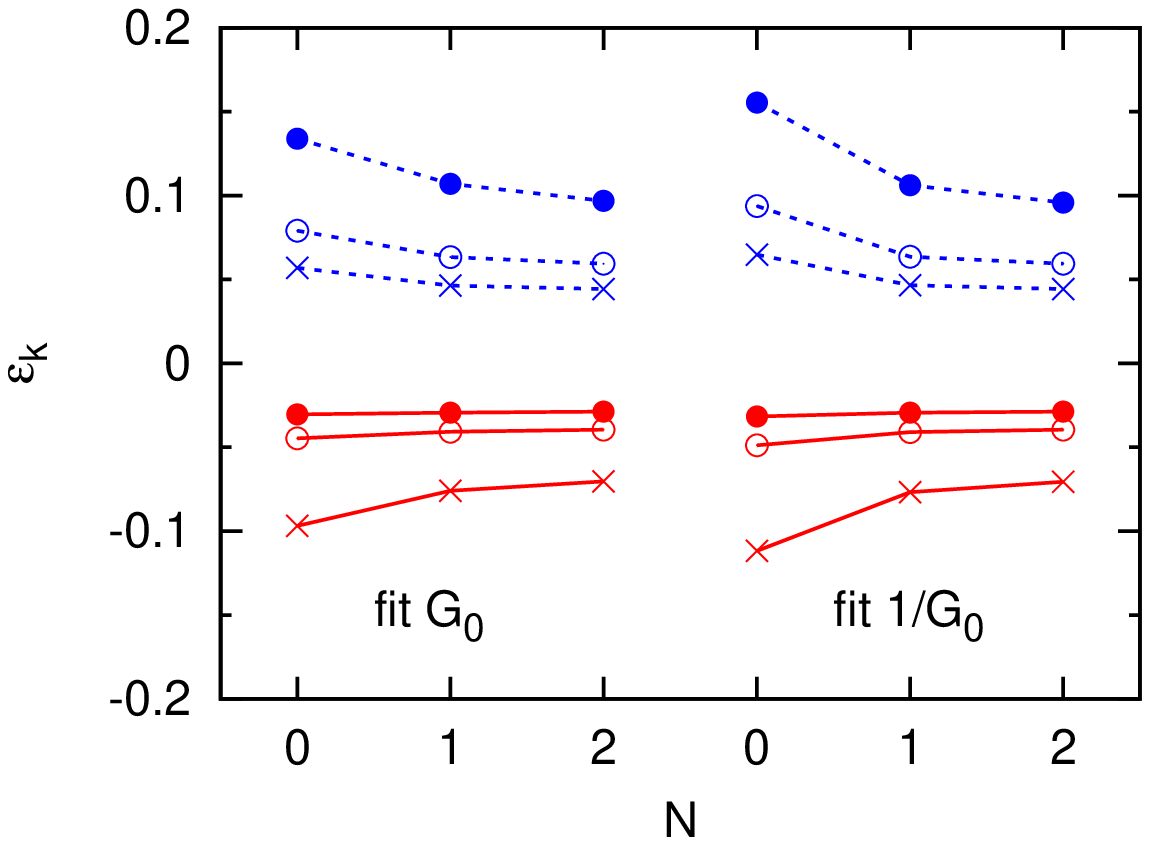} 
\includegraphics[width=4.5cm,height=6.5cm,angle=-90]{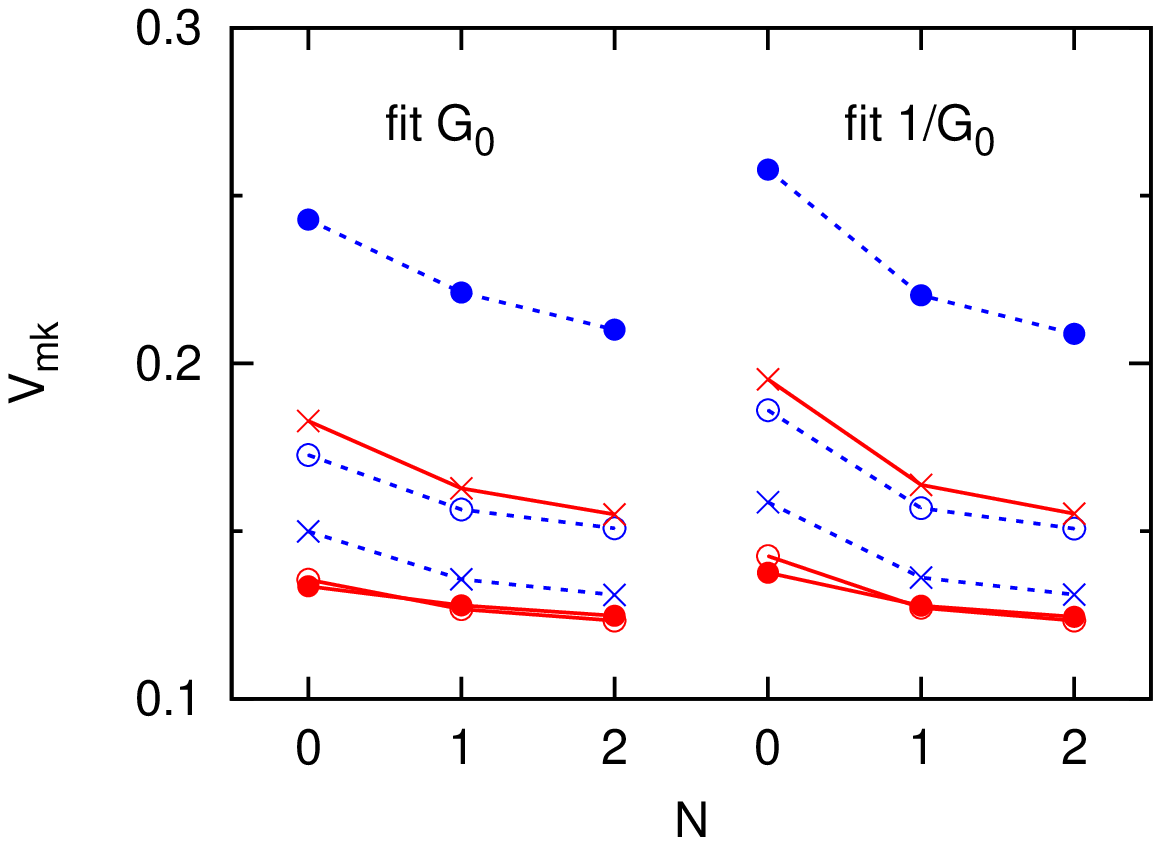} 
\end{center}
\caption{(Color online)
Bath levels $\varepsilon_k$ (top panel) and hybridization matrix elements $V_{mk}$ 
(lower panel) for fits in Figure~\ref{app.1}. Solid dots: $\mu=2$, empty dots:
$\mu=3$, x symbols: $\mu=4$. Left: minimization of $\vert G_0 - G_0^{cl}\vert$, 
see Eq.~(\ref{diff}); right: minimization of  $\vert 1/G_0 - 1/G_0^{cl}\vert$, 
 see Eq.~(\ref{diff'}). The two bath levels are denoted by the red and blue colors.  
}\label{epsk.1}\end{figure}

Figure~\ref{app.1} demonstrates that all six minimization procedures yield fits of 
good quality, in particular, for $N=1$ and $N=2$. Naturally, these fits imply 
slightly different bath parameters $\varepsilon_k$ and $V_{mk}$. As shown in 
Figure~\ref{epsk.1}, the differences between $N=1$ and $N=2$ are less pronounced than 
between $N=1$ and $N=0$. Also, fitting $G_0$ yields more stable $\varepsilon_k$ 
and $V_{mk}$ than fitting $1/G_0$. Since the bath parameters have only an auxiliary 
function, their variation does not have any physical significance. Their sole 
mathematical purpose is to provide an accurate representation of the frequency 
dependence of the lattice bath Green's functions $G_{0,m}(i\omega_n)$, Eq.~(\ref{G0}).
Figure~\ref{epsk.1} indicates that smaller values of $\varepsilon_k$ are associated 
with smaller values of $V_{mk}$. Because of this compensation, the quality of the 
various fits in Figure~\ref{app.1} is approximately the same. 

\begin{figure}  [t!] 
\begin{center}
\includegraphics[width=4.0cm,height=6.5cm,angle=-90]{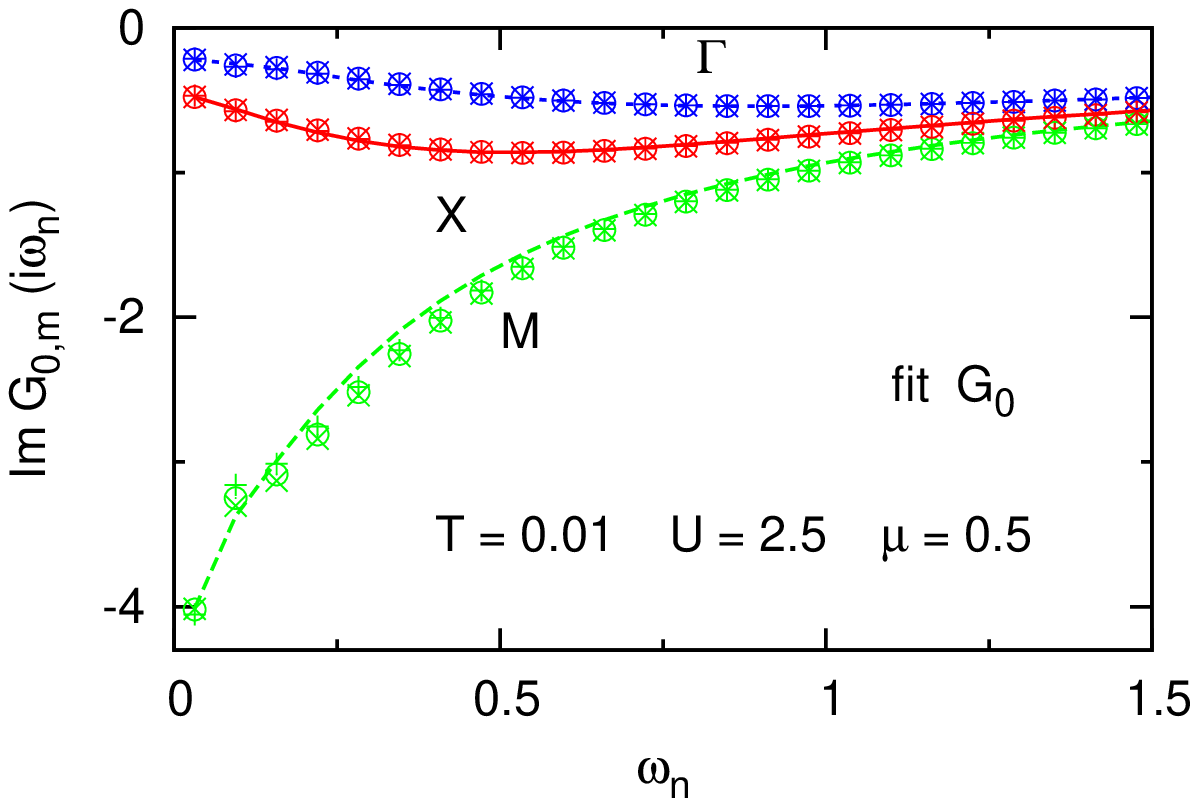} 
\includegraphics[width=4.0cm,height=6.5cm,angle=-90]{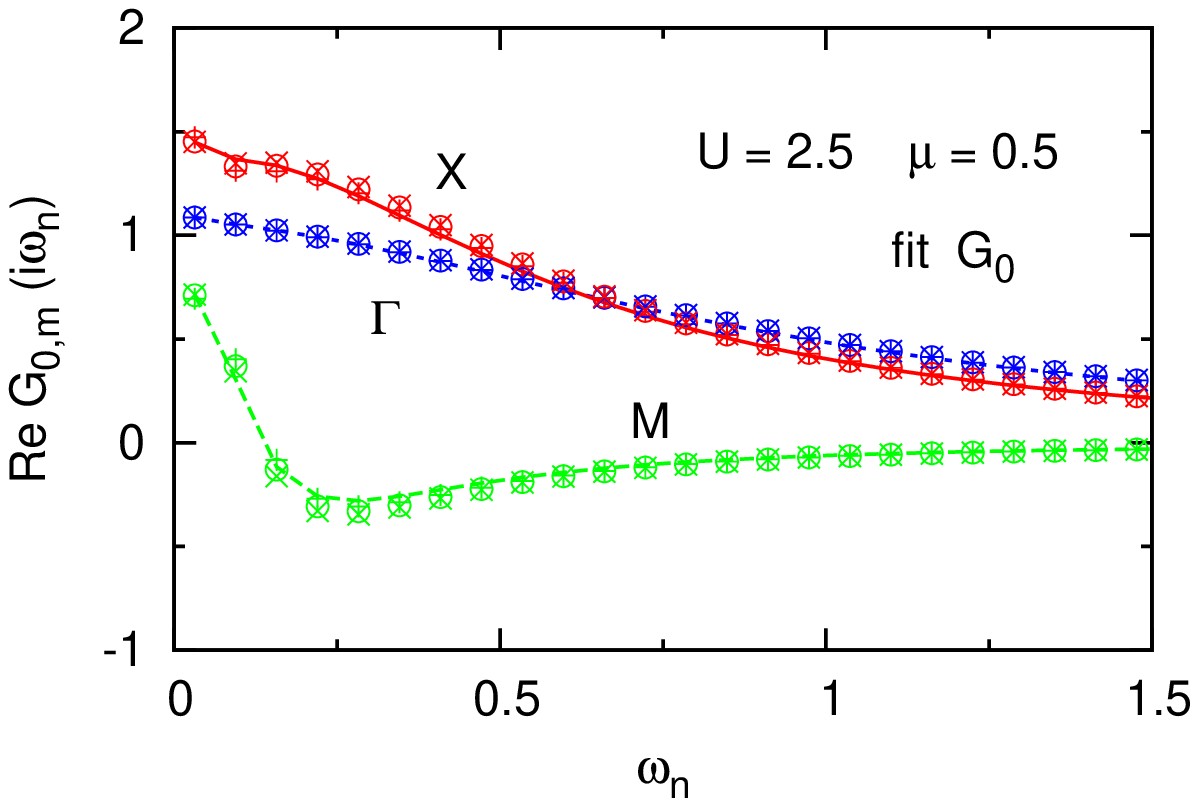} 
\includegraphics[width=4.0cm,height=6.5cm,angle=-90]{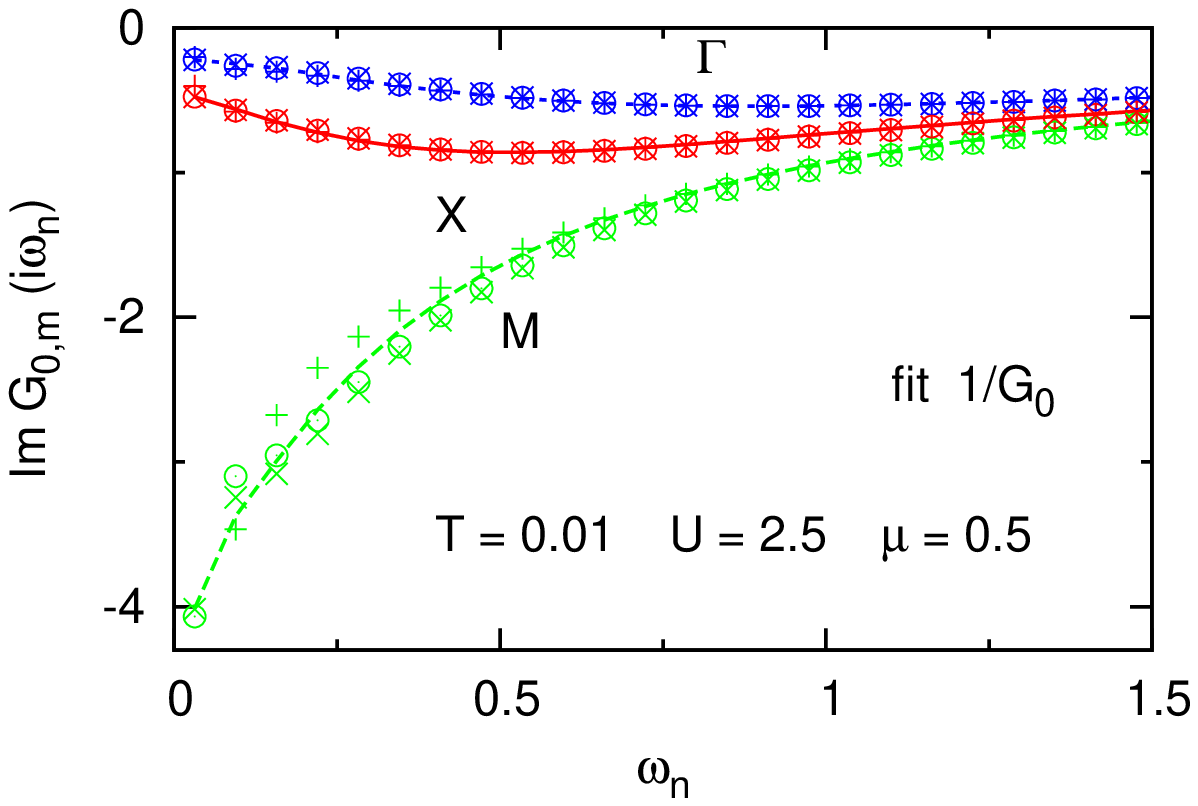} 
\includegraphics[width=4.0cm,height=6.5cm,angle=-90]{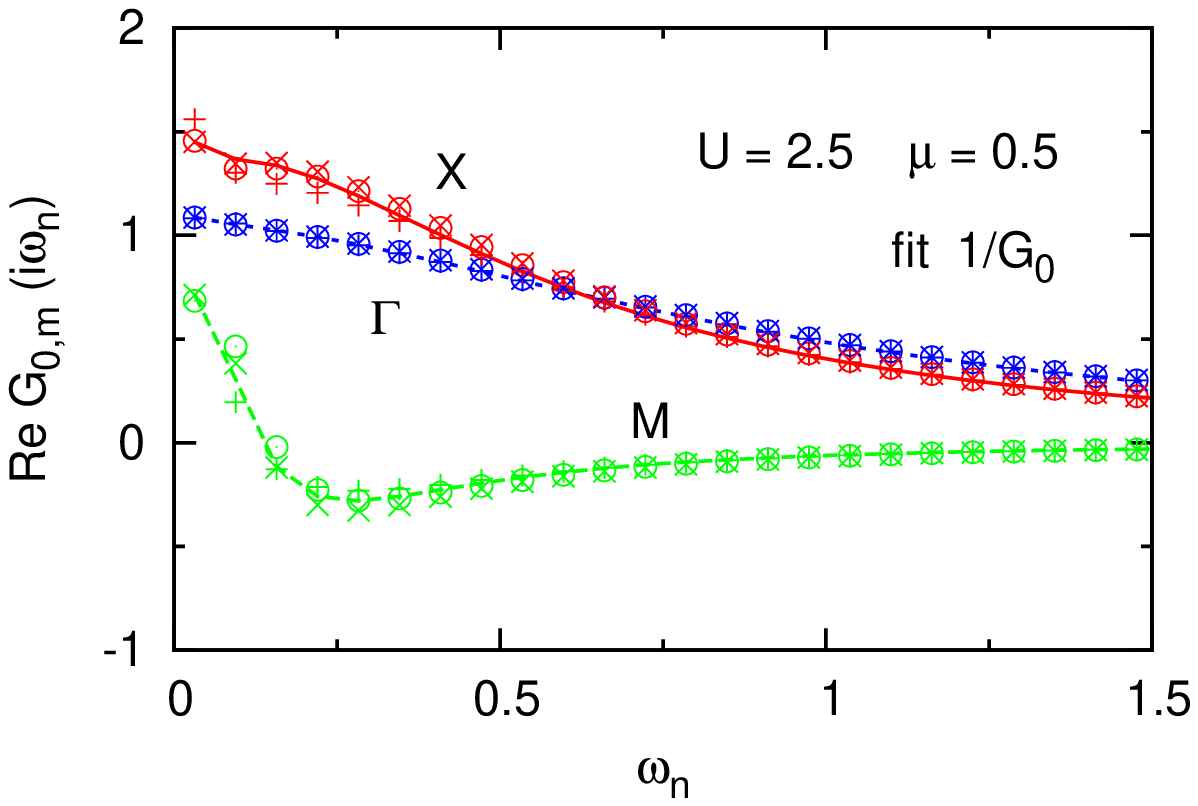} 
\end{center}\vskip-3mm
\caption{(Color online)
Discretization of bath Green's function for $2\times 2$ square lattice as in 
Section 3.5. Solid and dashed curves: lattice bath Green's function 
$G_{0,m}(i\omega_n)$ (as in Figure~\ref{DOPE4.Fig2}). 
Upper two panels: real and imaginary parts of cluster Green's function 
$G_{0,m}^{cl}(i\omega_n)$ obtained via minimization of $\vert G_0-G_0^{cl}\vert$, 
 see Eq.~(\ref{diff}), 
using two bath levels ($n_s=12)$ and weight functions $W_n=1/\omega_n^N$ with 
$N=0$ (+), $N=1$ (o), and $N=2$ (x).  
Lower two panels: analogous fits via minimization 
of $\vert 1/G_0 - 1/G_0^{cl}\vert$, see Eq.~(\ref{diff'}).  
}\label{app.2}\end{figure}

\begin{figure}  [t!] 
\begin{center}
\includegraphics[width=4.5cm,height=6.5cm,angle=-90]{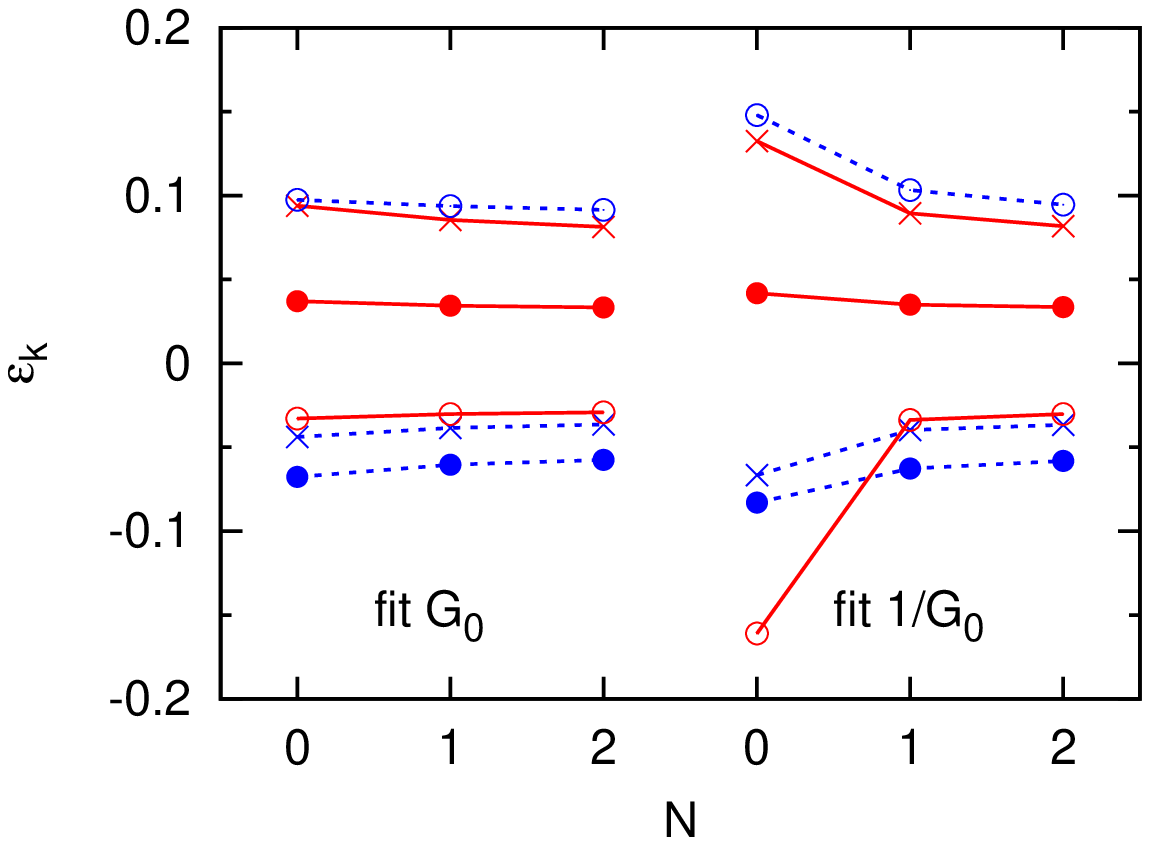} 
\includegraphics[width=4.5cm,height=6.5cm,angle=-90]{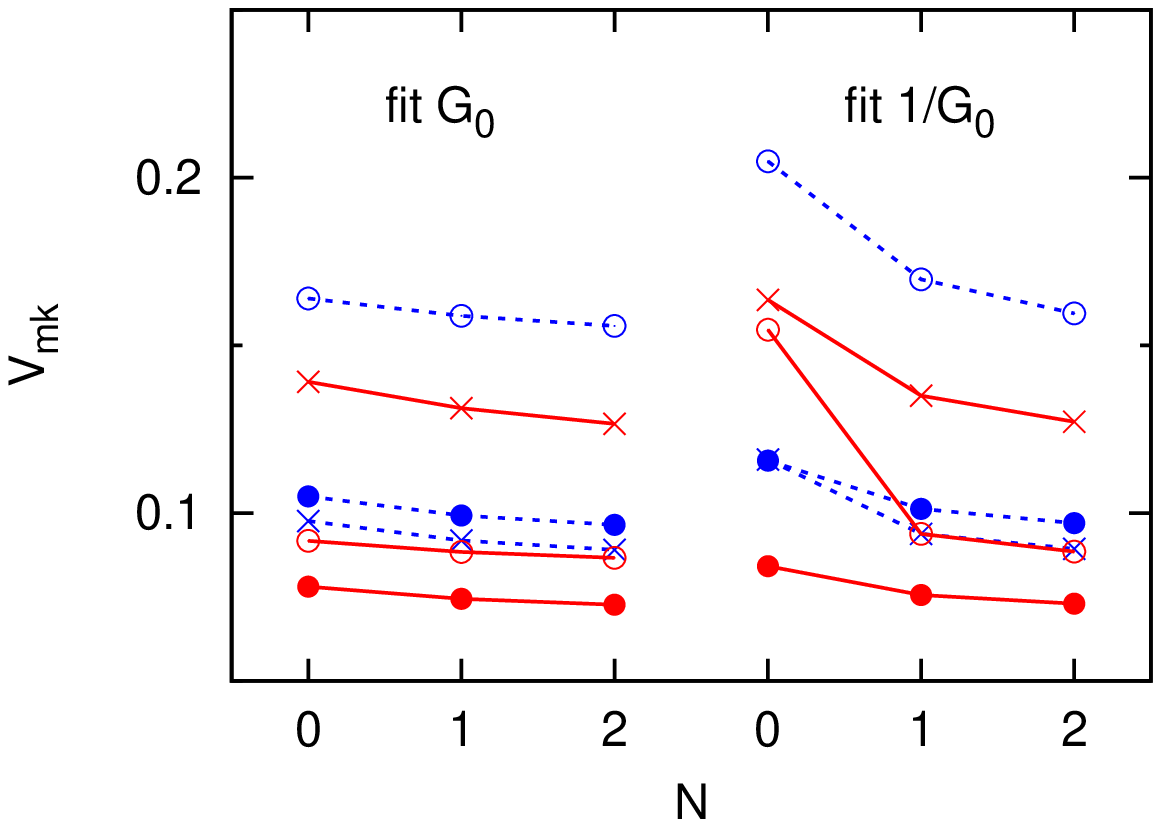} 
\end{center}
\caption{(Color online)
Bath levels $\varepsilon_k$ (top panel) and hybridization matrix elements $V_{mk}$ 
(lower panel) for fits in Figure~\ref{app.2}. Solid dots: $\Gamma$, empty dots:
$M$, x symbols: $X$. Left: minimization of $\vert G_0 - G_0^{cl}\vert$,  
see Eq.~(\ref{diff}); right: minimization of  $\vert 1/G_0 - 1/G_0^{cl}\vert$, 
 see Eq.~(\ref{diff'}). The two bath levels are denoted by the red and blue colors. 
}\label{epsk.2}\end{figure}

The fits shown in Figure~\ref{app.1} are for a fixed set of lattice bath Green's 
functions. We have also carried out complete self-consistency cycles using the 
six minimization procedures specified above. The resulting self-energies are in 
very good agreement. We therefore conclude that the arbitrariness caused by 
the discretization of the bath Green's function is of minor importance in ED DMFT.   
On the whole, iteration to self-consistency is found to be most reliable and 
efficient for the distance function in Eq.~(\ref{diff}), with $W_n=1/\omega_n$.

\vskip8mm
\centerline {\bf A.2. Square Lattice} 
\vskip5mm

In Figure~\ref{DOPE4.Fig2}, we have illustrated the quality of the discretization of 
the bath Green's function of the $2\times2$ square lattice for the three independent 
molecular orbitals corresponding to $\Gamma$, $M$, and $X$. The minimization 
was performed using Eq.~(\ref{diff}) with the weight function $W_n=1/\omega_n$.  
Figure~\ref{app.2} shows results of analogous discretizations using Eq.~(\ref{diff}) 
as well as Eq.~(\ref{diff'}), for weight functions $W_n=1/\omega_n^N$ with 
$N=0,1,2$. The impurity levels $\varepsilon_m$ are kept fixed at the values 
determined by the asymptotic behavior of $G_{0,m}$, as indicated in Eq.~(\ref{Gasym2}).
Thus, there are four fit parameters per molecular orbital component $G_{0,m}$. 

As in Figure~\ref{app.1}, there is little difference among the six minimizations.
Those for $N=1$ and $N=2$ are slightly more accurate than for $N=0$. 
Figure~\ref{epsk.2} shows the variation of the bath levels $\varepsilon_k$ and 
hopping matrix elements $V_{mk}$. Although there is considerable variation, in 
particular between $N=1$ and $N=0$, the net effect on the quality of the fit of 
$G_{0,m}$ using only two bath levels is small. Evidently, there is important 
compensation between the magnitudes of $\varepsilon_k$ and $V_{mk}$. 
As in Figure~\ref{epsk.1}, the differences between the results for $N=1$ and $N=2$ 
are much smaller than for $N=1$ and $N=0$. Also, fitting $G_0$ gives much more 
stable bath parameters than fitting $1/G_0$.

In Figs.~\ref{app.1} and \ref{app.2} the bath and cluster Green's functions are 
compared since these are the important quantities used as input in the exact 
diagonalization. If the lattice and cluster hybridization functions 
[see Eqs.~(\ref{Delta}) and (\ref{Delta0})] are compared instead, the differences
at large $\omega_n$ are slightly larger since, in contrast to $G_0$, these functions 
do not have unit asymptotic weight.



%

\end{document}